%% file: arxiv_v2.tex
\documentclass[nonblindrev]{informs3}

\usepackage[table, dvipsnames]{xcolor}
\usepackage{xargs}

\usepackage{tikz} 

\usepackage[linesnumbered,ruled,vlined]{algorithm2e}

\usepackage{float}

\usepackage{natbib} 
 \bibpunct[, ]{(}{)}{,}{a}{}{,}%
 \def\bibfont{\small}%
 \def\newblock{\ }%
\usepackage{multirow}
\usepackage{caption}
\usepackage{multirow}
\usepackage{longtable}
\usepackage{supertabular}

\usepackage{colortbl}
\usepackage[norefpage]{nomencl}

\usepackage{graphicx}
\usepackage{graphics}
\usepackage{qtree}
\usepackage{amsmath}
\usepackage{amsfonts,latexsym,amssymb}
\usepackage{ifthen}
\usepackage[gen]{eurosym}
\usepackage{colortbl}
\usepackage{wasysym}
\usepackage{multirow}
\usepackage[caption=false]{subfig}

\usepackage[colorinlistoftodos,prependcaption,textsize=small]{todonotes}
\theoremstyle{remark}

\newtheorem{innercustomthm}{Theorem}
\newenvironment{customthm}[1]
{\renewcommand\theinnercustomthm{#1}\innercustomthm}
{\endinnercustomthm}

\newenvironment{proof1}{\paragraph{Proof:}}{\hfill$\square$}
\makeatletter
\def\munderbar#1{\underline{\sbox\tw@{$#1$}\dp\tw@\z@\box\tw@}}
\makeatother

\DeclareMathOperator{\R}{\mathbb{R}}

\usepackage{graphicx}

\usepackage[ruled,vlined]{algorithm2e}
\usepackage{textcomp}
\usepackage{amssymb}
\usepackage{amsmath}
\usepackage{xparse}
\usepackage{todonotes}
\usepackage[american]{babel}
\usepackage{comment}
\usepackage{latexsym}
\usepackage{mathtools}
\usepackage{etoolbox}
\usepackage{siunitx}
\usepackage{xfrac}

\usepackage{xspace}

\DeclareMathOperator{\Ical}{\mathcal{I}}
\DeclareMathOperator{\Vcal}{\mathcal{V}}
\DeclareMathOperator{\Ocal}{\mathcal{O}}
\DeclareMathOperator{\Acal}{\mathcal{A}}
\DeclareMathOperator{\Scal}{\mathcal{S}}
\DeclareMathOperator{\bbbr}{\mathbb{R}}
\newboolean{includeHidden}
\setboolean{includeHidden}{false}
\newcommand{\hide}[1]{\ifthenelse{\boolean{includeHidden}}{{\tiny\textbf{HIDDEN:~}#1}}{}}

\TheoremsNumberedThrough     % Preferred (Theorem 1, Lemma 1, Theorem 2)
\EquationsNumberedThrough    % Default: (1), (2), ...
\MANUSCRIPTNO{}

\begin{document}
%%%%%%%%%%%%%%%%
\RUNAUTHOR{Bichler, Fichtl, Oberlechner}

\RUNTITLE{Computing Bayes Nash Equilibrium Strategies}

\TITLE{Computing Bayes Nash Equilibrium Strategies in Auction Games via Simultaneous Online Dual Averaging}
\ARTICLEAUTHORS{%
	\AUTHOR{
		Martin Bichler\footnote{Email: \texttt{bichler@in.tum.de}}, 
		Maximilian Fichtl\footnote{Email: \texttt{max.fichtl@tum.de}}, 
		Matthias Oberlechner\footnote{Email: \texttt{matthias.oberlechner@tum.de}}}
	\AFF{Department of Computer Science, Technical University of Munich, 85748 Garching, Germany}
}

\ABSTRACT{%
Auctions are modeled as Bayesian games with continuous type and action spaces. Determining equilibria in auction games is computationally hard in general and no exact solution theory is known. We introduce an algorithmic framework in which we discretize type and action space and then learn distributional strategies via online optimization algorithms. One advantage of distributional strategies is that we do not have to make any assumptions on the shape of the bid function. Besides, the expected utility of agents is linear in the strategies. It follows that if our optimization algorithms converge to a pure strategy, then they converge to an approximate equilibrium of the discretized game with high precision. 
Importantly, we show that the equilibrium of the discretized game approximates an equilibrium in the continuous game. In a wide variety of auction games, we provide empirical evidence that the approach approximates the analytical (pure) Bayes Nash equilibrium closely. This speed and precision is remarkable, because in many finite games learning dynamics do not converge or are even chaotic. In standard models where agents are symmetric, we find equilibrium in seconds. While we focus on dual averaging, we show that the overall approach converges independent of the regularizer and alternative online convex optimization methods achieve similar results, even though the discretized game neither satisfies monotonicity nor variational stability globally. The method allows for interdependent valuations and different types of utility functions and provides a foundation for broadly applicable equilibrium solvers that can push the boundaries of equilibrium analysis in auction markets and beyond. 
}
\KEYWORDS{auctions, Bayes-Nash equilibrium, online convex optimization}

%\acknowledgements{We gratefully acknowledge financial support from the German National Science Foundation through Grant DFG BI-1057/I-9.}

%Ali Abbas, 
%Manel Baucells, behavioral economist
%David Bell, HBS, MAUT, decision analysis
%David Brown, Duke, dynamic pricing, dynamic programs 
%Soo-Haeng Cho, CMU, operations management
%Yael Grushka-Cockayne, Virginia, MCDA
%Casey Lichtendahl, Virginia, Decision Analysis
%Peng Sun,Duke, Auctions and game theory
%Ilia Tsetlin, INSEAD, Forecasting 
%Canan Ulu.

\maketitle
{\begin{center}  \today \end{center}}

%\tableofcontents
%\newpage

\section{Introduction}

Auction games are arguably some of the most important applications of game theory and they can be analyzed as continuous-type, continuous-action Bayesian games. Bidders' valuations or types in such an auction game are drawn from some continuous distribution and they can choose from a continuous range of possible actions (or bids). Early on, Nobel Prize laureate \citet{vickrey1961counterspeculation} showed how to derive a Bayes-Nash equilibrium (BNE) strategy in a single-object first-price auction in the independent-private values (IPV) model with symmetric bidders and quasi-linear utility functions. The first-order conditions together with the assumption of symmetric bidding behavior lead to an ordinary differential equation, which has a closed-form solution for the BNE bidding strategy. 

The BNE provides a principled way to think about strategic interaction in auctions and a prescriptive model how rational bidders should behave. Unfortunately, deviations from the benchmark model by \citet{vickrey1961counterspeculation} lead to challenges in the equilibrium analysis \citep{mcafee1987auctions}. For example, when the valuations of potential bidders are interdependent, then the system of first-order partial differential equations that characterizes a BNE often becomes intractable \citep{campo2003asymmetry}. 
Computing Nash equilibria (NE) in complete-information finite games is already known to be PPAD-hard. However, computing of exact Bayesian Nash equilibria (BNE) can even be PP-hard, a complexity class that is clearly intractable \citep{cai2014simultaneous}. 
Overall, the analytical derivation of BNE strategies has been elusive for all but very simple auction games. 
Even existence of BNE has only been shown for a limited set of auction models \citep{jackson2005existence}. As a result, Bayes-Nash equilibrium analysis remained in the realm of academic research and it is rarely used by bidders in real-world auctions.

There have been a few approaches to develop numerical techniques for specific environments. 
For example, \citet{armantier2008approximation} introduced a BNE-computation method that is based on expressing the Bayesian game as the limit of a sequence of complete-information games.
\citet{rabinovich2013ComputingPureBayesianNash} study best-response dynamics in auctions with finite action spaces, while \citet{bosshardComputingBayesNashEquilibria2018} contribute an iterated best-response algorithm for combinatorial auctions with an elaborate empirical verification method. 
Recently, \citet{bichler2021npga} introduced a versatile technique to compute approximate Bayes-Nash equilibria (BNE) in a variety of auction models using neural networks and self-play. 
Their use of neural networks and evolutionary strategies leads to a relatively complex algorithm, which leverages massive parallelization on GPU hardware. In all prior approaches, numerical techniques are required to certify that the strategies found are indeed an approximate BNE and there are no guarantees that the process converges or that a BNE emerges if the algorithm converges. All these techniques are computationally expensive, even for simple symmetric auction models.

We introduce an algorithmic framework based on a discretization of the type and action space, in which we can use online convex optimization to learn \emph{distributional strategies} \citep{Milgrom1985}, which are a form of mixed strategies for Bayesian games. In contrast to learning algorithms for complete-information games, auction games require us to consider the prior type distributions.  
The distributional strategies allow us to derive gradients and implement gradient-based optimization algorithms without relying on neural networks with self-play as in \citet{bichler2021npga}. 
In \textit{Simultaneous Online Dual Averaging} (SODA) we focus on dual averaging as learning algorithm, which is one of the most effective online convex optimization algorithms. However, empirically we show that alternative algorithms such as mirror ascent or the Frank-Wolfe algorithm achieve very similar results in a wide variety of auction models and contests. 
SODA allows for interdependent types and different utility functions (e.g., risk aversion), which makes it a very fast and generic algorithm compared to existing approaches. It is straightforward to incorporate risk aversion or other behavioral motives in the utility function, which leads to complications in analytical derivations. Importantly, it does not make any assumptions on the parametric form of the bid function, allowing us to find non-smooth equilibria as well. 
An advantage of dual averaging is that the expected utility is linear in the distributional strategies as we show, which allows us to show that if the algorithm converges to a pure strategy, then it has to be an equilibrium of the discretized game. 
This is an advantage over prior numerical methods, which rely on numerical estimates of the utility loss to certify an approximate equilibrium. 
Importantly, we can show for single-object auctions that the distributional $\varepsilon$-BNE found in the discretized auction approximates a continuous equilibrium, if one exists. Note that there are examples where equilibria exist only in the discretized game and not in the continuous game \citep{jackson2005existence}.

Ex ante conditions that certify when gradient-based optimization algorithms converge to equilibrium even in finite, complete-information games turned out to be challenging. A number of recent results on matrix games showed that gradient-based algorithms either circle, diverge, or are even chaotic \citep{sanders2018prevalence}. Independent learning dynamics do not generally obtain a Nash equilibrium \citep{benaimMixedEquilibriaDynamical1999}. Actually, the study of gradient dynamics in games is akin to studying dynamical systems and characterizing environments where gradient dynamics converge to a Nash equilibrium (if one exists) can be arbitrarily complex \citep{andrade2021learning}. The analysis of Bayesian games with continuous type and action spaces is difficult: for a convergence analysis we need to study the properties of an expected utility function that is based on the characteristics of an unknown equilibrium bid function. While this is not the case in the discretized version of the game, ex-ante guarantees are still very challenging as we discuss in Section \ref{sec:convergence}. For example, we show that conditions such as monotonicity or variational stability do not hold globally in the discretized game. Yet, given that the algorithms are fast for standard auction and contest models, the ex-post verification we get with SODA is very useful. 

We provide extensive experimental results where we approximate the analytical pure BNE closely in a wide variety of auction games and contests. 
We could actually compute close approximations of the BNE with only a few bidders in seconds even for complex core-selecting combinatorial auctions. If we restrict ourselves to independent private values, we can solve large instances with dozens of bidders within seconds. 
This allows for a quick exploration of auction models with different priors or different utility functions.

The wide range of environments where SODA converges is remarkable. We illustrate results of SODA for environments where an analytical solution is known, but also provide equilibrium strategies for models where no Bayes Nash equilibrium was available so far. Experimental results are reported for single-object auctions with interdependent valuations, combinatorial auctions with independent and interdependent values, combinatorial split-award auctions, all-pay auctions and Tullock contests. 

Convergence of SODA to equilibrium is guaranteed, if the utility gradients are monotone or they satisfy relaxed notions such as variational stability \citep{geiger2013theorie, mertikopoulos2019learning, grossmann2007numerical}. With monotone utility gradients, the expected utility function is concave. Without knowing the parametric form of the bid function it is difficult to understand a priori whether concavity of the expected utility is satisfied in a specific Bayesian auction game. Numerical analysis with parametric assumptions on the bid function and the distribution function suggest that the expected utility function of several well-known auction games is concave or pseudo-concave for large ranges of the bid space, which explains the surprisingly positive results compared to several recent studies showing that gradient-based optimization algorithms and the resulting dynamics often do not converge in finite normal-form games. 

Overall, the paper shows that important applications of equilibrium computation problems in auctions and contests are tractable and we can find approximate equilibria quickly. This provides a foundation for numerical tools that allow us to push the boundaries of equilibrium analysis. Tools of this sort will prove useful for market designers to understand specific market rules, but also for bidders to study strategic interaction in high-stakes auctions.

Section \ref{sec:related} provides a brief overview of related literature. Then, Section \ref{sec:model_alg} introduces the notation and the algorithm, and discusses convergence and scalability. Section \ref{sec:auctions} reports results for various single-object and combinatorial auction models, before Section \ref{sec:conclusion} provides conclusions.

\section{Related Literature}\label{sec:related}

Our research primarily relates to the extensive economic literature on equilibrium in auctions and contests and to the literature on equilibrium learning.

\subsection{Equilibrium in Auction Games}
Our paper primarily deals with Bayesian auction games where type- and action-spaces are continuous. A first question is whether BNE always exist in such games. Auctions and contests are prime applications, central to economic theory. For finite, complete-information games, we know that a mixed Nash equilibrium exists \citep{nash1950equilibrium} and that the computation is generally PPAD-hard \citep{daskalakisComplexityComputingNash2009}. \citet{glicksberg1952further} extended the existence result to games with continuous and compact action sets. For Bayesian games with continuous action space, \citet{jackson2005existence} provide assumptions for the existence of equilibria in distributional strategies. For example, first-price and second-price single-unit auctions, all-pay auctions, double auctions, and multi-unit discriminatory or uniform price auctions were shown to have an equilibrium in distributional strategies. 
It is interesting to note that there are auction models where there is no Bayesian Nash equilibrium of the continuous game, but there are equilibria in the discretized game \citep{jackson2002communication}. 
Overall, we neither know of the existence of Bayes-Nash equilibria in general continuous-type and -action auction games, nor do we know how hard they are to find if they exist. 
\citet{cai2014simultaneous} showed that finding an exact BNE in specific simultaneous auctions for individual items is at least hard for PP, a complexity class higher than the polynomial hierarchy and close to PSPACE, and we know little about the complexity of finding BNE in other multi-item auctions. 

\subsection{Equilibrium Learning}
Our research is best situated in the literature on equilibrium learning. The theory of learning in games examines what kind of equilibrium arises as a consequence of a process of learning and adaptation, in which agents are trying to maximize their payoff while learning about the actions of other agents \citep{fudenberg2009learning}. Fictitious play is a natural method by which agents iteratively search for a pure Nash equilibrium and play a best response to the empirical frequency of play of other players \citep{brownIterativeSolutionGames1951}. 
Several algorithms have been proposed based on best or better response dynamics. Besides, gradient-based online optimization algorithms have been proposed for normal-form games \citep{singhNashConvergenceGradient2000, zinkevich2003online}.

While such online gradient ascent algorithms lead to zero regret for the participating agents, their strategies do not generally converge. Even in simple matching pennies games, the gradient dynamics circle \citep{bowling2005convergence}. Hence, no-regret learning algorithms do not find a BNE in general games. 
However, due to their simplicity, learning algorithms have been used to solve games for a long time. While there is no comprehensive characterization of games that are ``learnable,'' and one cannot expect that uncoupled dynamics lead to Nash equilibrium in all games \citep{hart2003uncoupled}, there are some important results regarding no-regret learners. First, one can distinguish between internal (or conditional) regret and a weaker version called external (or unconditional) regret. External regret compares the performance of an algorithm to the best single action in retrospect, while internal regret allows one to modify the online action sequence by changing every occurrence of a given action with an alternative one. 
For learning rules that satisfy the stronger no-internal regret condition, the empirical frequency of play converges to the game's set of correlated equilibria \citep{foster1997calibrated,hart2000simple,stoltz2007learning}.
The set of correlated equilibria (CE) is a nonempty convex polytope that contains the convex hull of the game's Nash equilibria. The coordination in CE can be implicit via the history of play \citep{foster1997calibrated,stoltz2007learning}. On the other hand, algorithms that are no-external-regret learners converge by definition to the set of coarse correlated equilibria (CCE). 
This set, in turn, contains the set of CE such that we get  $\text{NE} \subset \text{CE} \subset \text{CCE}$. In contrast to correlated equilibria, coarse correlated equilibria may contain strictly dominated (pure) strategy profiles with positive probability \citep{viossat2013no}, which makes them a relatively weak solution concept. 

Recent work shows that gradient dynamics often do not converge \citep{daskalakis2010learning, vlatakis2020no}. Standard learning algorithms can cycle, diverge, or even be chaotic in zero-sum games \citep{mertikopoulos2018cycles, baileyMultiplicativeWeightsUpdate2018, cheung2020chaos}. Actually, \citet{sanders2018prevalence} suggest that chaos is, in fact, typical behavior for more general matrix games. Simple examples where reasonable gradient-based methods cannot converge leave little hope for general gradient-based methods in the broader class of differential games \citep{letcher2019differentiable}. Notably, the dynamics of general matrix games can be arbitrarily complex and hard to characterize a priori \citep{andrade2021learning}. 
On the positive side, there is a long literature on monotonicity conditions that guarantee convergence to a Nash equilibrium \citep{kinderlehrer2000introduction, grossmann2007numerical, geiger2013theorie, mertikopoulos2019learning}. Apart from monotonicity, \citet{even2009convergence} introduce the notion of socially concave games. These are games where a convex combination of all agents' utilities exists that is concave, and each agent's utility is convex in the other agents' strategies. In contrast to monotonicity and variational stability, which suffices for convergence of the last iterate, social concavity only implies convergence of the mean of iterates to a BNE. However, it is also a strong assumption on the utility functions. 

A large part of the literature on equilibrium learning has focused on complete-information games \citep{foster1997calibrated, hart2000simple,jafari2001no,stoltz2007learning,hartlineNoRegretLearningBayesian2015,syrgkanis2015fast, foster2016learning}. 
Uncertainty about other players has also received attention. For example, there is work on Stackelberg games with uncertainty about the follower \citep{balcan2015commitment}, and there is a stream of literature on imperfect-information games as in Poker (see for example \cite{sandholm2015abstraction, brown2019SuperhumanAIMultiplayer}). The literature is too large to provide a comprehensive survey here. 
\textit{Bayesian games with continuous type and action spaces} as they are used to model auctions or contests are less well studied. Solving such problems is challenging because it requires learning a bid function over infinitely many types. Such problems can be formulated as systems of differential equations. We lack a solution theory for such problems in general. 
Given how hard it is to find Bayes-Nash equilibria even in simultaneous multi-object auctions in the worst case \citep{cai2014simultaneous}, it is far from obvious that gradient-based algorithms can find a BNE in continuous-type and -action Bayesian games. It is not even clear how gradient dynamics would be implemented in games with continuous type space. 

Neural Pseudogradient Ascent (NPGA) by \citet{bichler2021npga} was recently published to address equilibrium computation in auction games: it is the first numerical method to compute BNE in a wide variety of auction games, including multi-object auctions with interdependent types. Therefore, it will serve as our benchmark when we report our experimental results. NPGA uses neural networks as a bid function to be learned via self-play. The authors employ evolutionary strategies as a smoothing technique to deal with the discontinuities of the ex-post utility function, which allows them to compute BNE in a finite-dimensional parameter space of neural networks. However, the use of neural networks and specific training methods makes it hard to derive theoretical guarantees. Moreover, it takes an expensive empirical validation procedure to verify if the strategies found by the algorithm are approximate BNE. 

Our technique is quite different in that we discretize the type and action spaces and implement gradient dynamics in the discretized version of the game without using neural networks. We apply various well-known online learning algorithms to the discretized game. Our focus is on the dual averaging algorithm with entropy regularization, as it is often the method of choice in theoretical analyses \citep{mertikopoulos2019learning}, and it enjoys particularly good regret bounds \citep{shalev2007online}. While NPGA searches for pure Bayesian Nash equilibria, we compute distributional strategies in a discretized version of the game. Our technique is much faster for environments with a few players and items and can solve equilibria in symmetric auction games in seconds. Compared to NPGA, SODA is much easier to implement. Additionally, as a consequence of the no-regret property of our algorithm, if SODA converges, the limit point must necessarily be a Bayes-Nash equilibrium. Importantly, we can bound the approximation error to the original auction game with continuous type and action space. These theoretical guarantees are a significant advantage over NPGA because we do not require an expensive experimental verification of the solution.

\section{Model and Algorithm} \label{sec:model_alg}

We will first introduce the necessary notation before we discuss a small illustrative example, and then describe the algorithm more generally.

\subsection{Notation}\label{sec:notation}

An incomplete-information or \emph{Bayesian game} is given by a sextuplet $G = (\mathcal I, \mathcal V, \mathcal O, \mathcal A, f, u)$. Here $\mathcal I = \{1, \dots, n\}$ denotes the set of \emph{agents} participating in the game. Agent $i$'s private \emph{observation} is then given as a realization $o_i \in \mathcal O_i$, with $\mathcal O = \mathcal O_1 \times \cdots \times \mathcal O_n$ being the set of possible observation profiles. Similarly, $\mathcal V$ denotes the set of ``true'' but possibly unobserved valuations. Crucially, we make this distinction to model interdependencies in settings beyond purely private values or purely common values. Based on the observation $o_i$, the agent chooses an action, or \emph{bid}, $b_i \in \mathcal A_i$, and the set of possible action profiles is given by $\mathcal A = \mathcal A_1 \times \cdots \times \mathcal A_n $.
The joint probability density function $f: \mathcal O \times \mathcal V \rightarrow \mathbb{R}_{\geq 0}$ describes an atomless \emph{prior} distribution over agents' types, given by tuples $(o_i, v_i)$ of observations and valuations. We make no further restrictions on $f$, thus allowing for arbitrary correlations. $f$ is assumed to be common knowledge and we will denote its marginals by $f_{v_i}$, $f_{o_i}$, etc.; its conditionals by $f_{v_i\vert o_i}$, etc.; and its associated probability measure by $F$. 

For each possible action and valuation profile, the vector $u=(u_1, \dots, u_n)$ of $F$-integrable, individual \emph{(ex-post) utility} functions $u_i: \mathcal A \times \mathcal V_i \rightarrow \bbbr$ assigns the game outcome to each player.
\emph{Ex-ante}, before the game, agents neither have observations nor valuations, only knowledge about $f$. In the \emph{interim} stage, agents additionally observe $o_i$ providing (possibly partial or noisy) information about their own valuations $v_i$. Full access to the outcomes $u(v,b)$ is given only after taking actions (\emph{ex-post}). In our formulation, we do not assume explicit ex-post access to any values (e.g., $v_i, v_{-i}, b_{-i}$) beyond the outcome $u$ itself. An index $-i$ denotes a partial profile of all agents but agent $i$.

\input{tables/interdependent_models}

Taking an ex-ante view, players are tasked with finding strategies that link observations and bids. 
Instead of pure strategies, which are measurable functions $ \beta_i : \mathcal O_i \rightarrow \mathcal A_i $ that map observations to bids, we are interested in distributional strategies that induce a probability measure on the space of observations and actions \citep{Milgrom1985}.

\begin{definition} \label{def:distr_strat}
	In the private values model, a distributional strategy for player $ i $ is a probability measure $ \sigma $ on $ \mathcal O_i \times \mathcal A_i $ for which the marginal distribution on $ \mathcal O_i $ is $ f_{o_i} $. 
	Formally, the marginal condition can be written as $ \sigma(O \times \mathcal A_i) = F_{o_i}(O) $ for all measurable sets $ O \subset \mathcal O_i $. 
	When players adopt distributional strategies $ (\sigma_1,...,\sigma_n) $ the expected utility is given by
	\begin{equation}\label{eq:dist_strat}
		\tilde u_i(\sigma_1,...,\sigma_n) = \int u_i(b,o_i) \sigma_1(db_1 \vert o_1) ... \sigma_n(db_n \vert o_n) F(do)
	\end{equation}
\end{definition}
The strategy profile $ (\sigma_1,...,\sigma_n) $ is a $ \varepsilon $-Bayes-Nash equilibrium ($ \varepsilon $-BNE) if no bidder $ i $ can increase its utility by more than $ \varepsilon \geq 0 $ by unilaterally deviating from its distributional strategy $ \sigma_i $, i.e.,
\begin{equation}
	\tilde u_i(\sigma'_i,\sigma_{-i}) - \tilde u_i(\sigma_i,\sigma_{-i}) \leq \varepsilon \quad \forall \sigma'_i \text{ and } \forall i \in \Ical,
\end{equation}
where $ \sigma_{-i} $ denotes the partial strategy profile for all bidders but bidder $ i $. If $ \varepsilon = 0 $, the strategy profile corresponds to a Bayes-Nash equilibrium (BNE).

The primary Bayesian games we'll consider are \emph{sealed-bid auctions} on $I$ indivisible items. In general combinatorial auctions we thus have a set $\mathcal K$ of possible \emph{bundles} of items and the valuation- and action-spaces are therefore of dimension $\lvert{\mathcal K}\rvert=2^I$. In the \emph{private values} setting, we always have $o_i = v_i$; in the \emph{common values} setting, there is some unobserved constant $v_{c} = v_1 = \dots = v_n$ and the $o_i$ can be considered noisy measurements of $v_c$. Mixed settings are likewise possible. 
In any case, based on bid profile $b$, an \emph{auction mechanism} will determine two things: An allocation $x=x(b) = (x_1, \dots x_n)$ which constitutes a partition of the $m$ items, where bidder $i$ is allocated the bundle $x_i$; and a price vector $p(b) \in \bbbr^n$, where $p_i$ is the monetary amount bidder $i$ has to pay in order to receive $x_i$. Formally, one may consider the individual allocations to be one-hot-encoded vectors $x_i \in \{0,1\}^{|\mathcal K|}$.
In the standard risk-neutral model the utilities $u_i$ are then described by \emph{quasilinear} payoff functions $u_i^{QL}(b, v_i) = \left(x_i(b)\cdot v_i - p_i(b)\right)$, i.e., by how much players value their allocated bundle minus the price they have to pay.

An extension to this basic setting includes \emph{risk-aversion}. 
Here, we model risk-aversion via utilities $u^{RA} = \left(u^{QL}\right)^\rho$ where $\rho \in (0,1]$ is the risk attitude; $\rho=1$ describes risk-neutrality, smaller values lead to strictly concave, risk-averse transformations of $u^{QL}$.
Risk aversion is an established way to explain why in field studies of single-object first-price sealed-bid (FPSB) auctions, bidders  bid higher than their risk-neutral counterparts in analytical BNE \citep{bichler2015split}. However, different types of utility functions are possible.

%Due to the known computational hardness of computing NE and BNE, one is often interested in relaxations of equilibria that may be easier to find in some circumstances. For example, in \emph{local BNE}, the loss requirement is relaxed to only consider best responses from a neighborhood of the equilibrium strategy profile: We call a strategy profile $\beta^*$ a \emph{local ex-ante BNE}, iff there exists an open set $\emptyset \neq W_i \subset \Sigma_i$ such that $\beta_i^* \in W_i$ and $\tilde{u}_i(\beta_i^*,\beta^*_{-i}) \geq \tilde{u}_i(\beta'_i,\beta^*_{-i})$ for all agents and all alternative strategies $\beta_i' \in W_i$. If all utility functions $u_i$ are strictly concave in $i$'s action, the game admits a unique global BNE \cite{ui2016bayesian} and no other local BNE.

%Finally, symmetric models are prevalent in auction theory \cite{krishnaAuctionTheory2009}. We will call a Bayesian game \emph{symmetric}, if all players' $i,j\in\mathcal I$ marginal prior type distributions are identical, i.\,e.\ $f_{v_i, o_i} = f_{v_j,o_j}$ (but not necessarily independent), as are their individual utilities (almost surely, up to tie-braking): $u_i(\beta_i, \beta_{-i}) = u_j(\beta_i; \beta_{-i})$ with probability 1. The literature primarily discusses equilibria which are likewise symmetric, i.e., $\beta^* = (\beta^*_1, \beta^*_1, \dots \beta^*_1)$ \cite[Chapter 2.1]{krishnaAuctionTheory2009}. 

\subsection{An Illustrative Example}
Before we introduce the model and our algorithm in general, let us discuss a simplified setting: a single-object first-price sealed-bid auction with two symmetric bidders.
We focus on the IPV model, where both bidders $ i \in \{1,2\}$ observe their true valuation $ o_i = v_i $ for an item, which is drawn independently according to some prior marginal distribution $ F_{o_i} $ from a real interval $ \mathcal O_i \subset \bbbr $. 
Therefore, the common prior is the product of the two marginal distributions.
After both bidders submit their bids $ (b_1, b_2) $  the (ex-post) utility of player 1 (analogously for player 2) is given by
\begin{equation}
	u_1(b_1, b_2, o_1) = 
	\begin{cases} 
		o_1 - b_1 & \text{ if } b_1 > b_2 \\ \tfrac{1}{2}(o_1 - b_1) & \text{ if } b_1 = b_2 \\ 0 & \text{ else} 
	\end{cases}.
\end{equation}
The bidders are risk neutral and want to maximize their expected profits/utilities.
To analyze such auction formats, we are interested in equilibrium strategies, where no bidder has the incentive to deviate from the current strategy. 
Instead of pure strategies $ \beta: \Ocal_i \rightarrow \Acal_i $, we focus on distributional strategies $ \sigma $, which are probability measures over $ \Ocal_i \times \Acal_i $. 
This means rather than first observing $ o $ and then choosing an action $ b $ according to $ \beta $, distributional strategies assign probabilities to observation-action pairs. 

This idea becomes more tangible when we apply it in a discrete setting.
Let us consider the auction in discretized versions of the observation and action space i.e., $\Ocal^d_i = \lbrace o_1, ..., o_{K} \rbrace \subset \Ocal_i $ and $ \Acal^d_i = \lbrace b_1, ..., b_{L} \rbrace \subset \Acal_i $. 
The discrete distributional strategy $ s_i $ can be seen as a form of mixed strategy for Bayesian games over the discretized observation and action space, i.e., $ s_i \in \Delta(\Ocal^d_i \times \Acal^d_i) $. For each observation $ o_k $, the strategy $ s_i $ induces a mixed strategy $ s_i(\cdot \vert o_k) \in \Delta(\Acal^d) $ over the action space. 
This is similar to imperfect-information extensive-form games, where behavioral strategies induce mixed strategies at each information set \citep{Shoham2008}.
In distributional strategies, these different mixed strategies are now combined by weighting each mixed strategy with the probability  $ (f^d_{o_i})_k $ of the respective observation $ o_k $, induced by the prior distribution. That is, we obtain a matrix $ s_i \in \Delta(\Ocal^d_i \times \Acal^d_i) \subset \R^{K \times L}  $ in which each entry $ (s_i)_{k l} =  (f^d_{o_i})_k \cdot s_i(b_l \vert o_k)  $ indicates the probability that $ o_k $ is observed and $ b_l $ is played. By construction, this matrix satisfies the marginal condition as described in Definition \ref{def:distr_strat}. Note that the set of such distributional strategies, denoted by $ \Scal_i^d $, is convex.
Given a strategy profile $ (s_1, s_2) \in \Scal^d_1 \times \Scal^d_2$, the expected utility $ \tilde u_1 $ for player 1 is the sum of all outcomes weighted by their respective probability induced by the strategies:
\begin{align}
	\tilde{u}_1(s_1, s_2) 
	&= \sum_{k_1,l_1 = 1}^{K,L}  \sum_{k_2,l_2= 1}^{K,L} u_1(b_{l_1}, b_{l_2},  o_{k_1}) (s_1)_{ l_1 k_1} (s_2)_{l_2 k_2} \\
	&= \sum_{k_1,l_1 = 1}^{K,L} (s_1)_{k_1 l_1} \left( \sum_{k_2,l_2 = 1}^{K,L} u_1(b_{l_1},b_{l_2},o_{k_1}) (s_2)_{k_2 l_2} \right)
	=: \langle s_1, c_1 \rangle.
\end{align}
This linear structure of the utility function $ \tilde u_i $ allows for two things. First, the function is obviously differentiable with $ \nabla_{s_1} \tilde{u}_1(s_1; s_2) = c_1 $. And secondly, the best response $ s_1^{br}  = \argmax \{ \tilde u_1(s,s_2): \,s_1 \in \Scal^d \} $, given the opponents strategy $ s_2 $, is the solution of the following linear program
\begin{equation}\label{eq:LP}
	\begin{aligned} 
		\max \limits_{s \in \bbbr^{K \times L}} \langle  s, c_1 \rangle \, \text{ s.t. } \sum_{l=1}^L s_{k l} &= (f^d_{o_1})_{k} \quad \forall k \in \{1,\dots,K\} \\ s_{k l} &\geq 0 \quad \forall \,  k \in \{1,\dots,K\},\, l \in \{1,\dots,L\}.
	\end{aligned}
\end{equation}
This allows us to compute the utility loss, i.e., the utility gap $ \varepsilon $ of a $ \varepsilon $-NE.
%Note that we can even solve the LP directly with $s_{k l'} = (f^d_{o})_{k} $ for $ l' = \arg \max \lbrace (c_1)_{k l}:\, l = 1,...,L \rbrace  $, and $ s_{k l} = 0 $ for all other $ l \neq l' $ for all $ k = 1,...,N $.

Overall, we can define a complete-information game  $ \Gamma = (\Ical, \Scal^d_i, \tilde u_i) $ based on the discretized incomplete-information game. The distributional strategies correspond to a compact, convex action set, and the expected utility functions  $ \tilde u_i $ to differentiable utility functions.
This enables us to draw on standard equilibrium learning methods from online convex optimization. 
In this example we will focus on dual averaging (DA) \citep{nesterov2009primal}.
DA is based on two steps. Both players simultaneously use their gradients to update a variable in the dual space and then mirror this updated dual variable back to the feasible set $ \Scal^d_i $ to get an updated strategy $ s_i $. In this setting, this is equivalent to Follow-the-Regularized-Leader (FTRL) \citep{ShalevShwartz2012}.
We repeat these steps until the strategy profile is close enough to an equilibrium i.e., until the utility loss with respect to the best response in the current strategy profile is sufficiently small.
\begin{figure}
	\FIGURE
	{
		\includegraphics[width = .33\textwidth]{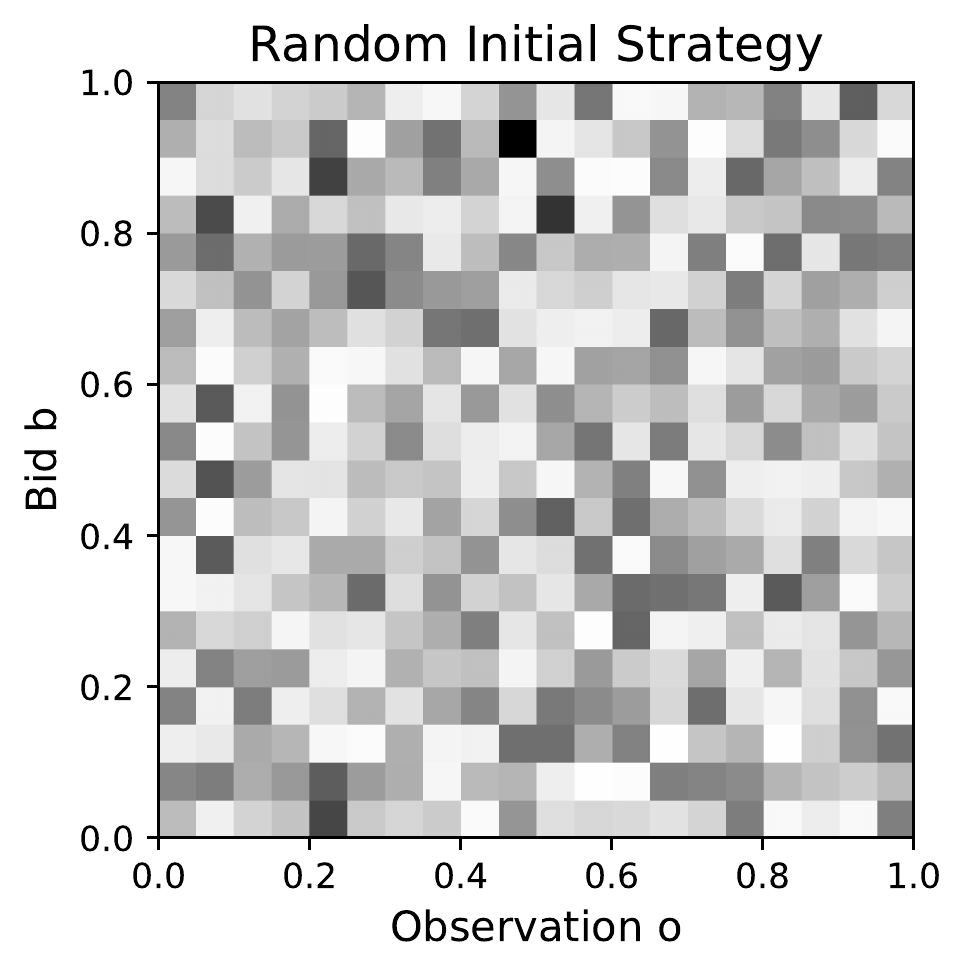}
		\includegraphics[width = .33\textwidth]{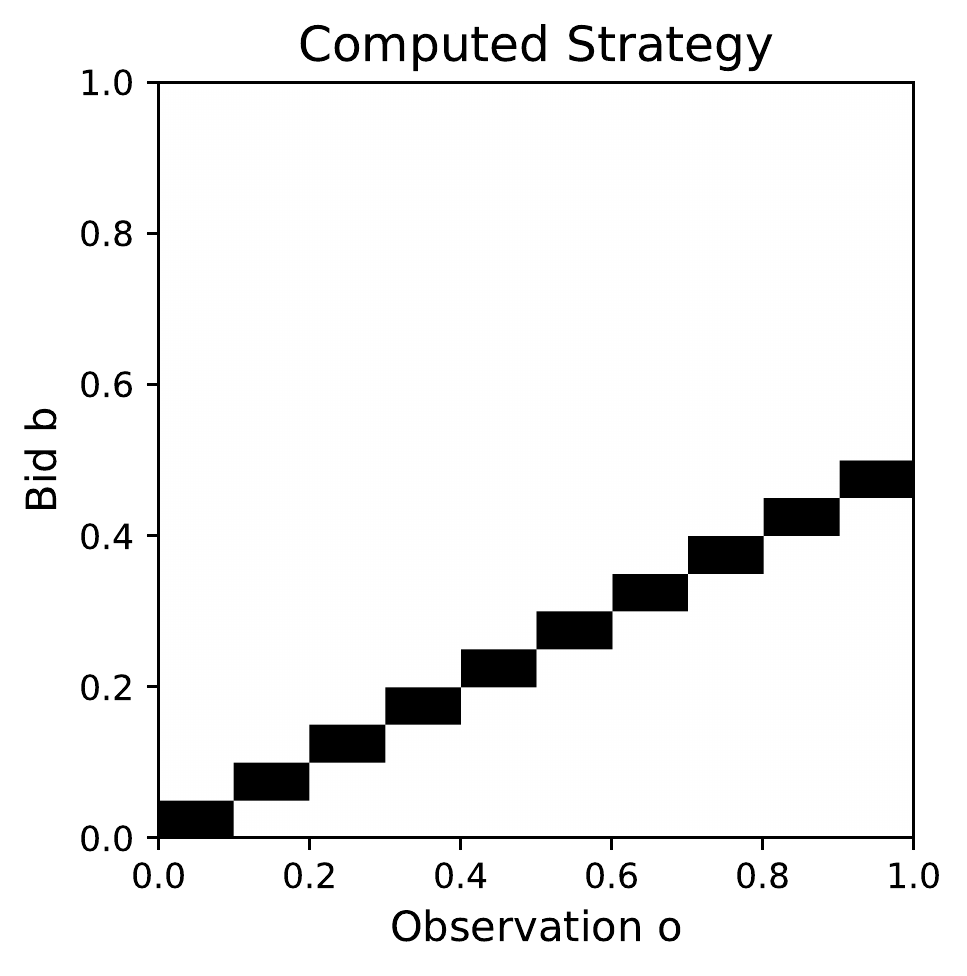}
		\includegraphics[width = .33\textwidth]{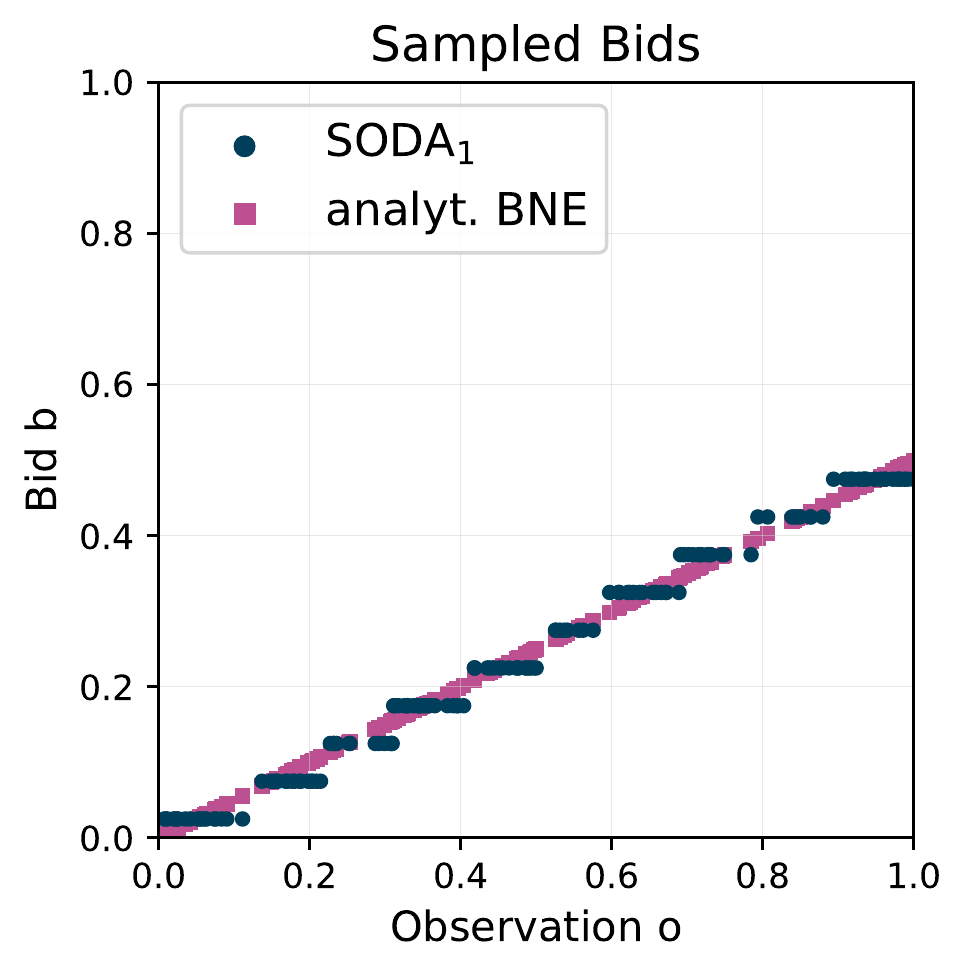}
	}
	{SODA applied to discretized FPSB auction with two symmetric bidders. \label{fig:ex}}
	{
		In the first plot we can see a random strategy $ s_i \in \mathcal S^d_i $.
		The color of each square represents the probability of the respective observation-action pair.
		The second plot shows the computed strategy using SODA. We can observe that the probabilities concentrate as expected near $ \tfrac{1}{2} o $.
		In the last plot we compare the computed strategy with the analytical BNE from the continuous setting. This is done by sampling 150 observations $ o $ according to the uniform prior. Bids from the computed strategy are then obtained by identifying the nearest discrete observation $ o_k $ and sample a bid from the induced mixed strategy $ s_i(\cdot \vert o_k) \in \Delta( \Acal^d_i) $ (blue dots). For the analytical BNE we simply plug in the sampled observations to the equilibrium function $ \beta_i(o) = \tfrac{1}{2} o $ (purple squares). This way, we can evaluate the approximation and compute the metrics as explained in Section \ref{sec:eval}.
	}
\end{figure}
In Figure \ref{fig:ex} we can see an application of this simultaneous online dual averaging (SODA) algorithm to our discretized first-price sealed-bid auction with two bidders as introduced above.
We consider uniformly distributed observations over $ \mathcal O_1 = \mathcal O_2 = [0,1] $ and allow for bids within the same interval, i.e., $ \mathcal A_1 = \mathcal A_2 = [0,1] $. Both spaces are discretized using $ K = L = 20 $ equidistant points.

After this illustration, let us now introduce the model and the algorithm formally.

\subsection{Discretization} \label{sec:model}

Our algorithms are based on a discrete version of the game and distributional strategies. 
As illustrated by the example in the previous section, these discrete distributional strategies are constructed by restricting ourselves to finite subsets of the observation, valuation, and action sets and considering finitely atomic measures as a counterpart to the distributional strategies in the continuous setting. This constitutes a specific discretized game formalization that the algorithms operate on, which we also refer to as approximation game. 

Formally speaking, we construct a discrete version $ G^d = (\Ical, \Vcal^d, \Ocal^d, \Acal^d, f^d, u) $ of the incomplete-information game $ G $. 
This is done by defining a set of discrete observations $ \mathcal O^d = \mathcal O^d_1 \times ... \times O^d_n $ where  $ \mathcal O_i^d := \lbrace o_1^i,...,o_K^i \rbrace \subset \mathcal O_i $. 
Similarly we define $ \mathcal A_i^d := \lbrace b_1^i,...,b_L^i \rbrace \subset \mathcal A_i $ and  $ \mathcal V_i^d := \lbrace v^i_1, ..., v^i_M \rbrace \subset \mathcal V_i $.
We further approximate the joint probability density function $ f $ by a discrete version $ f^d $ over $ \mathcal V^d \times \mathcal O^d $. The marginal distribution of $ f^d $ over $ \mathcal O_i^d $ can be written as $ f^d_{o_i} \in \Delta(\Ocal^d_i) \subset \R^K$. For simplicity we assume that the spaces are discretized with the same number of points for all agents. But this does not have to be the case.

The discrete version $ s_i $ of a distributional strategy $ \sigma_i $ for bidder $ i $ is now measure over $ O_i^d \times A_i^d $ and can be identified with a matrix $ s_i \in \Delta(O_i^d \times A_i^d) \subset \R^{K \times L} $.
The marginal condition for distributional strategies translates to $ \sum_{l} (s_i)_{k l} =(f^d_{o_i})_k $ for all $ k = 1,...,K $.
Therefore the set of all possible discrete distributional strategies for bidder $ i $ can be identified by matrices of the form:
\begin{equation}\label{disrete_strategies}
	\mathcal S^d_i := \big\lbrace s_i \in \bbbr^{K \times L} : \, (s_i)_{k l} \geq 0 \, \,\forall k,l, \,\text{ and } \, \sum_{l} s_{k l} = (f^d_{o_i})_k \,\, \forall k \big\rbrace
\end{equation}
For a given strategy profile $ (s_1,...,s_n) \in \mathcal S^d_1 \times ... \times \mathcal S^d_n $ we can compute the expected utility. This corresponds to the discretized version of equation (\ref{eq:dist_strat}).
\begin{align} \label{eq:utility}
	\tilde u_i (s_1,...,s_n) 
	&= \sum \limits_{k, l, m}  
	u_i(b_l,v_{m_i}) \prod_{j=1}^n (s_j)_{k_j l_j} \dfrac{(f^d)_{m,k}}{(f^d_{o_1})_{k_1} \cdots (f^d_{o_n})_{k_n} }\\
	&= \sum \limits_{k_i, l_i }  (s_i)_{ k_i l_i} \sum \limits_{m,k_{-i}, l_{-i} }
	u_i(b_l,v_{m_i}) \prod_{j \neq i} (s_j)_{k_j l_j} \dfrac{(f^d)_{m,k}}{\prod_{j'}  (f^d_{o_{j'}})_{k_{j'}} }\\
	\intertext{For all $ k_i,l_i $ we denote the second sum, which only depends on $s_{-i}$, as $ (c_i)_{k_i,l_i} $ and write}
	\tilde u_i (s_1,...,s_n) &= \sum \limits_{k_i, l_i }  (s_i)_{ k_i l_i} (c_i)_{k_i,l_i} = \langle s_i, c_i \rangle \label{eq:linear_util}
\end{align}
Note that in these equations, $ l = (l_1,...,l_n) $ is a multi-index and $ b_l = (b^1_{l_1},...,b^n_{l_n}) $ the action profile of all bidders (same for $ v $ and $ o $ respectively).
Since the second sum ($ c_i $) does not depend on $ s_i $, the expected utility function for bidder $ i $ is linear in the bidder's own strategy. 
Instead of considering the discretized incomplete-information game $ G^d $, we can use the expected utility $ \tilde u_i $ and the sets of discrete distributional strategies $ S^d_i $ to define a complete-information game.

\begin{definition} \label{def:approx_game}
	Given the Bayesian game $G = (\mathcal I, \mathcal V, \mathcal O, \mathcal A, f, u)$, we construct a discrete version $ G^d = (\Ical, \Vcal^d, \Ocal^d, \Acal^d, f^d, u) $ of the game by discretizing the respective spaces and probability distributions.
	The resulting sets of discrete distributional strategies $ \mathcal S^d_i $ and the expected utility $ \tilde u_i $ define a complete-information game  $ \Gamma = (\mathcal I, \mathcal S^d, \tilde u ) $, which we call the approximation game of $ G $.
\end{definition}
Observe that the Nash equilibria $ s \in \Scal^d $ of the approximation game $ \Gamma $, characterized by
\begin{equation}\label{eq:NE}
	\tilde u_i(s_i,s_{-i}) \geq \tilde u_i (s'_i,s_{-i}) \quad \forall s' \in \Scal^d_i, \, \forall i \in \Ical,
\end{equation}
correspond to Bayes Nash equilibria in the discretized Bayesian game $ G^d $.

% Algorithms
\subsection{Algorithm} \label{sec:algorithm}
The approximation game  $ \Gamma = (\mathcal I, \mathcal S^d, \tilde u ) $ is a well-behaved complete-information game with linear (in $ s_i $) utility functions $ \tilde u_i $ and compact, convex action sets $ \mathcal S^d_i \subset \R^{K \times L} $. 
This structure allows us to use algorithms from online convex optimization, where all agents simultaneously compute the gradient given the current strategy profile and update their strategies according to some chosen method (Algorithm \ref{alg:soga}). 
In particular, we focus on Dual Averaging (DA) \citep{nesterov2009primal} since  \cite{mertikopoulos2019learning} provide an ex-post certificate for the computed strategies if we converge (see Corollary \ref{cor:conv_da}). While DA is our baseline algorithm, we also analyze alternative gradient-based algorithms. This will help us understand whether convergence in these games is restricted to a specific type of algorithm or regularizer. Specifically, we provide results for gradient-based methods such as Mirror Descent (MD) \citep{nemirovskij1983problem} and the Frank-Wolfe Algorithm \citep{Frank1956}. 

Let us briefly summarize the gradient-based methods we are considering.
% Mirror Descent
{Mirror Descent} can be interpreted as a generalized projected gradient descent, where the projection is with respect to the Bregman divergence induced by a distance-generating mirror map $ g $ \citep{Beck2003}. Commonly used mirror maps are strongly convex functions such as the negative entropy $ g_1(x) = \sum_i x_i \log x_i $ with $ g_1(0) = 0 $ and $ g(x) = \infty $ for all $ x \notin \R^n_{\geq0} $, and the Euclidean distance squared $ g_2(x) = \Vert x \Vert_2^2 $. While $ g_2 $ leads to the standard projected gradient algorithm, the update step generated by $ g_1 $ is known as the entropic descent algorithm \citep{Beck2003}.

% Dual Averaging
In Dual Averaging one distinguishes between dual and primal iterates. It is considered to be a lazy version of Mirror Descent since the gradient update is only done in the dual space. To get the next iterate in the primal space, the updated dual variable is projected onto the feasible set in the primal space. The projection is done with respect to some regularizer $ h $ which again is induced by a strongly convex function. For our examples the mirror maps $ g_a $ with $ a \in \{1,2\} $ induce regularization functions $ h_a $ by $ h_a(x) = g_a(x) + I_{\Scal_i}(x) $, where $ I_{\Scal_i}(x) = 0 $ if $ x \in \Scal_i $ and $ + \infty $ else.
For $ h_1 $ we get the same update step as in MD, namely the entropic descent algorithm. But for the Euclidean regularizer, DA leads to a lazy version of the projected gradient descent which is equal to the (linearized) FTRL with Euclidean regularizer. A pseudo-code can be found in Algorithm \ref{alg:soga}.

\begin{algorithm}[h]
	% \SetAlgoNoLine
	\DontPrintSemicolon
	\KwIn{Approximation game $\Gamma = (\mathcal I, \mathcal S^d, \tilde u ) $, initial strategies $ s_{1} \in \mathcal S^d $}
	
	\For{$t = 1, 2, \dots, T$}
	{
		\For{each agent $i\in \mathcal I$}
		{	
			\tcp{calculate gradient}
			$ c_{i,t} \leftarrow \nabla_{s_i} \tilde u_i(s_{i,t}, s_{-i,t}) $\;	
			\tcp{update strategy (using Dual Averaging)}
			$ y_{i, t+1} \leftarrow y_{i,t} + \eta_t \cdot c_{i,t} $\;
			$ s_{i,t+1} \leftarrow \nabla h^*(y_{i,t+1}) $ \;
		}
	}
	\caption{Simultaneous Online Dual Averaging (SODA)}
	\label{alg:soga}
\end{algorithm}

MD and DA are widely used and no-regret learners \citep{ShalevShwartz2012}. 
% MD vs. DA
\cite{Juditsky2022} provides a detailed analysis of Mirror Descent and Dual Averaging, unifying both approaches and explaining the differences between mirror maps (MD) and regularizers (DA). They also provide intuition for the cases where both methods coincide, as we can observe for the negative entropy.

% Frank Wolfe
Another method we consider is the Frank-Wolfe (FW) algorithm, also known as conditional gradient. This method uses gradient feedback to solve the linear program induced by the first-order approximation of the objective function. The next iterate is a convex combination of this optimal solution and the previous iterate. Since the feasible set is convex, one avoids the potentially expensive projection which has to be computed in the other methods. \cite{Hazan2012} introduced an online version of the Frank-Wolfe, where the solution of the linear program is computed with respect to the aggregated objective functions of all previous iterates. In contrast to the standard version, the online version also has the no-regret property. But due to better performance in our experiments, we stick with the standard Frank-Wolfe algorithm.

An overview of the different update rules we used in our experiments is provided in Table \ref{tab:update-steps}. 
\input{tables/learning_algorithms}
Interestingly, we find that all algorithms in Table \ref{tab:update-steps} converge to equilibrum quickly and the results are similar.
We also report results for Fictitious Play (FP), as an algorithm that is not gradient-based. FP is the oldest and best-known equilibrium learning technique \citep{brownIterativeSolutionGames1951}. At each round, each player  best responds to the empirical frequency of play of their opponent. Also FP converges in the analyzed model, but it is not as efficient as the gradient-based algorithms.

\subsection{Approximation via Discretization}

Next, we show that approximate BNEs of the discrete game $G^d$ naturally induce approximate BNEs of the continous game $G$, where the quality of the approximation depends on the coarseness of the discretization. Thus, if our algorithm finds a good solution to the discretized setting, this also induces a good solution for the continuous setting, where the quality depends on the coarseness of the discretization. We only consider some specific single-object auctions here. Apart from that, we do not postulate any strong assumptions, such as symmetry or independence. The precise formal statement of the following theorem, together with its proof, can be found in Appendix \ref{app:thm}. The assumptions for the proof include single-object auctions such as the first-price and second-price sealed bid auctions as well as first-price and second-price all pay auctions (e.g., war of attrition).

\begin{theorem} \label{prop:approx}
	Let $s \in \Scal^d$ be an $\varepsilon$-BNE of the discrete game $G^d$ of a single-object auction. Let $\sigma \in \Scal$ be the strategy profile, where each $\sigma_i$ is the strategy induced by $s_i$. Then $\sigma$ is an $ \varepsilon + \Ocal(\delta_\alpha + \delta_\tau)$-BNE of the continuous game $G$.
\end{theorem}

Here $\delta_\tau$ and $\delta_\alpha$ denote the coarseness of the discretization of the valuation and the action space. The central message of the proposition is that if we find an approximate BNE for the discrete game, we also find an approximate BNE for the continuous game with an additional error term decreasing linearly with the coarseness of the discretization. 

The idea of the proof is as follows. Given an arbitrary strategy profile $s \in \Scal^d$ of the discrete game, we show that $s$ naturally induces a feasible strategy profile $\sigma \in \Scal$ of the continuous game and that the difference of utilities of these two solutions is small. Conversely, we can construct a feasible discrete strategy profile $s$ from a given continuous strategy profile $\sigma$. Our central argument is that if we start with a continuous strategy profile $\sigma \in \Scal$, and consider the induced discrete strategy profile $s \in \Scal^d$, which in turn induces a continuous strategy $\tilde \sigma_i \in \Scal_i$ for each agent $i$, the loss of utility is in $\mathcal{O}(\delta_\tau + \delta_\alpha)$. Now suppose we find an $\varepsilon$-BNE $s^* \in \Scal^d$ of the discrete game and consider the induced continuous strategy profile $\sigma^*$. Let $\sigma_i$ be a best response to $\sigma^*_{-i}$. Then the discrete strategy $s_i$ induced by $\sigma_i$ cannot be much better than $s^*_i$, since $s^*$ is an $\varepsilon$-BNE. But by the result mentioned above, the utility of the continuous strategy $\tilde \sigma_i$ neither differs by much from $s_i$, nor from $\sigma^*_i$. Thus, the gain of utility from switching to $\sigma_i$ is in $\varepsilon + \mathcal{O}(\delta_\tau + \delta_\alpha)$. 

In all our experiments not only the utility loss converged with finer discretization, but also the strategies converged. However, this is not necessarily the case. There are auction models where there is an approximate equilibrium in the discretized auction, but not in the continuous case \citep{jackson2005existence}. However, there are also cases where we know that pure, symmetric equilibria exist, but there may be no corresponding equilibrium in the discretized case. \citeauthor{rasooly2021} show that for different tie-breaking rules there are only asymmetric pure equilibria for some simple first-price sealed bid auction settings and only mixed equilibria for some all-pay auction settings. Although such situations can happen, we can observe that SODA approximates the continuous equilibria well due to the richer class of symmetric distributional strategies that are being learned. 

\subsection{Ex-Post Certificates} \label{sec:convergence}

An advantage of SODA over earlier methods for equilibrium computation in auctions \citep{bosshardComputingBayesNashEquilibria2018, bichler2021npga} is that SODA does not need an empirical verifier, which is computationally expensive. This insight relies on \citet{mertikopoulos2019learning}, who prove in their Theorem 4.1 that if a sequence of pure strategy profiles resulting from dual averaging converges to a strategy profile for all players, then this profile is a Nash equilibrium. 
A consequence of the distributional strategies that we learn is that the expected utility $\tilde{u}(s_1,\cdots,s_n)$ is linear in the bidder's own strategy, satisfying the assumption of the theorem, that the utility functions are (pseudo-) concave in the bidders' own strategies. Consequently, if SODA converges to a pure strategy, it also converges to a Nash equilibrium. 

\begin{corollary}[to \citet{mertikopoulos2019learning}, Theorem 4.1] \label{cor:conv_da}
	Suppose that SODA is run with a step-size sequence that is square summable but not summable and produces the sequence $(s^t)_{t\in T}$ of action profiles. If the sequence of strategy profiles $(s_i^t)_{t\in T}$ converges to $s_i^* \in \Scal^d_i$ for all $i \in \mathcal{I}$, then $s^*$ is a Nash equilibrium.
\end{corollary}

Of course, checking empirically whether an infinite sequence of iterates converges by inspecting finitely many of them is not possible. However, we believe that the rapidly decreasing distance between consecutive iterates we observe in our experiments strongly indicates that we indeed approximate exact BNEs with high precision. 

\section{Experimental Evaluation \label{sec:auctions}}

We illustrate the versatility of our method by analyzing a number of very different auctions and contests.
We report results on single-object auctions with interdependent valuations, combinatorial auctions with single-minded bidders and multi-minded bidders, single-object auctions with risk-averse bidders, and Tullock contests with a randomized contest success function. 
In some of these models the analytical BNE is given, which provides an unambigous baseline to compare against. However, we also explore models where no BNE was known so far, which includes all-pay auctions with risk-averse bidders and the Tullock contest.
With only a few bidders we can compute BNE within a few minutes or seconds. 
We compare our results to those in \citet{bichler2021npga} on NPGA to illustrate the performance increase we get for these environments. 
%Due to space constraints, we focus on a selected set of models and algorithms, namely SODA$_1$, SODA$_2$, NPGA. %In Appendix \ref{app:add_exp} we provide additional results on mirror ascent, the Frank-Wolfe algorithm, and Fictitious Play. All algorithms converged quickly in all models analyzed.

%\todo[inline]{The AE is stressing point 3 of referee 1. If we want to address this point beyond any doubt, we should consider another environment, maybe in the appendix. }

\subsection{Parameter and Evaluation Criteria} \label{sec:eval}
% Approximation Game
We start by constructing the approximation game by discretizing each dimension of the respective spaces with $ K=L=M=64 $ equidistant points, if not stated otherwise.
The discrete prior distribution is computed by evaluating the density function at these discrete points and normalizing the resulting probability vector. Since ties happen with a positive probability in the discretized game, we also have to define a tie-breaking rule. Due to better performance in our experiments, we deviate from the standard random tie-breaking and implement a rule where no agent wins if the maximal bid is not unique. 

% Learning Algorithm
Given the constructed approximation game, we apply the learning algorithms as defined in Section \ref{sec:algorithm}.
The algorithms stop either after a fixed number of iterations ($ T=1\thinspace000 $) or whenever the stopping criterion is satisfied, i.e., $ \ell_i < \varepsilon_\text{tol} = 10^{-4} $, for each agent $ i $. We use the relative utility loss $ \ell_i $ as the stopping criterion, which denotes the relative improvement of the expected utility $ \tilde u $ an agent can achieve in the approximation game when fixing the opponents strategies $ s_{-i} $ and playing the best response $ s_i^{br} $ instead of $ s_i $. The best response is the solution of a simple LP (see Equation \ref{eq:LP}). A low relative utility loss means that we are in some approximate NE in the approximation game. If bidders are symmetric, we learn a single strategy for all of them.
% Metrics
After computing an approximate discrete distributional equilibrium strategy, we want to evaluate the computed solution within the initial continuous setting of the auction game.
To do this, we sample observations according to the prior distribution and determine the corresponding bids from our strategies.
Note that, unlike to pure strategy functions $ \beta_i $, we cannot simply plug in the sampled observations $ o_i $ and get a bid $ b_i = \beta_i(o_i) $. Instead, we have to identify the closest discrete observation $ o_{k_i} $ and sample a discrete bid $ b_i \sim s_i(\, \cdot \, \vert o_{k_i}) $ according to the induced mixed strategy by $ s_i $.
To compare our results with NPGA from \citet{bichler2021npga}, we choose the same approach and focus on two metrics.
First, given the opponents' strategies $ \beta_{-i} $ we estimate the ex-ante utility using the sample-mean of the ex-post utilities $ \hat u_i (\cdot, \beta_{-i}) := \tfrac{1}{n_o} \sum_{o} u_i( \cdot, \beta_{-i}((o_{-i}))) $.
We then compare the outcome of a player bidding according to the computed strategy $ s_i $ versus bidding according to the known equilibrium strategy $ \beta_i $, while all opponents $ j $ play the equilibrium strategy $ \beta_j $.
This leads to the \textit{relative ex-ante utility loss} $ \mathcal L(s_i; \beta) = 1 - \tfrac{\hat u_i (s_i, \beta_{-i})}{\hat u_i (\beta_i , \beta_{-i})} $.
Secondly, we report the probability-weighted root mean squared error of the sampled bids from $ s_i $ and the bids from the equilibrium strategy $ \beta_i $, which approximates the \textit{$ L_2 $ distance} of two functions or in our case between the function and the sampled bids $ b_i $, i.e., $ L_2(s_i, \beta_i) = \big( \tfrac{1}{n_o} \sum_{o_i} (s_i(\cdot \, \vert o_i) - \beta(o_i))^2 \big)^\frac{1}{2} $. This metric ensures that we not only achieve a low utility loss but also approximate the equilibrium strategies.
Similar to \cite{bichler2021npga} we sample $ n_o = 2^{22} $ observation (or valuation) profiles for both metrics.
We report the mean and standard deviation of all metrics over ten runs with random initial strategies. 

All experiments are run on a computer with an Intel Core i7-8565U CPU @ 1.80 GHz and 16GB of RAM. The implementation of the algorithm uses Python 3.8.5.

% Experiments on Single-Object Auctions
\subsection{Single-Object Auctions\label{sec:mineral-rights}}

We start with single-item auctions with interdependencies. The most well-known examples of interdependencies are the common value model (with independent observations $o$) and the affiliated value model for single-item auctions \citep{krishnaAuctionTheory2009}. 
\begin{figure}[h]
	\FIGURE
	{
		\includegraphics[width = 0.45\textwidth]{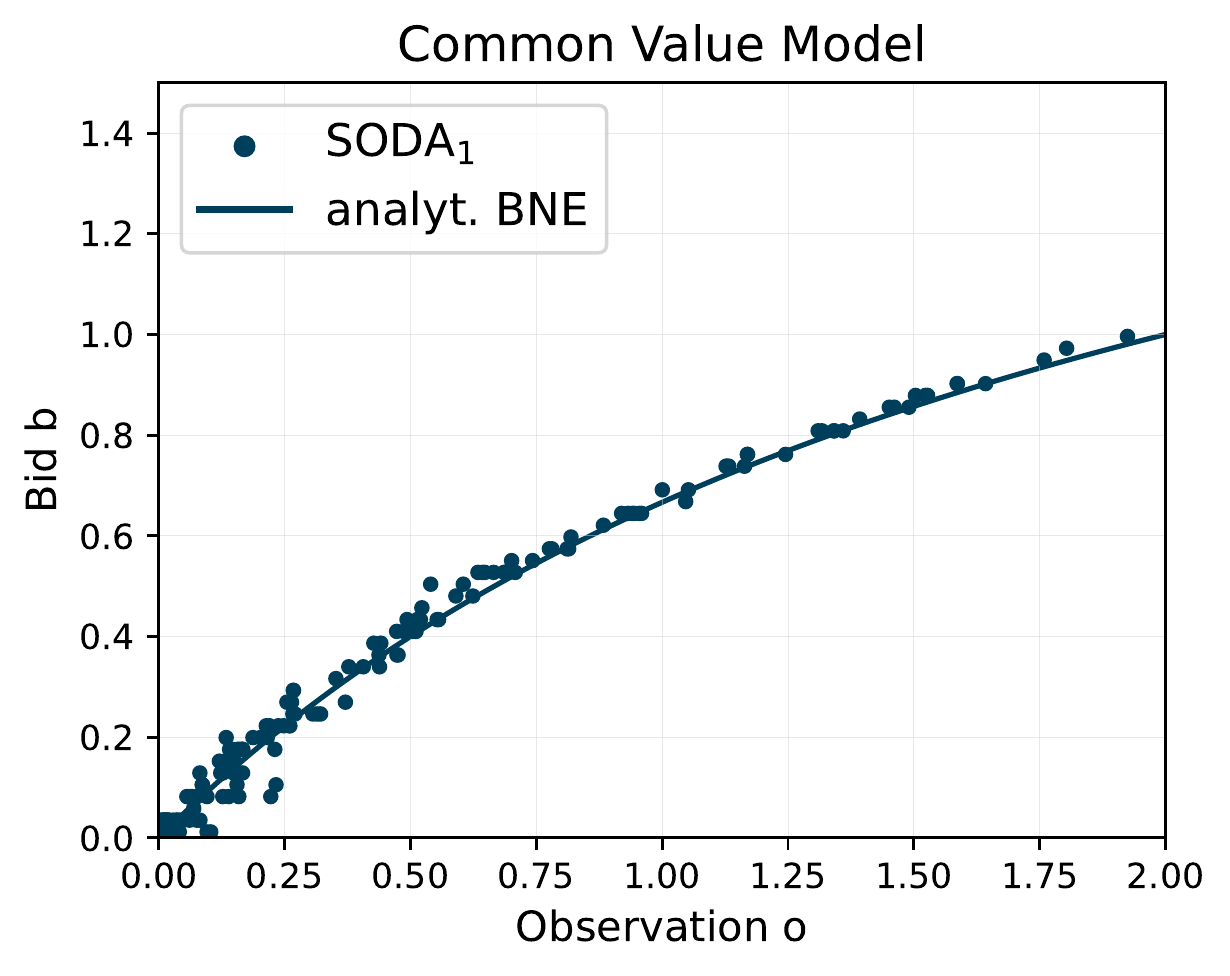}
		\includegraphics[width = 0.45\textwidth]{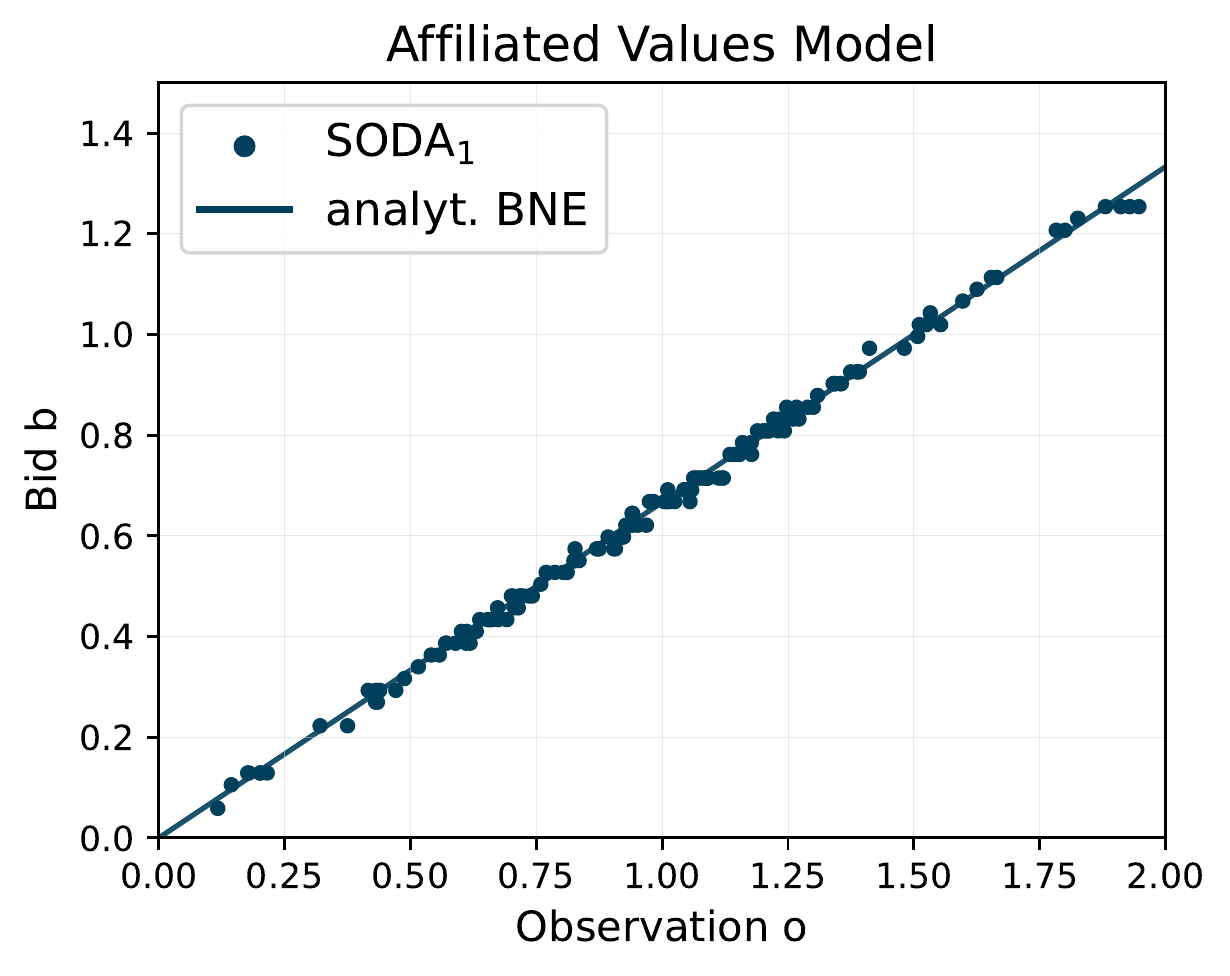}
	}
	{Computed strategies for single-item auctions with interdependencies. \label{fig:interdep}}
	{We draw 150 observations according to the prior distribution and sample the corresponding bids from the computed discrete distributional strategies (blue dots). The colored lines indicate the analytical equilibrium strategies in these settings.}
\end{figure}
% Common Value Model
We explore the second-price auction in an environment where there is one pure common value that is the same among all bidders. Three bidders $i \in \{1,2,3\}$ share a common $\mathcal{U}(0,1)$-distributed value for the item of interest. Conditioned on this value, the observation $o_i$ of bidder $i$ is uniformly---and independently from the other observations---distributed on the interval from zero to two times the common value. Formally, we can define the joint prior probability density function $f$ with a four-dimensional uniformly distributed random variable $\Omega = [0,1]^4$. For a draw $\omega \sim \mathcal{U}(\Omega)$ we set each player's type to $v_i(\omega) = \omega_4$ and each observation to be $o_i(\omega) = 2 \cdot \omega_i \cdot \omega_4$. Notice, all agents have the same value (or type), but they learn their value only if they win the auction. In this model, the symmetric BNE strategy profile can be stated in closed form as $ \beta^*_i(o_i) = \frac{2o_i}{2+o_i} $.
For our algorithms we restrict the spaces to the intervals $ \Ocal_i = [0,2] $, $ \Vcal_i = [0,1] $, and $ \Acal_i = [0,1.5] $ and discretize them. Since all bidders are symmetric we learn a single strategy for all of them. We observe (see Table \ref{tab:interdep_results_cv}) that the standard projected gradient ascent ($ \text{SOMA}_2 $) converges within seconds, while all other methods run for several minutes. The strategies computed with SOFW deviate significantly from the equilibrium strategy for low valuations, which explains the high $ L_2 $ norm. But since this only happens for low valuations with low bids, there is little effect on the utility loss $ \mathcal L $.
\input{tables/interdep_results_cv}

% Affiliated Values Model
In the \textit{affiliated values model} the individual observations are correlated. In a model with two bidders (see also \citet[Example~6.2]{krishnaAuctionTheory2009}), we can set $\Omega = [0,1]^3$ and bidder $i \in \{1, 2\}$ then makes the observation $ o_i(\omega) = \omega_i + \omega_3 $ and both have a common value of $v(\omega) = \frac{1}{2}(\omega_1 + \omega_2) + \omega_3$. The symmetric BNE strategy for both agents under a second-price payment rule is to bid truthfully and for a first-price payment rule to bid according to $\beta_i^*(o_i) = \frac{2}{3} o_i$. 
In contrast to the common value model, we do not need an additional valuation space and only discretize the spaces $ \Ocal_i = [0,2] $ and $ \Acal_i = [0,1.5] $.
Together with fewer bidders (i.e., two symmetric bidders), this leads to significantly faster computations of the equilibrium strategies as we can see in Table \ref{tab:interdep_results_av}. 
\input{tables/interdep_results_av}

\subsection{Combinatorial Auctions in the Local-Local-Global Model} \label{sec:llg}
Bayesian Nash equilibria are rarely available for multi-object auctions. Combinatorial auctions have received significant attention due to their use in spectrum sales and other applications \citep{bichler2017handbook}. The local-local-global (LLG) model has received significant attention in the analysis of core-selecting combinatorial auctions \citep{goeree2016impossibility}. The \emph{core} of an auction game describes the set of outcomes such that no \emph{coalition} of bidders (and possibly the auctioneer) can profitably deviate given the bids. This LLG model is simple enough to allow for the derivation of analytical results \citep{ausubel2019CoreselectingAuctionsIncomplete}. 
At the same time, core-selecting auction mechanisms are challenging and among the most complex auction formats used today, which provides an interesting benchmark for equilibrium computation. 

The LLG model consists of two objects $\{1,2\}$, two local bidders $i\in \{1,2\}$ and one global bidder $i=3$, each being interested only in one specific bundle (of the single object $i$ (locals) or both objects (global)), and we denote the valuation of each bidder's single bundle by $v_i\in \bbbr$. We consider a private values (but not \emph{independent} private values) setting with $o_i = v_i$ which allows for correlation. 
It was shown that with independent private values and risk-neutral bidders, core-selecting payment rules lead to significant inefficiencies in equilibrium \citep{goeree2016impossibility} in combinatorial auctions. Essentially, the two local bidders attempt to free-ride on each other. Depending on the prior value distributions, it can happen that both local bidders bid too low in total and they fail to outbid the global bidder, even if their combined valuations are higher than the global bidder's. This results in an inefficient outcome and it has been used as an argument against core-selecting combinatorial auctions \citep{bichler2017handbook}. 
Now, it is interesting to understand equilibria with different assumptions. For example, it is reasonable to believe that bidder valuations in spectrum auctions are correlated because telecoms face the same downstream market. 

%Recently, the model was analyzed with different types of correlation \cite{ausubel2019CoreselectingAuctionsIncomplete}. However, with standard core-selecting payment rules, it turns out that correlation alone cannot mitigate the efficiency and revenue loss encountered with independent private values. Risk aversion has not yet been analyzed, although it plays a role in the revenue ranking of single-object auctions. In contrast to single-object auctions, it has been unclear how risk-aversion plays out in equilibrium. If one local bidder knows that the other is risk averse and might thus bid higher, he might bid even lower as a result of this knowledge. The environment is not symmetric as there are two local and a global bidder. However, the global bidder has a simple dominant strategy to bid truthful and the two local bidders can indeed be considered symmetric whenever $f_{v_1} = f_{v_2}$.

\citet{ausubel2019CoreselectingAuctionsIncomplete} investigate two models of correlation among local bidders' private values and derive analytical BNE. We will focus on the \emph{Bernoulli weights model} and use it as a baseline in our experiments in addition to the results of NPGA. Let's define the joint prior $f$ to be the five-dimensional uniform distribution of a latent random variable $\omega \sim \mathcal U [0,1]^5$. Then let $v_3 = 2\omega_3$ be the valuation of the global bidder and
\begin{equation}
	v_1(\omega) = w \omega_4 + (1-w)\omega_1, \quad v_2(\omega) = w \omega_4 + (1-w)\omega_2
\end{equation}
be the valuations of the local bidders where the \emph{weight} $w$ is a random variable depending on $\omega_5$ only. The valuations of the local bidders can be thought of as a linear combination of an individual component $\omega_i$ and a common component $\omega_4$. Now given an exogenous correlation parameter $\gamma \in [0,1]$, \citet{ausubel2019CoreselectingAuctionsIncomplete} choose $w$ such that $\text{corr}(v_1,v_2) = \gamma$ via the Bernoulli weights model: $ w(\omega) = 1 $ if $ \omega_5 < \gamma $ and $ w(\omega) = 0 $ else.
\begin{figure}[h]
	\FIGURE
	{
		\includegraphics[width = .33\textwidth]{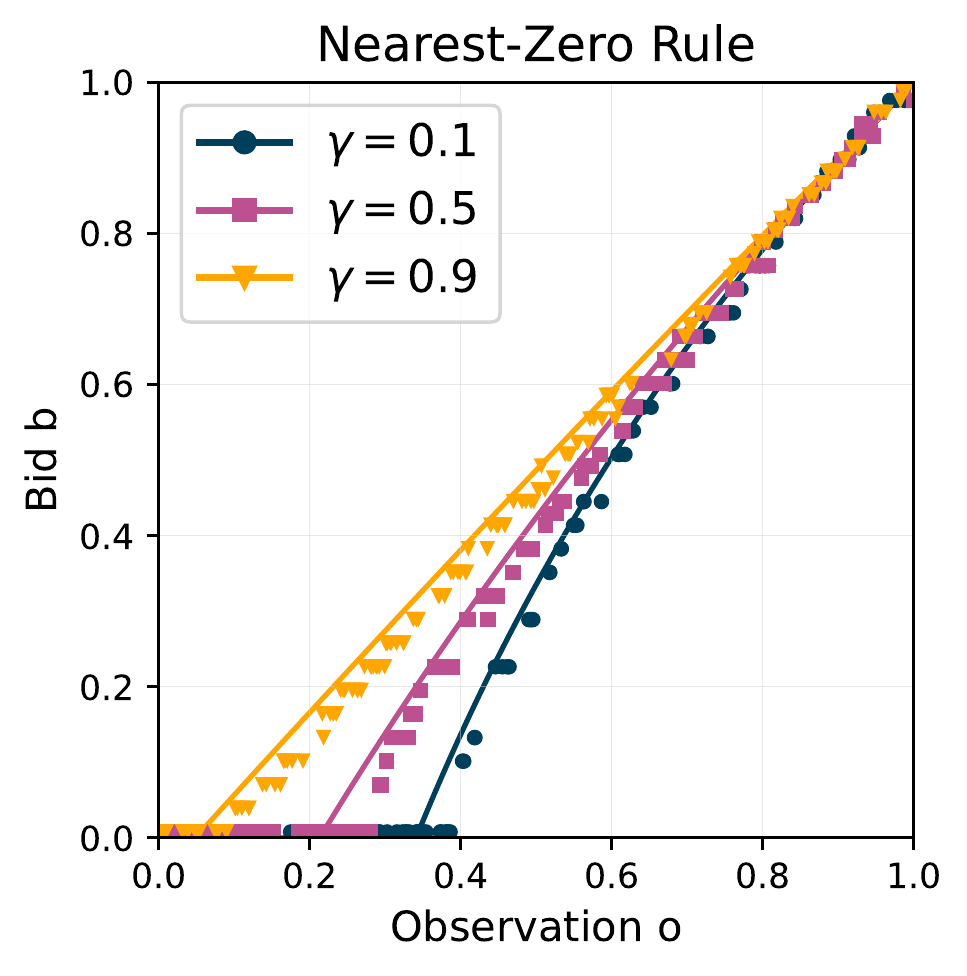}
		\includegraphics[width = .33\textwidth]{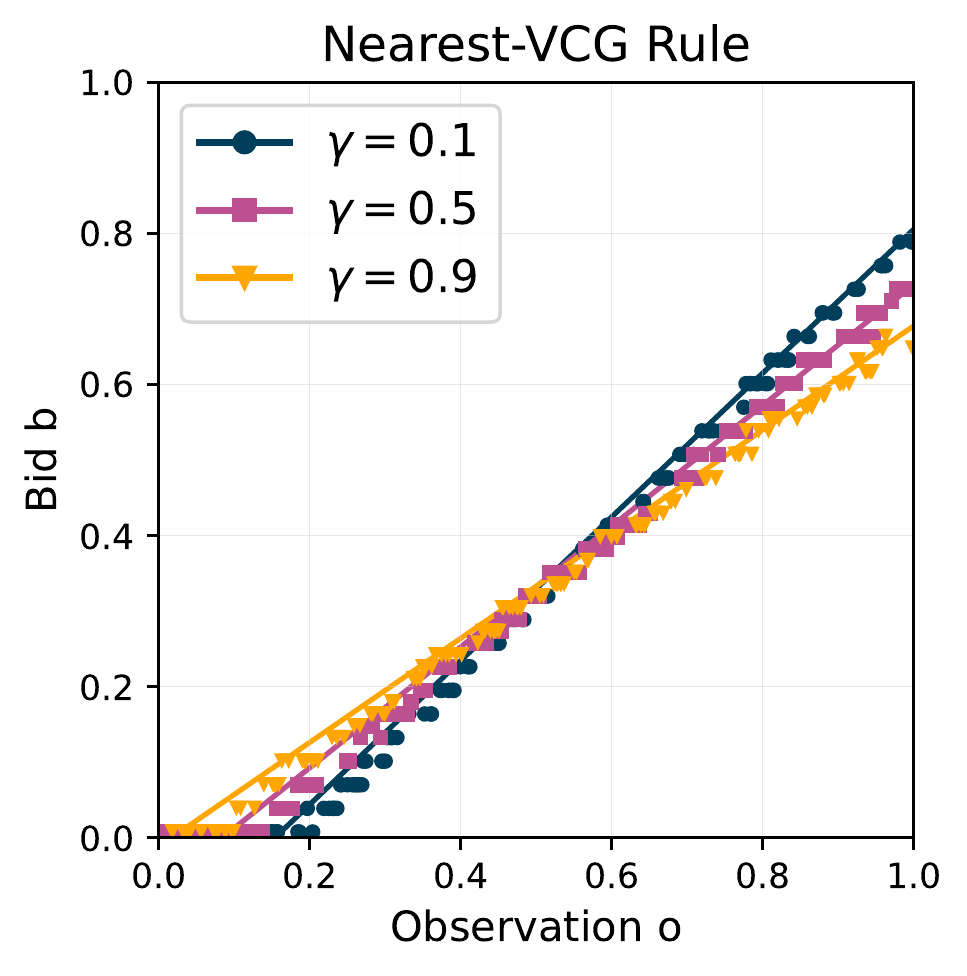}
		\includegraphics[width = .33\textwidth]{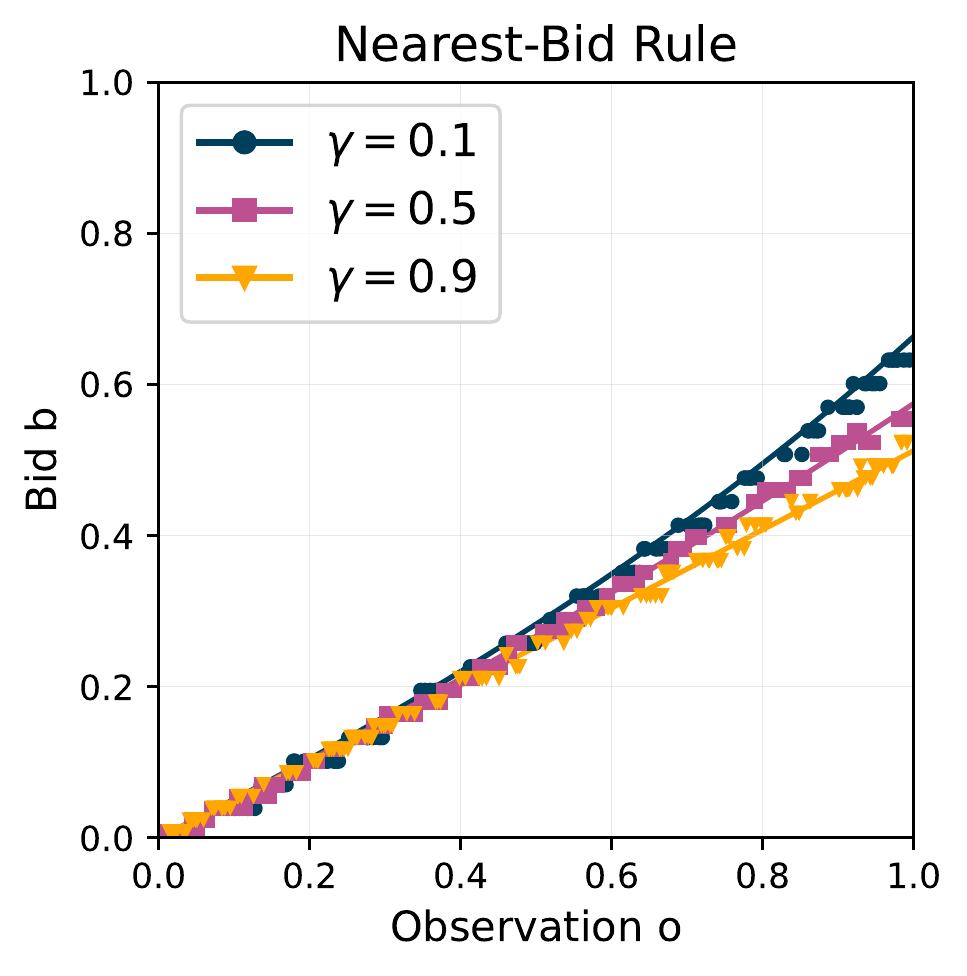}
	}
	{Computed strategies for the local bidders in the LLG model. \label{fig:LLG}}
	{We draw 150 observations according to the prior distribution and sample the corresponding bids from the computed discrete distributional strategies using $ \text{SODA}_2 $ (colored shapes). The colored lines indicate the analytical equilibrium strategies for these settings.
		We consider the three core-selecting payment-rules and different correlations according to the Bernoulli weights model with parameter $ \gamma \in \{0.1, 0.5, 0.9\} $.}
\end{figure}
The authors analytically derive the unique symmetric BNE strategies for multiple bidder-optimal core-selecting payment rules including the nearest-zero (NZ), nearest-VCG (NVCG), and nearest-bid (NB) rule in the Bernoulli weights model. These rules all choose the efficient allocation $x$ (according to the submitted bids) but select different price vectors $p$ from the set of core-stable outcomes. 
For example, the nearest-VCG rule picks the point in the core that minimizes the Euclidean distance to the (unique) Vickrey-Clarke-Groves payments. Similarly, the nearest-zero point takes the origin of the coordinate system as a reference point, while the nearest-bid rule minimizes the distance to the vector of submitted bids $b$. 
We report the results for these core-selecting payment rules with different Bernoulli weights $ \gamma \in \{0.1, 0.5, 0.9\} $ in  Table \ref{tab:llg_nz}-\ref{tab:llg_nb}. Since truthful bidding is a dominant strategy for the global bidder, which is easier to approximate and leads to more accurate results in all instances, we only report the results for the local bidders. For $ \gamma=0.5 $ we compare our results to NPGA.

We construct the approximation game by discretizing the spaces $ \Ocal_L = [0,1] $ and $ \Ocal_G = [0,2] $, according to the prior distribution, and the action spaces $ \Acal_i = \Ocal_i, \, i \in \{L,G\} $. Since the local bidders are symmetric we learn a single strategy for both. 
For each update method, we use a single step rule for all different settings, i.e., $ \text{SODA}_1 $ ($ \beta=0.05, \eta_0 = 100 $), $ \text{SODA}_2 $ ($ \beta=0.05, \eta_0 = 50 $), and $ \text{SOMA}_2 $ ($ \beta=0.05, \eta_0 = 50 $).

Overall, we can observe that SODA shows an comparable low utility loss to NPGA. However, NPGA was again run for 15 minutes while $ \text{SODA}_1 $ converged in less than 0.5 minutes and often even within a few seconds.
\input{tables/llg_results_nz}
Across all experiments all methods except for fictitious play converge, i.e., achieve a relative utility loss of less than $ 10^{-4} $ in the discretized game within the $ 1000 $ iterations. Especially Frank-Wolfe and the standard projected gradient ascent ($ \text{SOMA}_2 $) only need a few iterations until the stopping criterion is satisfied. Nevertheless, all computed strategies perform well when compared to the analytical BNE in the continuous setting.
\input{tables/llg_results_nvcg}
\input{tables/llg_results_nb}

\begin{figure}[h]
	\FIGURE
	{
		\includegraphics[width = .33\textwidth]{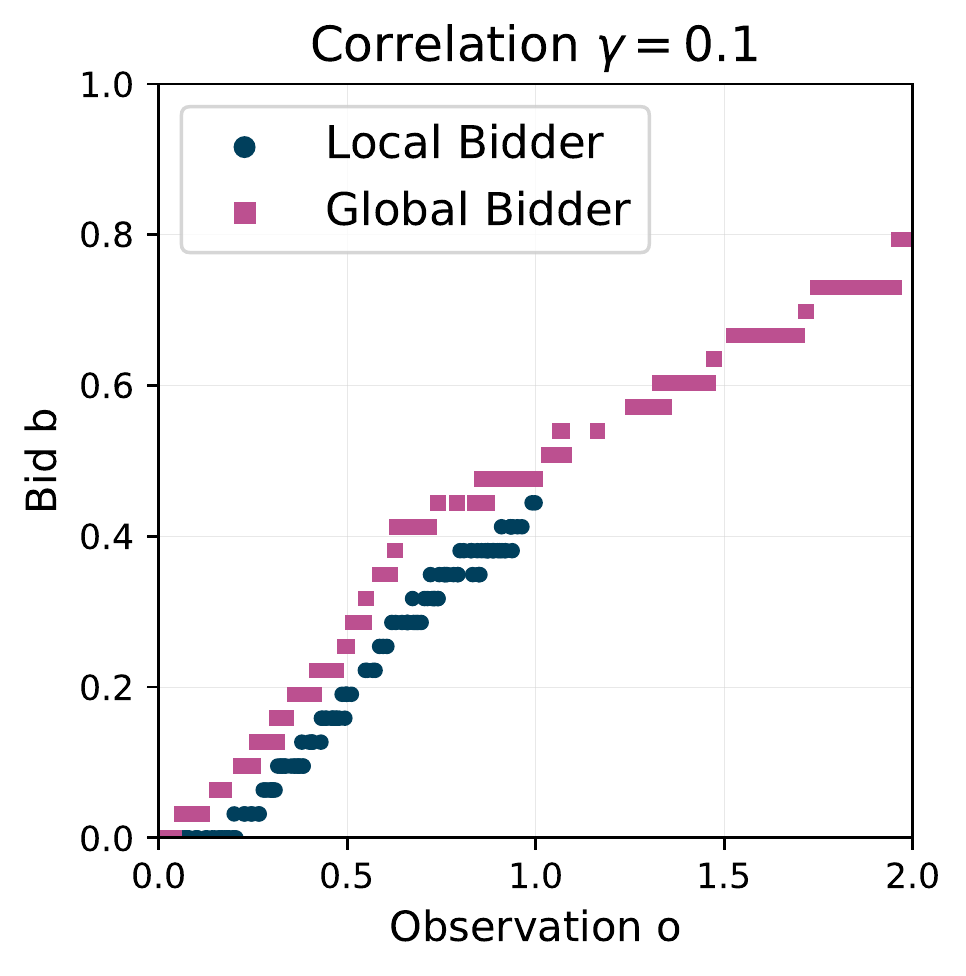}
		\includegraphics[width = .33\textwidth]{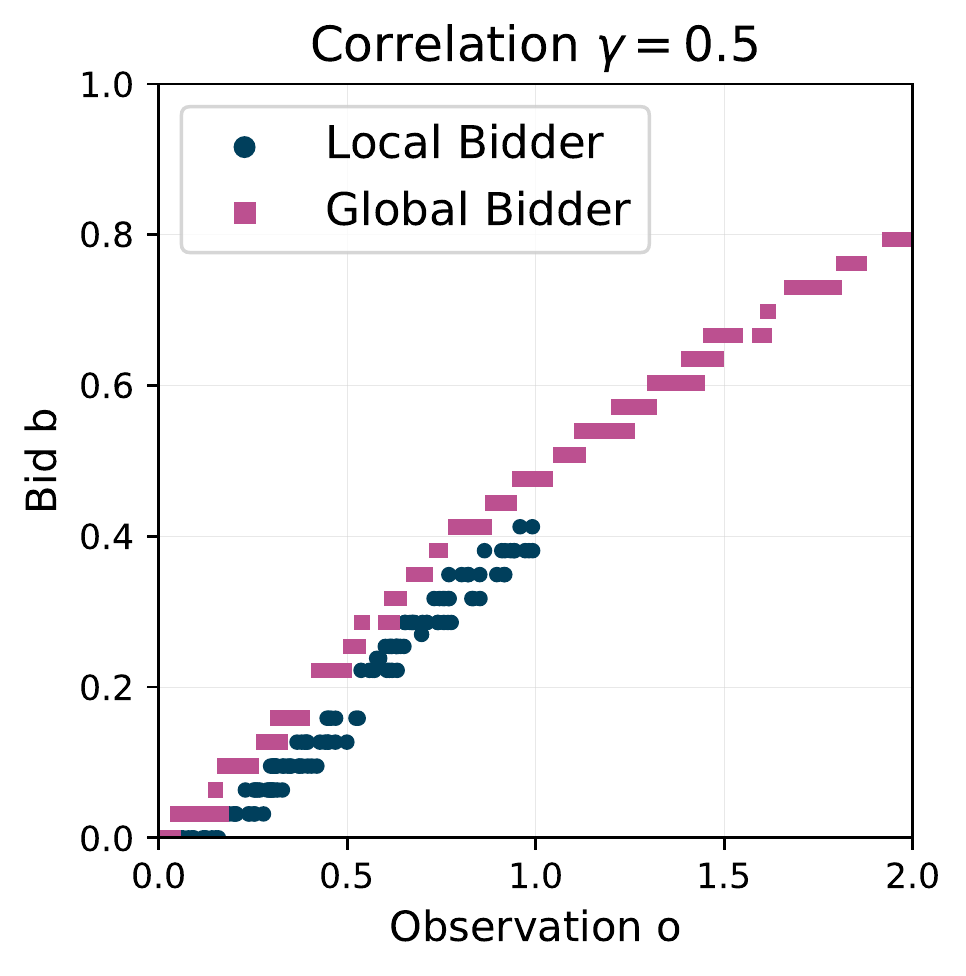}
		\includegraphics[width = .33\textwidth]{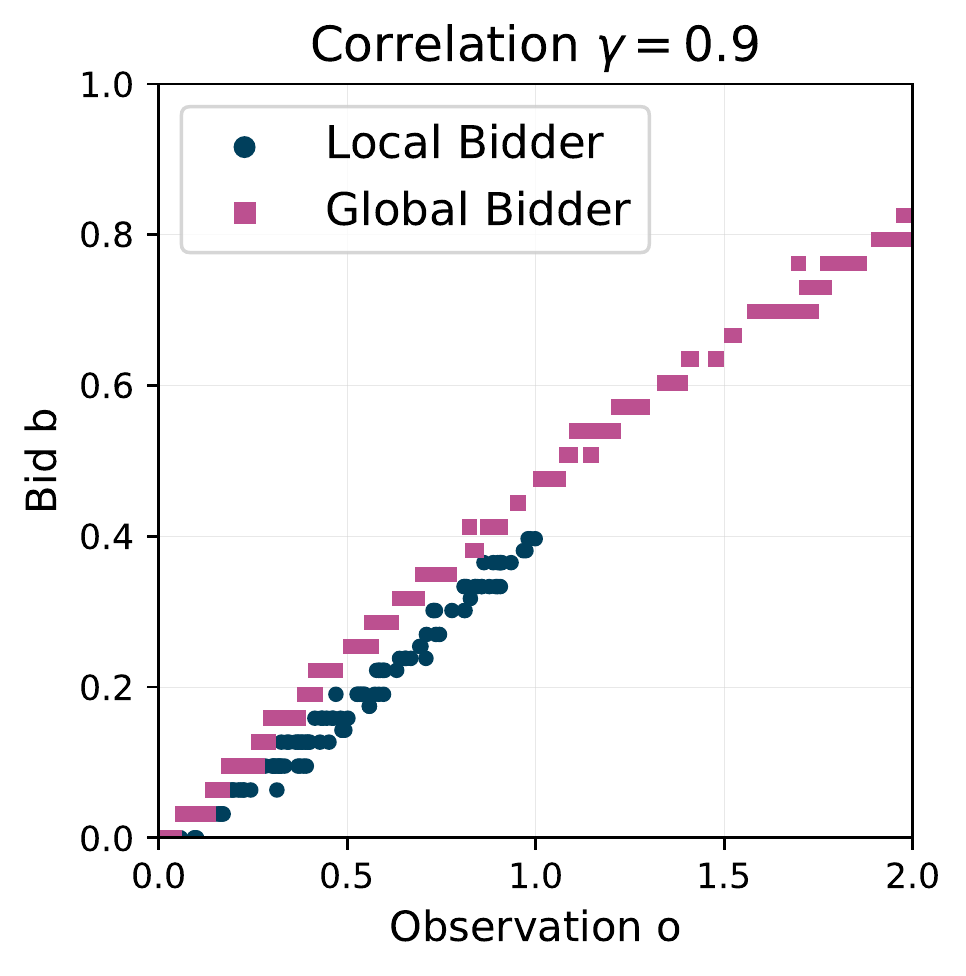}
	}
	{Computed strategies in the LLG model with a first-price payment rule. \label{fig:LLG_FP}}
	{We draw 150 observations according to the prior distribution and sample the corresponding bids from the computed discrete distributional strategies using SOFW (colored shapes) for different correlation parameters $ \gamma $.}
\end{figure}

A setting where no analytical equilibria are known is the LLG model with a first-price payment rule. This auction format is  important as a number of countries used first-price combinatorial auctions in high-stakes spectrum auctions \citep{bichler2017handbook}. Using the Frank-Wolfe algorithm, we converge within 30 seconds in the discretized game. The corresponding equilibrium strategies are visualized in Figure with different levels of correlation \ref{fig:LLG_FP}. 
%The quality of the approximation seems to be worse compared to other experiments, which might be explained by the asymmetry in bidders. 
In contrast to the other settings, the global bidder has no simple dominant strategy. The resulting equilibrium strategy is not as smooth as in other models, but the relative ex-ante utility loss is very small as in other models ($ \ell < 10^{-4}$). 

\subsection{Combinatorial Split-Award Auction} \label{sec:split_award}

Another combinatorial auction environment for which the BNE strategies are known is that of combinatorial split-award procurement auctions 
\citep{kokottBeautyDutchExpost2019}. In contrast to the LLG model, bidders are not single-minded but they are interested in either one share of a contract or the entire contract. 
Importantly, there are two pure BNE for the two symmetric bidders in the FPSB combinatorial procurement market, which makes the analysis interesting. 
A specific version with two suppliers and two lots has been analyzed by \citet{anton1992coordination}. 
Here, suppliers $ i \in \lbrace 1, 2 \rbrace $ can bid on a 100\% and a 50\% share. 
With dis-economies of scale, we have the economically inefficient ``winner-takes-all'' (WTA) equilibrium where one bidder wins both lots (the 100\% share) and a continuum of efficient ``pooling equilibria'' where both suppliers coordinate and each bidder wins one good (a 50\% share) at a high pooling price. The equilibrium with the highest bids on one lot out of all the efficient pooling equilibria is the payoff-dominant strategy for each bidder.
\begin{figure}[h]
	\FIGURE
	{
		\includegraphics[width = 0.45\textwidth]{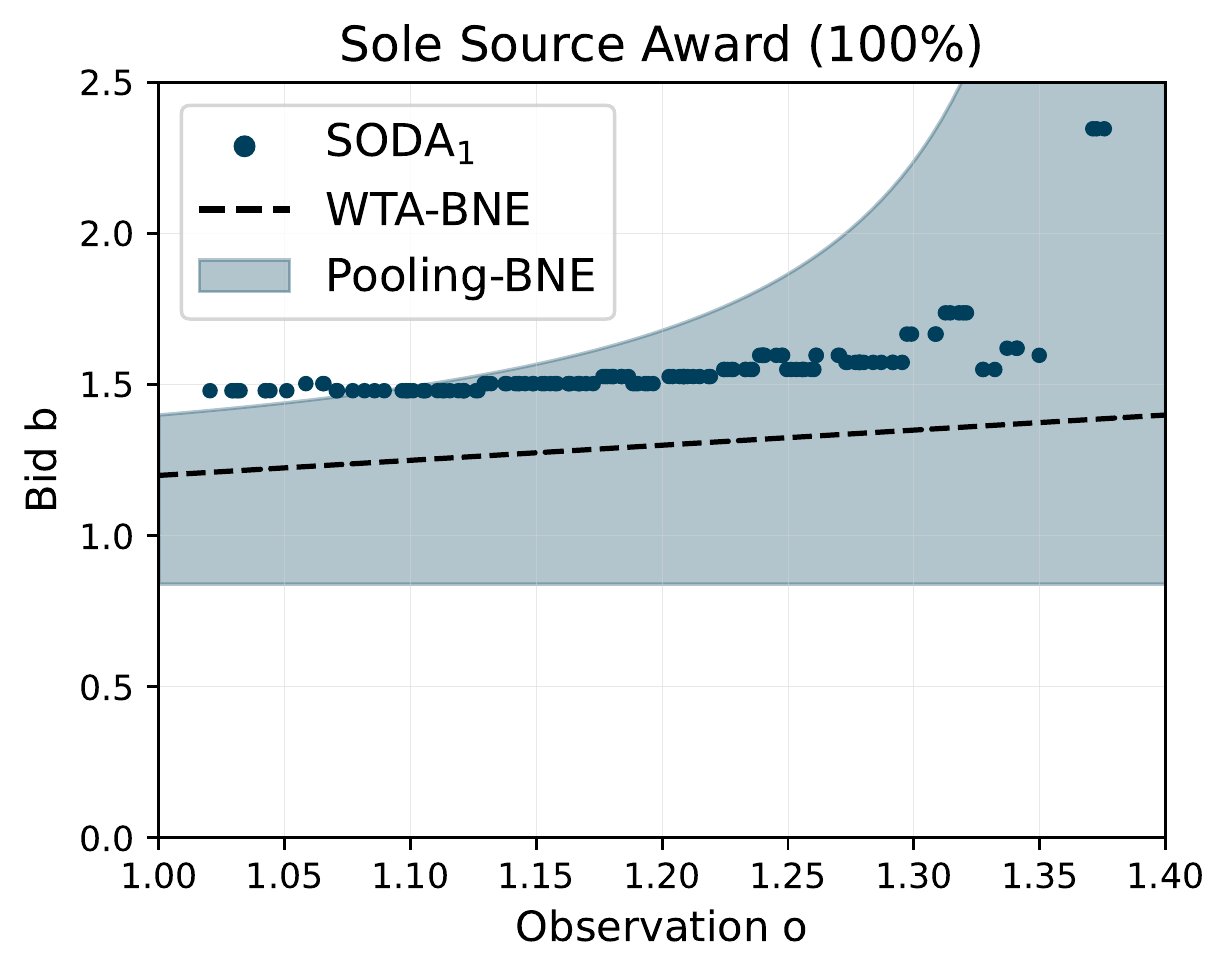}
		\includegraphics[width = 0.45\textwidth]{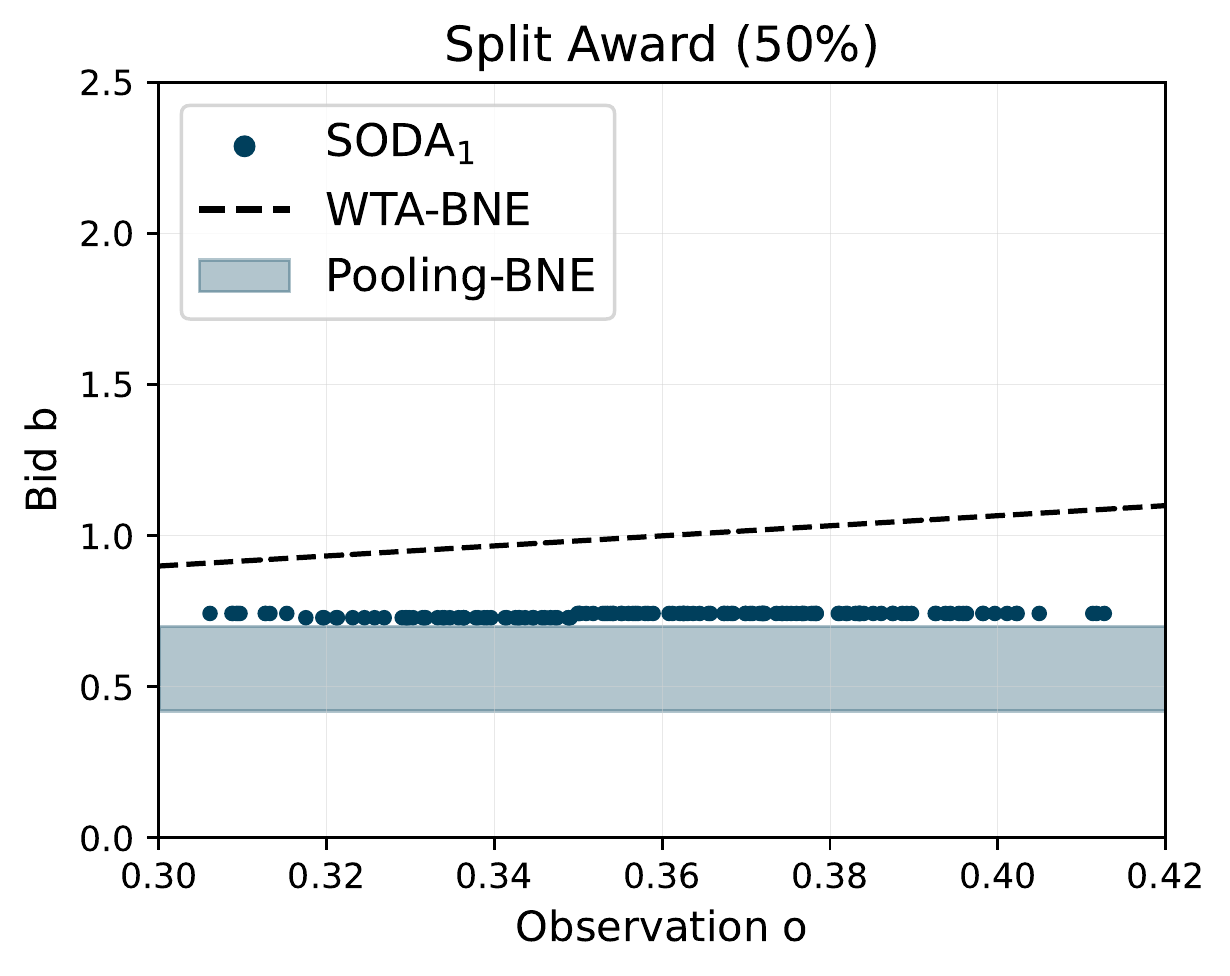}
	}
	{Computed strategy for the FPSB Split-Award Auction with a truncated Gaussian prior. \label{fig:SA}}
	{We draw 150 observations from a truncated Gaussian prior distribution and sample the corresponding bids from the computed discrete distributional strategies (blue dots). The dashed lines indicate the winner-takes-all (WTA) equilibrium, while the shaded area denotes all possible pooling equilibria.}
\end{figure}

We applied SODA to this setting with uniform and Gaussian (truncated with $ l=1.2 $ and $ \sigma=0.1 $) distributed observations. We consider dis-economies of scale and choose marginal costs for the split source of $ C = 0.3 $. The parameters are consistent with experiments from \citet{kokottBeautyDutchExpost2019}. To compare our results to the analytical BNE (Table \ref{tab:sa_gaussian}) we consider a truncated version of the Gaussian prior since the equilibrium analysis requires bounded observations. We can observe that for both priors, Gaussian (Figure \ref{fig:SA}) and uniform, SODA always finds the efficient equilibrium. This is remarkable, because coordination is strategically more challenging than in the WTA equilibrium in which bidders just compete on the 100\% share similar to a single-object auction. In the pooling equilibrium bidders bid high on the 50\% share, but they also need to find a bid on the bundle of both lots (the 100\% share) such that it is not profitable for the opponent to deviate from the pooling equilibrium. 

\input{tables/sa_results_gaussian}

The algorithms take several minutes since we have a two-dimensional action space  $ \mathcal A_i = [1.0, 2.5] \times [0.3 , 1.2] $ where each interval is discretized using $ L=64 $ equidistant points and the observation space $ \mathcal O_i = [1.0, 1.4] $ which is discretized using $K=32$ points. We choose a lower discretization for the observation space because otherwise we would run into memory issues for the computation of the gradients. In Figure \ref{fig:SA} we can observe that in the case of the Gaussian prior, the agents bid slightly above the analytical BNE for the winning bid. This leads to a higher utility compared to the utility in BNE and thereby to a negative utility loss. This also explains the rather large $L_2$ distance in Table \ref{tab:sa_gaussian}. We get more accurate results for the uniform prior, as we can see in Table \ref{tab:sa_uniform}.
Note that we do not consider the $L_2$ distance for the bid on the 100\% share, since the strategy is spread within the continuum of pooling BNE.

\input{tables/sa_results_uniform}

We observe that $ \text{SODA}_1 $ and $ \text{SODA}_2 $ converge within 5 minutes in all instances and achieve results similar to NPGA, which takes around 15 min to get $ \mathcal L = 0.019 $ and a $ L_2 = 0.025 $ for the uniform prior \citep{bichler2023asym}. $ \text{SOMA}_2 $ and the Frank-Wolfe algorithm perform worse, especially for the uniform prior, and Fictitous Play doesn't achieve a sufficient accuracy in any setting. Nevertheless, all methods approximate the payoff-dominant equilibrium.

\subsection{Single-Object Auctions with Risk-Averse Bidders}
In addition to single-object auctions with the standard quasi-linear utility functions, we can also consider extensions such as risk-aversion. As described in Section \ref{sec:notation}, risk aversion can be modeled using a risk attitude $\rho \in (0, 1]$ by transforming the standard quasi-linear utility $u_i^{QL}$ into (strictly) concave payoff functions of the form $u_i^{RA} = (u_i^{QL})^\rho$. This model is also known as \textit{constant relative risk aversion (CRRA)}.

\begin{figure}[h]
	\FIGURE
	{
		\includegraphics[width = .33\textwidth]{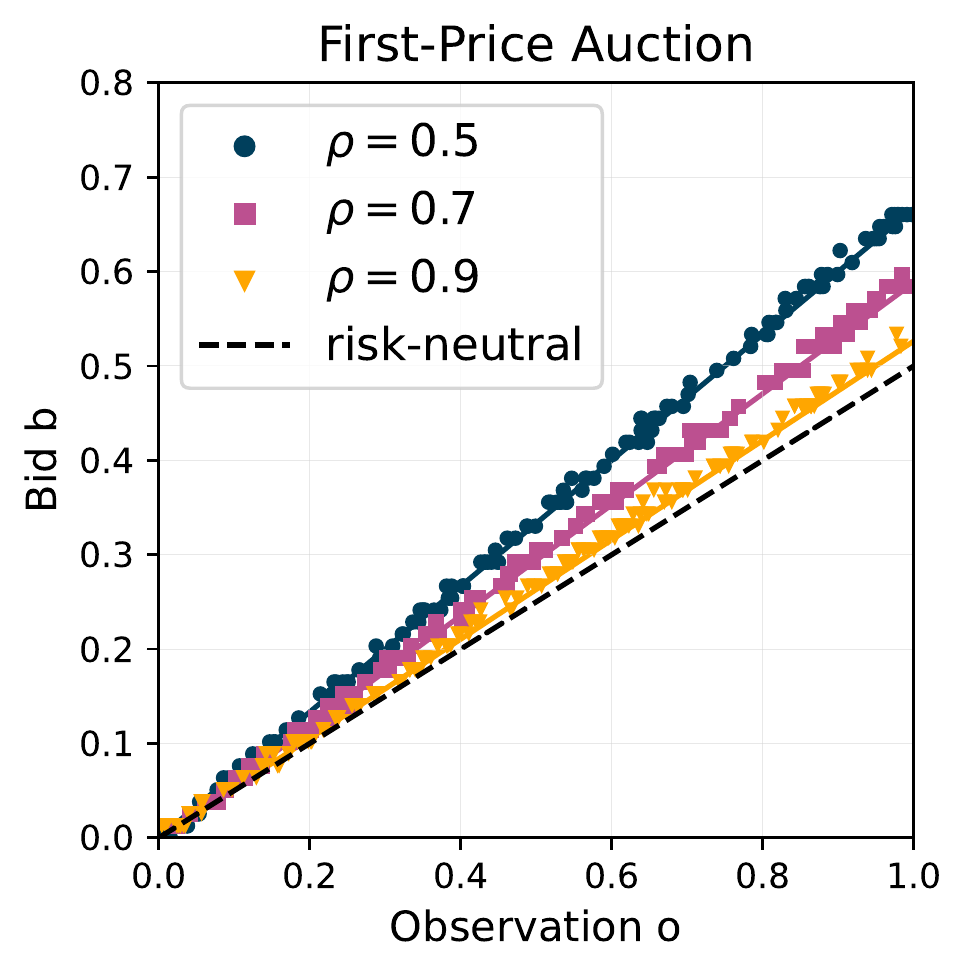}
		\includegraphics[width = .33\textwidth]{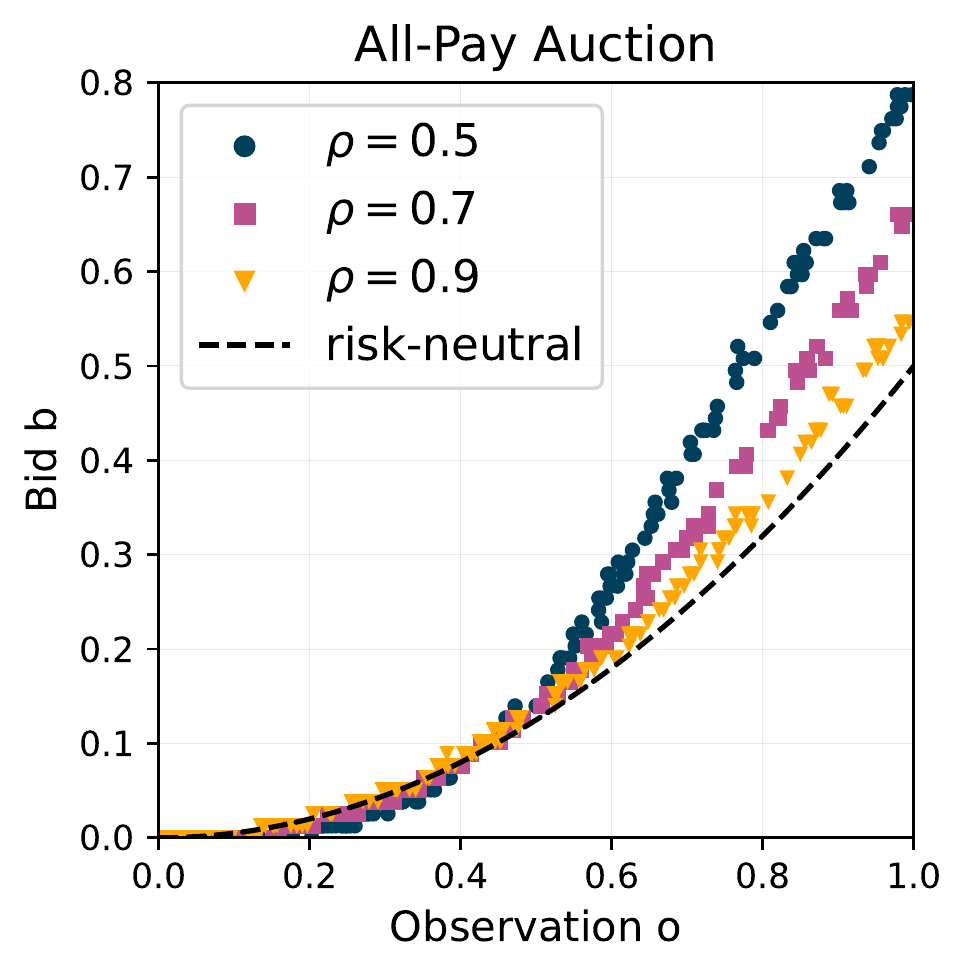}
		\includegraphics[width = .33\textwidth]{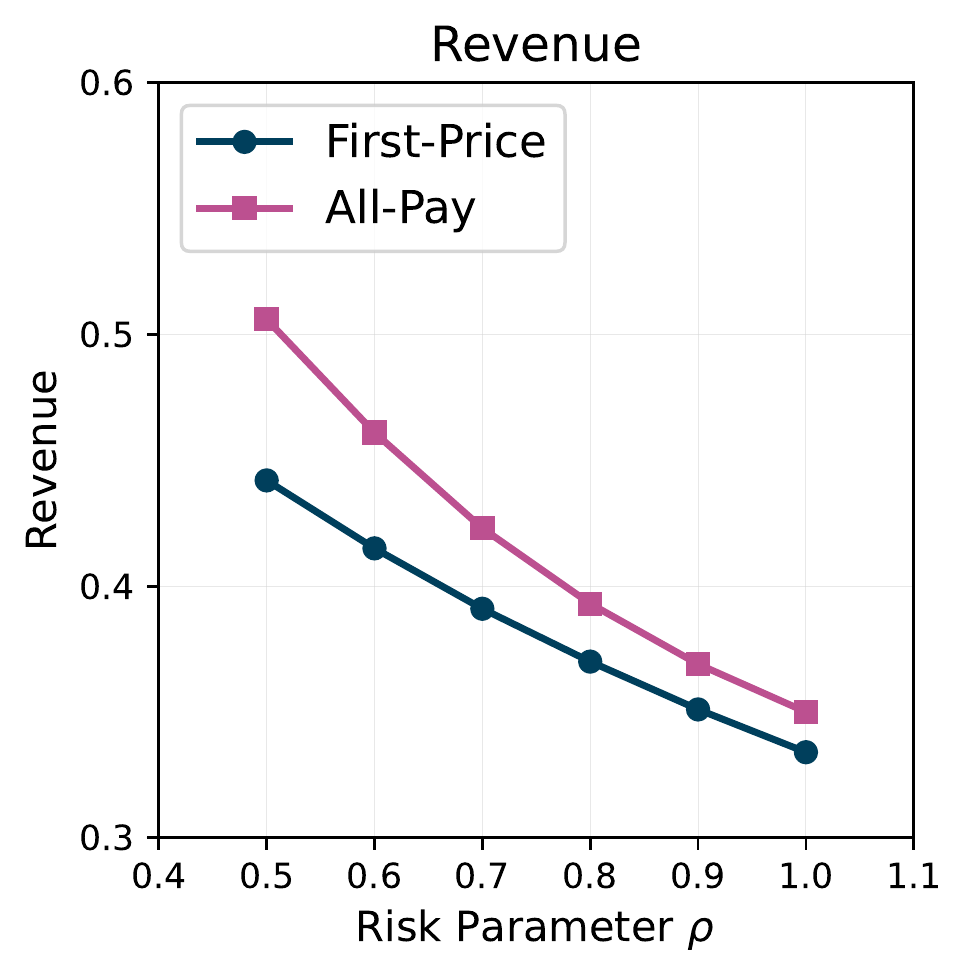}
	}
	{Computed strategies and revenue for the first-price and all-pay auction with risk-averse bidders. \label{fig:risk}}
	{The first two plot shows the equilibrium strategies for the first-price and all-pay auction under risk-aversion compared to the risk-neutral equilibrium strategy (black line). The computed strategies are illustrated by drawing 150 observations according to the prior distribution and sampling the corresponding bids. In the last plot we visualize the approximated expected revenue under different risk parameters.}
\end{figure}

We consider settings with two symmetric bidders who observe their uniformly distributed and private valuations independently.
For the first-price sealed-bid auction it is well known that risk-averse bidders ($ \rho \in (0,1) $) bid higher than risk-neutral bidders ($ \rho=1 $), which leads to a higher revenue for the seller \citep{Maskin1984}. 
For all-pay auctions on the other hand, results are much more limited. \citet{Fibich2006} analyze the first-order conditions and show that in the independent private value setting, risk-averse bidders bid lower for low valuations and higher for high valuations compared to the risk-neutral equilibrium strategy. But they are not able to derive explicit equilibrium strategies or to make statements about how risk-aversion affects the expected revenue. 
Here, our methods can add to the existing literature. While we observe the effects in the equilibrium strategies predicted by \citet{Fibich2006}, we can also observe that risk aversion, similar to first-price auctions, increases the expected revenue in the all-pay auction (Figure \ref{fig:risk}).

\input{tables/fpsb_risk}

For the numerical experiments we consider first-price and all-pay auctions with two symmetric bidders. They independently observe their uniformly distributed valuations from $ \Ocal_i = [0,1] $. We restrict the action space to $ \Acal_i = [0, 0.8] $. The strategies in Figure \ref{fig:risk} are computed using  $ \text{SODA}_1 $ with parameters $\beta=0.05, \,\eta_0 = 25$ for the all-pay and $\beta=0.05, \, \eta_0 = 20$ for the first-price auctions. The revenue is the mean over $2^{22}$ simulated auctions using the computed strategies over ten runs. 
For risk-averse bidders in the first-price auction we can use the analytical solution to evaluate the computed strategies. The results are reported in Table \ref{tab:risk}. 
The parameters for the learning algorithms, i.e., $ \text{SODA}_2 $ with $ \beta=0.05, \, \eta_0=0.1 $ and $ \text{SOMA}_2 $ with $ \beta=0.5,\, \eta_0 = 0.5 $, are constant over the different risk parameters. Note that for our learning algorithm we have to extend the definition of \textit{CRRA} to negative numbers. This is done by $u_i^{RA} = \text{sign}(u_i^{QL}) \cdot \vert u_i^{QL} \vert ^\rho$.

\subsection{Tullock Contests}

Finally, we consider Tullock contests. In contests, agents invest efforts toward winning one or more prizes, and these efforts are costly and irreversible. One distinguishes between perfectly discriminating contests, such as all-pay auctions, where the bidder with the highest effort wins the prize with certainty, and imperfectly discriminating contests, where the probability of winning is a monotonically increasing function of one's own effort (bid). Contests occur in various contexts such as rent-seeking, warfare conflicts, R\&D competition, and the labor market \citep{vojnovic2016}.
The Tullock lottery \citep{tullock1980efficient} is the best known example of such a contest, where the probability of winning the prize is proportional to agents effort. We will focus on the slightly more general $ r$-Tullock contest with parameter $ r>0 $, where the (ex-post) utility of player $ i $ is given by
\begin{equation}
	u_i(b_i,b_{-i},o_i) = \begin{cases}
		o_i \tfrac{b_i^r}{\sum_{j=1}^n b_j^r} - b_i &\text{if } \sum_{j=1}^n b_j > 0 \\
		o_i \tfrac{1}{n} &\text{else}
	\end{cases}.
\end{equation}
If $ r=1 $, the contest corresponds to the aforementioned Tullock lottery. Due to the discontinuity at zero, the model is hard to analyze in the incomplete-information setting. Existence of pure BNE in the IPV model ($ o_i = v_i $ independent for all $ i $) is only known for the concave case, i.e., with $ r\leq1 $, while we only get existence in behavioral strategies for $ r>1 $ \citep{haimanko2021bayesian}. But even in the symmetric, concave case, no analytical equilibrium strategy is known. For $ r \in \{0.5, 1\} $ numerical approximations were obtained by discretizing the integral in the first order condition and iteratively following the best response until convergence is reached \citep{fey2008rent,ryvkin2010contests}. 
In contrast, our method does not rely on the first-order condition and can be easily adapted to more general contests. 
Especially in settings with asymmetric bidders, where the first order condition becomes a system of non-linear ODEs, our approach becomes even more valuable.

\begin{figure}[h]
	\FIGURE
	{
		\includegraphics[width = .33\textwidth]{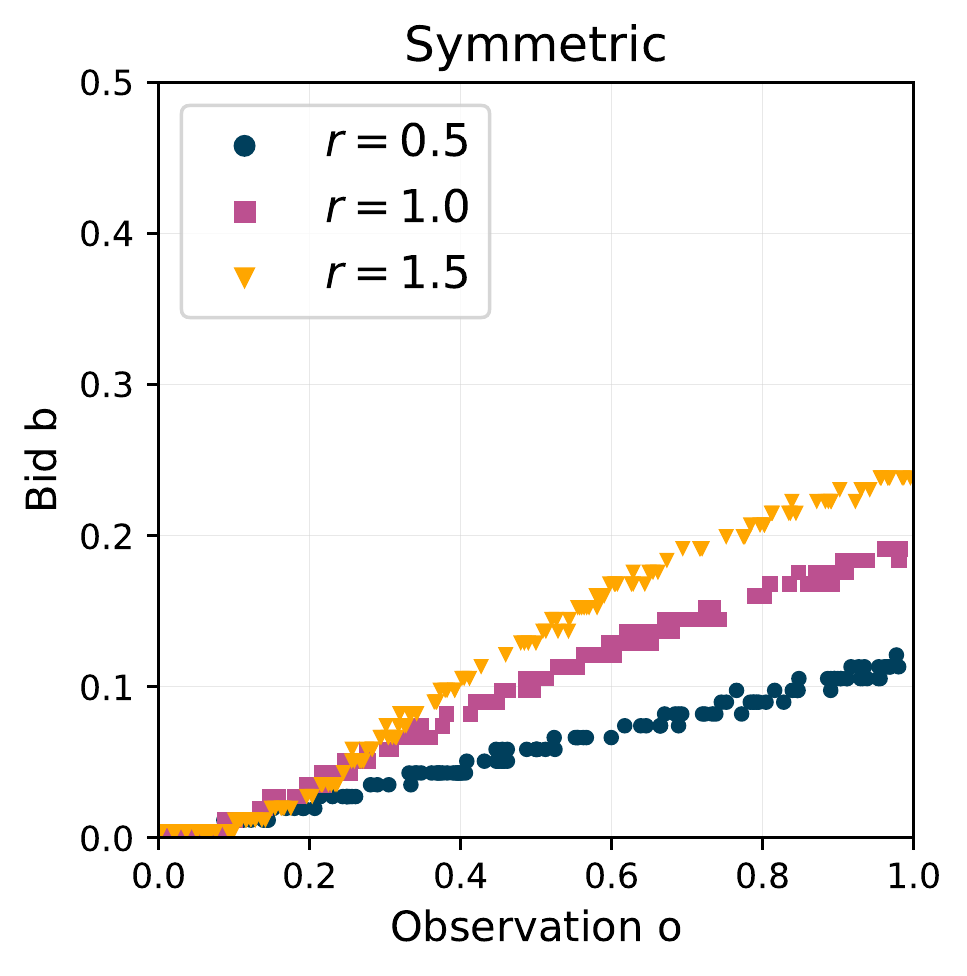}
		\includegraphics[width = .33\textwidth]{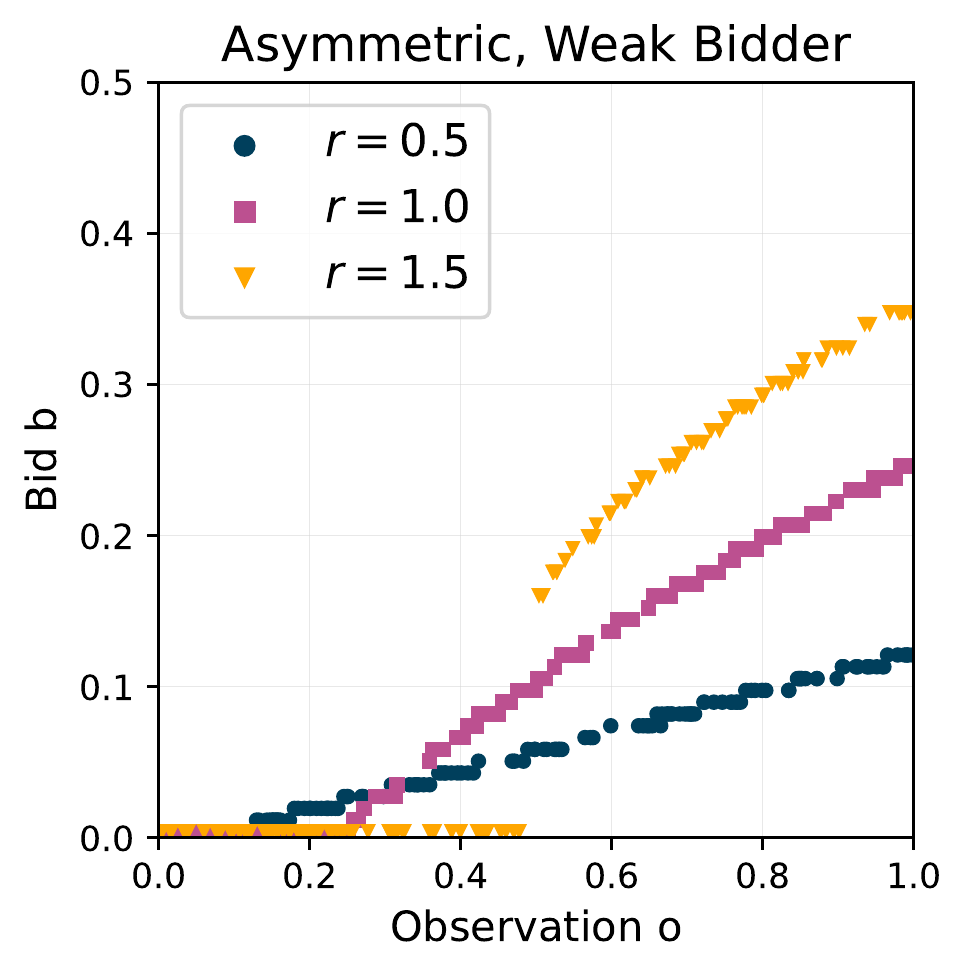}
		\includegraphics[width = .33\textwidth]{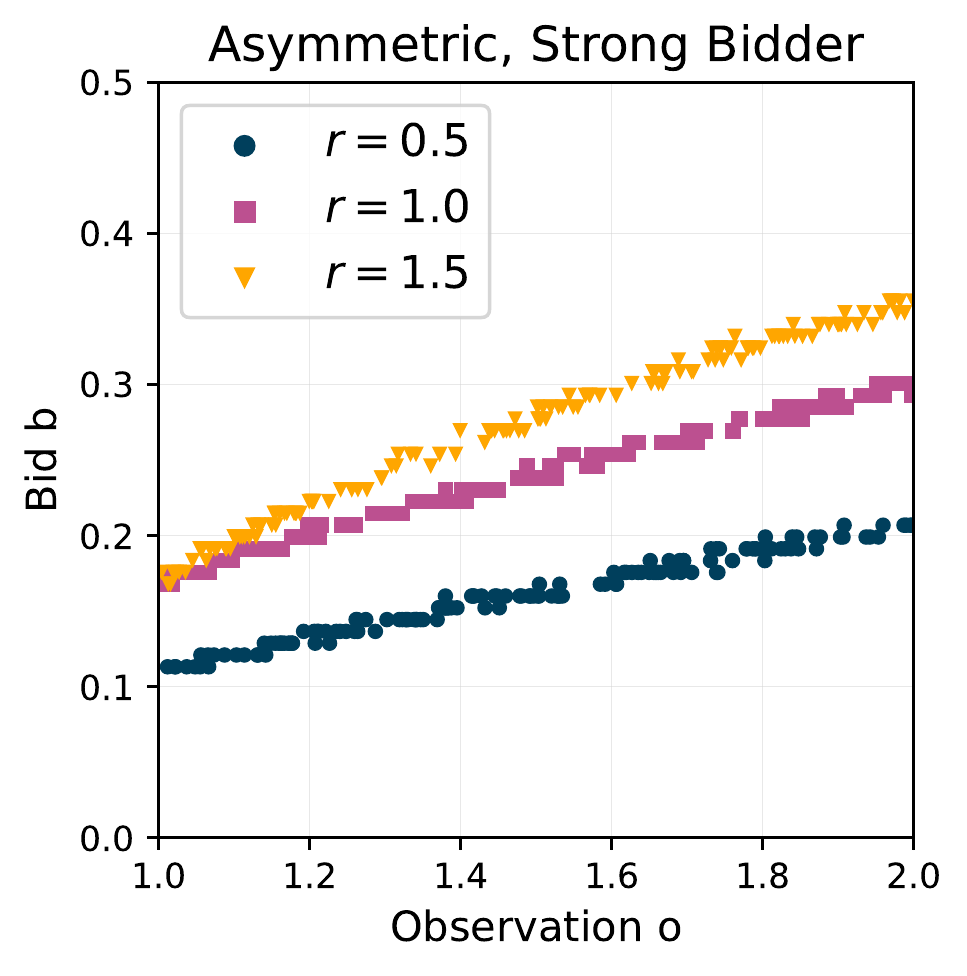}
	}
	{Computed strategies for the generalized Tullock Contest with two symmetric and asymmetric bidders \label{fig:contest}}
	{We draw 150 observations according to the prior distribution and sample the corresponding bids from the computed discrete distributional strategies using $ \text{SODA}_1 $ (colored shapes).
		The first plot shows the equilibrium strategies in the symmetric setting, while the other two plots depict the weak and strong bidder in the asymmetric case.}
\end{figure}
In Figure \ref{fig:contest} we show the computed equilibrium strategies for 2 player r-Tullock contests with $ r \in\{0.5, 1.0, 1.5\}$. Note that this also includes a non-concave setting ($ r=1.5 $) where existence of pure BNE has not been shown yet.
We consider a symmetric version where the valuations of both contestants are uniformly distributed  on $ [0,1] $ and an asymmetric setting, where we have a weak bidder with $ o_{\text{weak}} \sim U([0,1]) $ and a strong bidder with $ o_{\text{strong}} \sim U([1,2]) $. We restrict the actions to $ \Acal_i = [0, 0.5] $ and discretize all spaces with $ K = L = 64 $ equidistant points. 
For dual averaging and mirror descent we used the following parameter:
$ \text{SODA}_1 $: $ \eta=100, \beta=0.05  $, $ \text{SODA}_2 $: $ \eta=10, \beta=0.05  $, $ \text{SOMA}_2 $: $ \eta=100, \beta=0.5$.
It takes all methods less than 0.1s in the symmetric and less than 2s in the asymmetric settings to converge, i.e., achieve a relative utility loss $ \ell < 10^{-4} $ in the discretized game.
\section{Discussion}

SODA converges in a wide range of environments as illustrated in the previous section. In this section, we discuss what is known about convergence and scalability of the approach.

\subsection{Convergence}
Although we can certify equilibrium ex post, an intriguing question remains: why do gradient dynamics converge to an equilibrium in such a wide variety of auctions and contests, even though gradient dynamics don't converge in many finite games \citep{sanders2018prevalence}? 
This is a notoriously challenging question. \citet{andrade2021learning} write that there is little hope for a general understanding of the behaviors arising from optimization-driven dynamics even in normal-form games. 
Whether learning algorithms converge or not depends on the properties of the game being played. Apparently, a wide variety of auctions and contests have properties that allow for SODA to converge to equilibrium. 

There is a long literature on variational inequalities and how they are used to model equilibrium problems \citep{kinderlehrer2000introduction, grossmann2007numerical, geiger2013theorie}. We know that projection algorithms converge if a complete-information game with continuous action spaces satisfies monotonicity or the weaker variational stability condition, but that they do not converge if there are only mixed equilibria \citep{mertikopoulos2019learning, flokas2020no}. 
Variational stability coincides with the existence of (one or more) sharp equilibria in complete-information games \citep{mertikopoulos2019learning}. Unfortunately, it is not easy to assess ex ante whether a specific game has a sharp or even only a pure Nash equilibrium. 

The early theorems of \citet{nash1950equilibrium} and \citet{debreu1952social} reveal that games possess a pure strategy Nash equilibrium if (1) the strategy spaces are nonempty, convex, and compact, and (2) players have continuous and quasi-concave payoff functions. These assumptions are necessary for the fixed-point theorems that the authors draw on. However, in many economic models, the payoffs are discontinuous. Bidders in an auction experience a discontinuous jump in their utility when their bid on some unit increases to the point where it is no longer a losing bid. This led to a literature on equilibrium existence in discontinuous games (see the survey by \citet{reny2020nash}). \citet{athey2001single} introduced the single-crossing property: whenever each opponent uses a non-decreasing strategy in the sense that higher types choose higher actions, a player’s best response strategy is also non-decreasing. When the property holds, a pure-strategy Nash equilibrium exists in every finite-action game. Further, for games with discontinuous payoffs and a continuum of actions, there exists a sequence of pure-strategy Nash equilibria to finite-action games that converges to a PSNE of the continuum-action game. The condition was shown to hold for first-price, multi-unit, and all-pay auctions, as well as pricing games with incomplete-information about costs. \citet{reny2011existence} generalizes these results and also covers more general multi-unit auctions with risk-averse bidders. However, these ex-ante characteristics are not easy to verify and 70 years after Nash's original work understanding whether a game has a pure or even a strict equilibrium is still a challenge. 

One could try to analyze the monotonicity of the continuous ex-ante game. But for this it is important to understand the individual utility functions and their gradients. However, the agents' utility functions are based on an unknown bid function. Without strong assumptions on the functional form of the bid function, it is hard to characterize the payoff gradient explicitly. Appendix \ref{app:plot} summarizes a number of plots, where we do make parametric assumptions on the prior distribution and the bid function. The plots suggest that under a variety of assumptions the resulting expected utility function is quasi-concave or at least unimodal. However, the parametric assumptions are hard to justify.  
SODA is based on the discretized approximation game, not the continuous ex-ante game. Unfortunately, we can show that the approximation game satisfies neither monotonicity nor variational stability globally. Yet, we find convergence in a wide variety of games. A longer discussion and definitions are provided in Appendix \ref{app:variational}.

\subsection{Scalability} \label{sec:scale}

Although convergence is difficult to analyze, we want to provide some drivers for the computational complexity of SODA. 
The main factors are the number of players, the number of items or bundles (which drives the number of strategies), and the level of discretization. If the number of strategies is exponential in the number of items (as in a combinatorial auction with general valuations), then gradient-based optimization as in SODA explores all exponentially-many strategies. As a result, an algorithm learning even only approximate $\varepsilon$-BNE cannot be polynomial in the number of items. \citet{cai2014simultaneous} showed with a similar argument that computing approximate $\varepsilon$-BNE in combinatorial auctions is NP-hard. 

In most auction-theoretical models, the number of items or strategies per agent is small. Examples include single-minded bidders in combinatorial auctions or split-award auctions with two or three items only. Apart from this, a standard assumption in auction theory is that of symmetric priors and symmetric equilibrium strategies, which leads to the fact that we only need to explore the strategies of a single and not of multiple players. For example, if we further assume that the bidders are independent, the computational effort can be further reduced. In such a first-price sealed-bid auction, the expected utility can be written as
\begin{equation} \label{eq:util_sym}
	\tilde u_i (s_1,...,s_n) = \sum \limits_{k, l } (s_i)_{k l} (o_{k} - b_{l}) \mathbb{P}(b_{l} \text{ is highest bid}; s_{-i}).
\end{equation}
Compared to the very general formulation (\ref{eq:linear_util}), where we sum over all combination of bids which grows exponentially in the number of bidders $n$, we compute the first order statistic. This way the complexity does not increase with the number of bidders, which allows us to analyze much larger settings (see Appendix \ref{sec:runtimes}).
So, while we know that the complexity of finding $\varepsilon$-BNE in general is NP-hard, computation is not necessarily a limiting factor in most of the models analyzed in auction theory, where we focus on small markets with a few players only. 

\section{Conclusions} \label{sec:conclusion}

Computing Bayesian Nash equilibria for continuous-type and -action auction games was considered intractable. Sixty years after Vickrey's seminal work on single-object auctions, we still only know equilibrium strategies for very restricted environments such as single-object auctions. These equilibrium problems can be modeled as systems of differential equations and for many model assumptions we don't have a complete mathematical solution theory. 

SODA is a new numerical technique that relies on distributional strategies and a discretization of the type and action spaces that takes the prior distributions into account. 
The method is very fast for auction models with symmetric bidders. In first-price environments with independent private values, SODA computes approximate equilibrium also for large numbers of bidders in seconds, which makes SODA a convenient numerical tool for analysts. 
We analyzed very different types of auctions and contests and SODA converged in all of them. Ex-post verification upon convergence is very useful, because the algorithms are very fast for standard models and analysts are not required to perform costly numerical validation. While these formal convergence results have only been shown for SODA, we demonstrated empirically that other first-order methods are as effective in finding equilibrium strategies. 


\newpage
\begin{APPENDICES}
	
	\section{Running Time with Different Discretizations and Symmetry Assumptions}\label{sec:runtimes}
	
	In what follows, we report the impact of different levels of discretization and number of bidders on the running time, and we explore the performance gains from symmetric models.
	
	First, we investigate the effect of finer discretizations in the approximation game on the accuracy of the approximation. We apply $ \text{SODA}_1 $ ($ \eta_0 = 10, \beta=0.05 $) to a FPSB auction with two symmetric bidders and independent, uniformly distributed priors. The computations are repeated using different numbers of discrete points. More precisely, we discretize the action and observation spaces with $ K = L \in \{16, 32, 64, 128 \} $ equidistant points. We run SODA and stop the algorithm after $ 1 \thinspace 000 $ iterations.
	Afterwards we compare the computed strategies with the analytical BNEs in the continuous setting (i.e., approximate $ \mathcal L $ and $ L_2 $) as described in section \ref{sec:eval}. The results are reported in Table \ref{tab:discr_approx}.
	
	\input{tables/discr_approx}
	
	As expected, increasing the number of discretization points leads to better approximations. 
	But obviously this has a huge effect on the runtime of our algorithm.
	The computation of the gradient requires computing the weighted sum over $ K \cdot L^n $ elements (all possible combinations of valuation and action profiles) in the general formulation as described in Section \ref{sec:model} for one-dimensional spaces. 
	The number of possible outcomes increases exponentially in the number of bidders $ n $ (or the dimension of the spaces).
	Therefore, our method is limited to models with a small number of bidders or items.
	
	\begin{figure}[h]
		\FIGURE
		{
			\includegraphics[width = .45\textwidth]{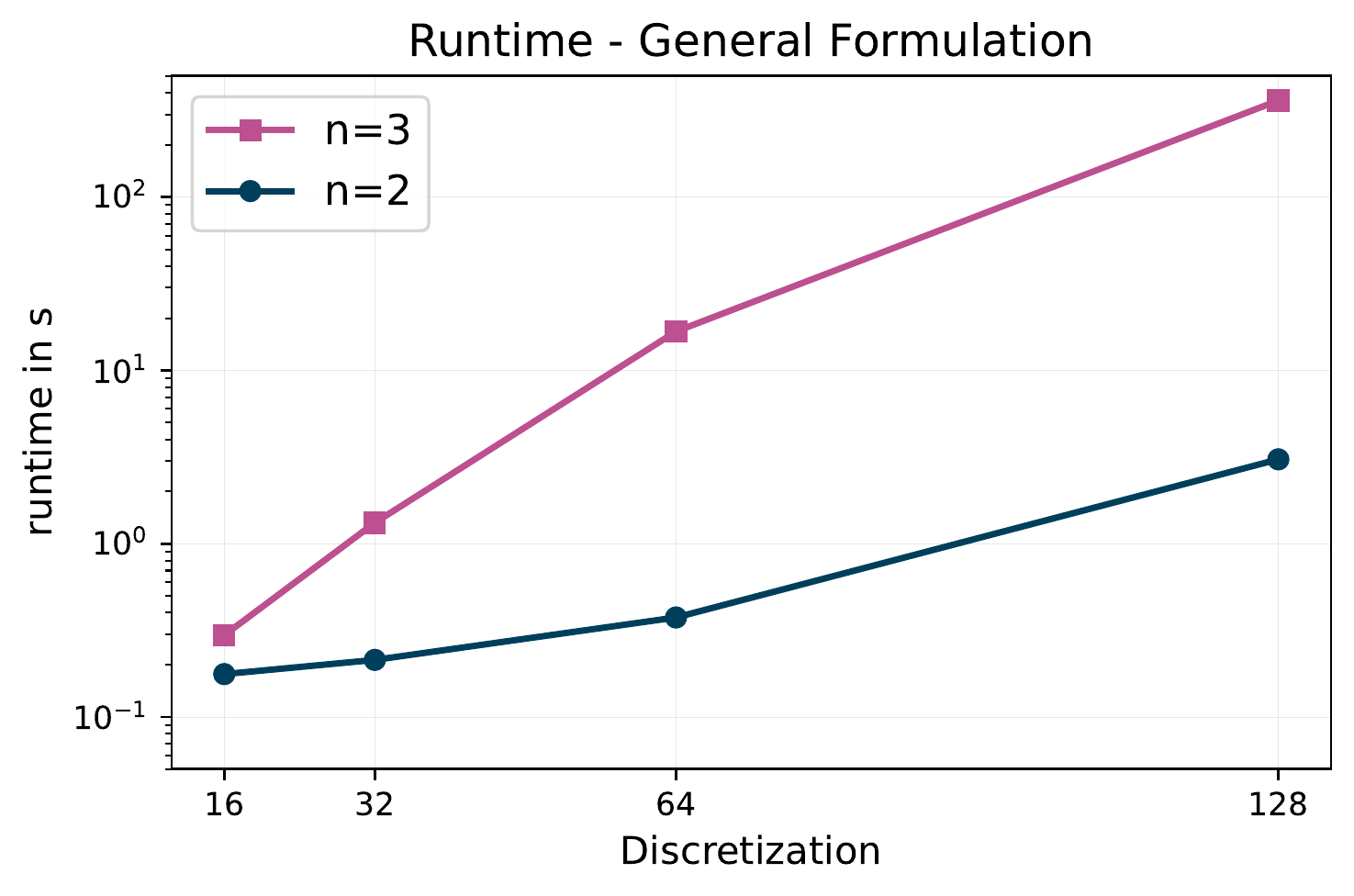}
			\includegraphics[width = .45\textwidth]{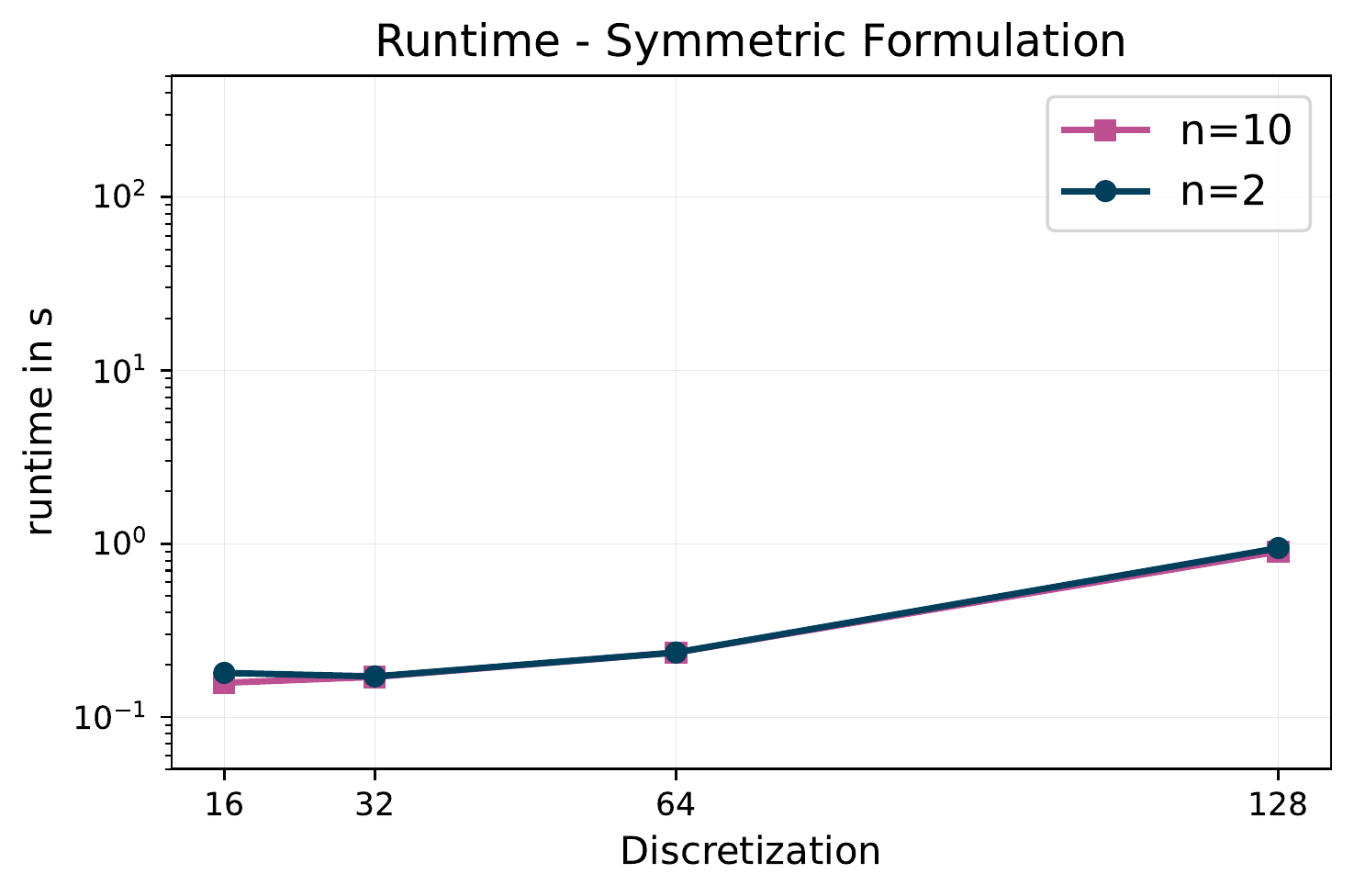}
		}
		{Runtime for a FPSB using the general and symmetric formulation.\label{fig:runtime}}
		{We report the mean runtime for 1000 iterations over 10 runs for a single-item FPSB with uniform prior.}
	\end{figure}
	
	But there are settings, where our method can be used even for a very large number of bidders. As described in Section \ref{sec:scale}, one often considers independent symmetric agents in single-item first-price auctions. 
	This allows us to use an alternative way of calculating the gradient, where the computational effort does not depend on the number of bidders. 
	In Figure \ref{fig:runtime} we can see that increasing the number of bidders from two to just three already increases the runtime for higher discretizations from a few seconds to minutes in the general formulation. 
	Using the symmetric formulation on the other hand allows us to consider any number of bidders.
	
	\section{Proof of Theorem \ref{prop:approx}}\label{app:thm}
	Given the auction game $G = (\Ical, \Vcal, \Acal, f, u)$, we make following assumptions.
	
	\begin{assumption}
		The type spaces $\Vcal_i$ and action spaces $\Acal_i$ are compact intervals of $ \R $.
	\end{assumption}
	\begin{assumption}
		The associated probability measure of the common prior $F$ is absolutely continuous with respect to its marginals $F_i$, with $L_f$-Lipschitz continuous Radon-Nikodym derivative $f$:
		$$
		F(V) = \int_V f(v) dF_1(v_1)\dots dF_n(v_n), \quad V \subset \Vcal \text{ measurable}.
		$$
		We assume $ \Vcal_i $ to be the support of $ F_i $. Since $\Vcal$ is compact, there is also $M > 0$ such that $ F(V_1\times \dots \times V_n) \leq M F_1(V_1)\cdot \dots \cdot F_n(V_n) $.
	\end{assumption}
	\begin{assumption}
		With each agent $i$ there are associated two payment (or transfer) functions $t_i^l: \Acal \rightarrow \R$ and $t_i^w : \Acal \rightarrow \R$, determining the agent's payment when they lose ($t_i^l$) or win ($t_i^w$) the good. All $t_i^l$ and $t_i^w$ are $L_t$-Lipschitz continuous. Moreover, for fixed bids $b_{-i}$ of the other agents, they are nondecreasing in $b_i$.
	\end{assumption}
	\begin{assumption}
		Each agent has a nondecreasing von Neumann-Morgenstern utility function $U_i: \R \rightarrow \R$. Thus, the agent's utility for winning the good is $U_i(v_i - t_i^w(b))$, and for losing it is $U_i(-t_i^l(b))$. The $U_i$ are $L_U$-Lipschitz-continuous.
	\end{assumption}
	\begin{assumption}
		The allocation function $x_i: \Acal \rightarrow [0,1]$ denotes the probability of agent $i$ winning the good, given the bids of all agents. Only maximal bids are winning, i.e., $x_i(b) > 0 \Rightarrow b_i \geq b_j \, \forall j$. We assume that $x_i$ is nondecreasing in $b_i$ for fixed $b_{-i}$, and $\sum_{i=1}^n x_i(b) \leq 1$ for all $b \in \Acal$.
	\end{assumption}
	The ex-post utility of agent $i$ can thus be written as 
	\[ u_i(b,v_i) = x_i(b)U_i(v_i - t^w_i(b)) + (1-x_i(b))U_i(-t^l_i(b)).\]
	It is easy to see that the Lipschitz-continuity of $ U_i $ results in the Lipschitz-continuity of $ u_i $, e.g., for $ v_i, v'_i \in \Vcal_i $
	\begin{align*}
		\vert u_i(b,v_i) - u_i(b,v'_i) \vert 
		&=  \vert x_i(b) (U_i(v_i - t^w_i(b)) - x_i(b) U_i(v'_i - t^w_i(b))) \vert \\
		&\leq \vert U_i(v_i - t^w_i(b)) -  U_i(v'_i - t^w_i(b)) \vert \leq L_U \vert v_i - v_i' \vert.
	\end{align*}

	\begin{assumption}
		For each agent $i$, there is a function $p_i: \Acal_i \rightarrow  \R$, determining the marginal payment at ties: formally, if $b \in \Acal$ is a bid vector such that $b_i$ is a maximal bid and there is a $j \neq i$ with $b_i = b_j$, then $t_i^w(b)-t_i^l(b) = p_i(b_i)$. Hence, at ties marginal payments depend only on agent $i$'s bid $b_i$. Note that the $p_i$ are $L_t$-Lipschitz continuous.
	\end{assumption}
	
	These assumptions include single-object auction formats such as the first-price and the second-price sealed bid auctions, and first-price as well as second-price all-pay auctions (war of attrition) \cite{jackson2005existence}.

	To formally describe our discretized game $ G^d(\Ical, \Vcal^d, \Acal^d, F^d, u) $, we use the following definitions.
	\begin{definition}
		The discrete type space of agent $i$ is a finite subset $\Vcal_i^d \subseteq \Vcal_i$. There is a function $\tau_i: \Vcal_i \rightarrow \Vcal_i^d$, mapping each $v_i \in \Vcal_i$ to its discrete representant $\tau_i(v_i)$ and mapping each $v_i^d \in \Vcal_i^d$ to itself. Denote by $\delta_{\tau} = \max_i \sup_{v_i \in \Vcal_i} |v_i - \tau_i(v_i)|$.
	\end{definition}
	
	If we discretize the valuation space, for instance, using $ N $ equally sized sub-intervals and $ \tau_i $ maps $ v_i \in \Vcal_i $ to the midpoint of the respective interval. Then we get $ \delta_\tau = \frac{1}{2N} \vert \Vcal_i \vert $.  
	
	\begin{definition}
		The discrete action space of agent $i$ is a finite subset $\Acal_i^d \subseteq \Acal_i$. $\Acal_i^d$ contains the minimal and maximal element of $\Acal_i$. Denote by $\alpha_i^+: \Acal_i \rightarrow \Acal_i^d$ the function mapping each $b_i \in \Acal_i$  to the minimal element in $\Acal_i^d$ not smaller than $b_i$. Similarly, denote by $\alpha_i^-: \Acal_i \rightarrow \Acal_i^d$ the function mapping $b_i \in \Acal_i$ to the maximal element in $\Acal_i^d$ not greater than $b_i$. Denote by $\delta_{\alpha} = \max_{s \in \{+,-\}} \max_i \sup_{b_i \in \Acal_i} |b_i - \alpha^s_i(b_i)|$.
	\end{definition}
	\begin{definition}
		The valuations in $\Vcal^d = \Vcal^d_1 \times \dots \times \Vcal^d_n$ are distributed according to probability measure $F^d$ on $\Vcal^d$, given by 
		\[ F^d(\{v^d_1\}\times \dots \times \{v_n^d\}) = F(\tau_1^{-1}(v_1^d) \times \dots \times \tau_n^{-1}(v_n^d)) \text{ for all }  v_i^d \in \Vcal^d_i . \] 
		Consequently, $F^d$ has marginals $F_i^d(\{v_i^d\}) = F_i(\tau_i^{-1}(v_i^d))$ and density $f^d(v_1^d,\dots,v_n^d) = F^d(\{v^d_1\}\times \dots \times \{v_n^d\})/\Pi_i F_i^d(\{v_i^d\})$ with respect to the marginals $F_i^d$. Note that $F^d$ can also be interpreted as a probability measure on $\Vcal$ via $F^d(V) = F^d(V \cap \Vcal^d)$ for $V \subseteq \Vcal$ measurable.
	\end{definition}
	We denote distributional strategies in $G^d$ for agent $i$ by $s_i$, and distributional strategies in $G$ by $\sigma_i$.
	Given a discrete strategy $s_i$ for agent $i$, possibly computed by our algorithm, it is straightforward to construct a corresponding distributional strategy $\sigma_i$ which is feasible for the game $G$: Given an arbitrary type $v_i \in \Vcal_i$, compute its discrete representant $\tau_i(v_i)$. Then choose strategy $b_i^d \in \Acal_i^d \subseteq \Acal_i$ with the same probability as $b_i^d$ is chosen in the discrete game when agent $i$ has type $\tau_i(v_i)$. Formally, we set
	\[
	\sigma_i(V_i \times \{b_i^d\}) = \sum_{v_i^d \in \Vcal_i^d} F_i(V_i \cap \tau_i^{-1}(v_i^d))\frac{s_i(\{v_i^d\} \times \{b_i^d\})}{F_i^d(\{v_i^d\})}
	\]
	for $V_i \subseteq \Vcal_i$ measurable. We call this $\sigma_i$ the strategy induced by $s_i$. Since $s_i$ has $\Vcal_i^d$-marginal $F_i^d$, $s_i(\{v_i^d\}\times \Acal_i^d) = F_i^d(\{v_i^d\})$, and $\sigma_i(V_i \times \Acal_i) = \sum_{v_i^d \in \Vcal_i^d} F_i(V_i \cap \tau_i^{-1}(v_i^d)) = F_i(V_i)$, so $\sigma_i$ is indeed feasible for the game $\Gamma$.
	
	\begin{lemma} \label{lem:disc_cont}
		Let $s = (s_1,\dots,s_n)$ be a strategy profile of the discretized game $G^d$ and $\sigma = (\sigma_1,\dots,\sigma_n)$ the strategy profile of the continuous game $G$, where the $\sigma_i$ are induced by $s_i$. Then the difference in the expected utilities is $|\tilde u_i(\sigma)- \tilde u_i(s)| \leq L_U \delta_{\tau}$.
	\end{lemma}
	\begin{proof1}
		Consider fixed $b_i^d \in \Acal_i^d$ and $v_i^d \in \Vcal_i^d$ for all agents $i$. Set $V_i = \tau_i^{-1}(v_i^d)$ and define $V = V_1 \times \dots \times V_n$ and $A = \{b_1^d\} \times \dots \times \{b_n^d\}$. Using the definitions of $ \sigma_i $ and $ f^d $ we get
		\begin{align*}
			\int_{V \times A} f(v) &d\sigma_1(v_1,b_1)\dots d\sigma_n(v_n,b_n) \\ 
			&= \Pi_i \frac{s_i(\{v_i^d\}\times \{b_i^d\})}{\Pi_i F_i^d(\{v_i^d\})} \int_{V \times A} f(v) dF_1(v_1)\dots dF_n(v_n) \\
			&= \Pi_i s_i(\{v_i^d\}\times \{b_i^d\}) \frac{F(V)}{\Pi_i F_i^d(\{v_i^d\})}
			= \Pi_i s_i(\{v_i^d\}\times \{b_i^d\}) f^d(v_i^d) \\
			&= \int_{V \times A} f^d(v) ds_1(v_1,b_1)\dots ds_n(v_n,b_n).
		\end{align*}
		It follows that
		\[
		\int_{V \times A} u_i(b,v_i)f^d(v) ds(v,b) = \int_{V \times A} u_i(b,v_i^d)f(v)d\sigma(v,b), 
		\]
		where $b^d = (b_1^d,\dots,b_n^d)$. Now
		\begin{align*}
			&\left|\int_{V \times A} u_i(b,v_i)f(v)  d\sigma(v,b) - \int_{V \times A} u_i(b,v_i^d)f(v) d\sigma(v,b)\right| \\
			&\leq L_U \delta_{\tau}\int_{V \times A} f(v) d\sigma(v,b).
		\end{align*}
		In the last step we used that $ u_i $ is Lipschitz continuous and non-decreasing in $ v_i $, i.e., $ u_i(b,v_i) - u_i(b,v_i^d) \leq L_U \delta_\tau $.
		Hence, summing over all such sets $V$ and $A$, we get
		\[
		|\tilde u_i(\sigma)-\tilde u_i(s)| \leq L_U \delta_{\tau}.
		\]
	\end{proof1}
	
	In the next step, we want to compare the utility of a continuous strategy $ \sigma $ compared to the strategy $ \tilde \sigma_i $ induced by the discrete strategy $ s_i $, which was in return induced by $ \sigma_i $. To do so, we have to define the discrete strategy $ s_i $ which is induced by $ \sigma_i $.
	
	Define a function $\psi: \Vcal_i \times \Acal_i \rightarrow  \Vcal_i^d  \times \Acal_i^d$ by
	\[
	\psi(v_i, b_i) = \begin{cases}
		(\tau_i(v_i), \alpha_i^+(b_i)) \text{ if } v_i -p_i(b_i) \geq 0 \\
		(\tau_i(v_i), \alpha_i^-(b_i)) \text{ else. }
	\end{cases}
	\]
	Thus, we define the discrete strategy $s_i$ by $s_i(\{v_i^d\}\times \{b_i^d\}) = \sigma_i(\psi^{-1}(v_i^d,b_i^d))$.
	
	\begin{lemma} \label{lem:cont_cont}
		Let $\sigma$ be a strategy profile in the continuous game $G$ and $i$ an arbitrary agent. Then there is a strategy $\tilde \sigma_i$ that is induced by a strategy $s_i$ of the discrete game $G^d$ such that $\tilde u_i(\tilde \sigma_i, \sigma_{-i}) \geq \tilde u_i(\sigma) - L_U(4L_t \delta_{\alpha} + \delta_{\tau})$.
	\end{lemma}
	\begin{proof1}
		The proof is similar to the proof of Lemma 7 in \cite{jackson2005existence}.
		We denote $\tilde \sigma_i$ the continuous strategy induced by $s_i$. 
		Let $V_i = \tau_i^{-1}(v_i^d)$ and $A_i = \{b_i^d\}$.\\
		First, we are going to show that for $(v_i,b_i) \in \psi^{-1}(v_i^d, b_i^d)$ and for arbitrary $(v_{-i},b_{-i}) \in \Vcal_{-i} \times \Acal_{-i}$, we have that $|u_i(b,v_i)-u_i(b_i^d,b_{-i},v_i^d)|$ is small:\\
		By the Lipschitz continuity of the payment functions, we have $|t^s_i(b_i,b_{-i})-t^s_i(b_i^d,b_{-i})| \leq L_{t}\delta_{\alpha}$ for $s \in \{w,l\}$, so
		\begin{align*}
			|U_i(v_i-t_i^w(b_i,b_{-i})) - U_i(v_i^d - t_i^w(b_i^d,b_{-i}))| &\leq L_{U}(\delta_{\tau} + L_t\delta_{\alpha}) \\
			|U_i(-t_i^l(b_i,b_{-i}))-U_i(-t_i^l(b_i^d,b_{-i}))| &\leq L_UL_t\delta_{\alpha}.
		\end{align*} 
		We distinguish two cases: either the allocation for agent $i$ changes when the bid changes from $b_i$ to $b_i^d$, or it does not change. If it does not change, i.e., $x_i(b_i,b_{-i}) = x_i(b_i^d, b_{-i})$, then
		\begin{align*}
			u_i(b_i^d,b_{-i},v_i) 
			&= x_i(b_i^d,b_{-i})U_i(v_i-t_i^w(b_i^d,b_{-i})) + (1-x_i(b_i^d,b_{-i}))U_i(-t_i^l(b_i^d,b_{-i}))\\
			&\leq x_i(b_i,b_{-i})U_i(v_i-t_i^w(b_i,b_{-i})) + (1-x_i(b_i,b_{-i}))U_i(-t_i^l(b_i,b_{-i}))\\
			& \quad + x_i(b_i,b_{-i}) L_U (\delta_\tau + L_t \delta_\alpha) + (1-x_i(b_i,b_{-i})) L_U L_t \delta_\alpha \\ 
			&\leq u_i(b,v_i) + L_U(\delta_{\tau} + L_t\delta_{\alpha}).
		\end{align*}
		Now consider the case where allocations differ, i.e., $x_i(b_i,b_{-i}) \neq x_i(b_i^d,b_{-i})$. Let us consider the case $b_i^d > b_i$, i.e., $b_i^d = \alpha_i^+(b_i)$ - the case $b_i^d < b_i$ can be treated similarly. Then there exists  some bid $\tilde b_i \in [b_i,b_i^d]$ such that there is a tie between bidder $i$ and some other bidder. Consequently, we have $t_i^w(\tilde b_i, b_{-i})-t_i^l(\tilde b_i, b_{-i}) = p_i(\tilde b_i)$, so
		\begin{align*}
			& |(t_i^w(b_i,b_{-i})-t_i^l(b_i,b_{-i}))-p_i(b_i)| \\ 
			& = | t_i^w(b_i,b_{-i}) -t_i^w(\tilde b_i,b_{-i}) -t_i^l(b_i,b_{-i}) -t_i^l(\tilde b_i,b_{-i}) - p_i(b_i)- p_i(\tilde b_i)| \\
			%\leq& | t_i^w(b_i,b_{-i}) -t_i^w(\tilde b_i,b_{-i}) \vert + \vert t_i^l(b_i,b_{-i}) -t_i^l(\tilde b_i,b_{-i}) \vert +  \vert p_i(b_i)- p_i(\tilde b_i)| \\
			&\leq  3 L_t |b_i - \tilde b_i|.
		\end{align*}
		Since $v_i - p_i(b_i) \geq 0$, this implies 
		\[
		v_i - t_i^w(b_i,b_{-i}) \geq - t_i^l(b_i,b_{-i}) + 3L_t|b_i - \tilde b_i|,
		\]
		and therefore
		\begin{align*}
			U_i(v_i-t_i^w(b_i,b_{-i}))
			&\geq U_i( - t_i^l(b_i,b_{-i}) + 3L_t|b_i - \tilde b_i|) \\
			&\geq  U_i( - t_i^l(b_i,b_{-i})) - 3L_U L_t|b_i - \tilde b_i|
		\end{align*}
		and thus, using that $x_i(b_i^d,b_{-i}) \geq x_i(b)$,
		\begin{align*}
			u_i(b,v_i)
			&= x_i(b)U_i(v_i-t_i^w(b_i,b_{-i})) + (1-x_i(b))U_i(-t_i^l(b_i,b_{-i})) \\ 
			&= x_i(b_i^d,b_{-i})U_i(v_i-t_i^w(b_i,b_{-i})) + (1-x_i(b_i^d,b_{-i}))U_i(-t_i^l(b_i,b_{-i})) \\
			& \quad + (x_i(b) - x_i(b_i^d,b_{-i}))(U_i(v_i-t_i^w(b_i,b_{-i})) - U_i(-t_i^l(b_i,b_{-i})) )\\
			&\leq x_i(b_i^d,b_{-i})U_i(v_i-t_i^w(b_i,b_{-i})) + (1-x_i(b_i^d,b_{-i}))U_i(-t_i^l(b_i,b_{-i})) + 3L_U L_t \delta_{\alpha} \\ 
			&\leq x_i(b_i^d,b_{-i})U_i(v_i-t_i^w(b^d_i,b_{-i})) + (1-x_i(b_i^d,b_{-i}))(U_i(-t_i^l(b^d_i,b_{-i})) + 4L_U L_t\delta_{\alpha}) \\
			&= u_i(b_i^d, b_{-i}, v_i) + 4L_UL_t \delta_{\alpha} \\ 
			&\leq u_i(b_i^d, b_{-i}, v^d_i) + 4L_UL_t \delta_{\alpha} + L_U\delta_{\tau}.
		\end{align*}
		By using an analogous argument, we arrive at the same bound for the case $b_i^d = \alpha_i^-(b_i)$. \\
		Let us now evaluate the expected utilities with respect to $\sigma_i$ and $\tilde \sigma_{i}$. For $(v_i^d,b_i^d) \in \Vcal_i^d \times \Acal_i^d$ and fixed $v_{-i}, b_{-i}$, we have that
		\begin{align*}
			\int_{\psi^{-1}(v_i^d,b_i^d)} u_i(b,v_i)f(v)d\sigma_i 
			&\leq \int_{\psi^{-1}(v_i^d,b_i^d)} (u_i(b_i^d,b_{-i},v^d_i)+4L_UL_t\delta_{\alpha}+L_U\delta_{\tau})f(v)d\sigma_i \\
			&= \int_{\psi^{-1}(v_i^d,b_i^d)} (u_i(b_i^d,b_{-i},v^d_i)+4L_UL_t\delta_{\alpha}+L_U\delta_{\tau})f(v)d\tilde \sigma_i.
		\end{align*}
		By summing the integral over all sets $\psi^{-1}(v_i^d,b_i^d)$ and integrating with respect to $\sigma_{-i}$, we see that $ \tilde u_i(\sigma) \leq \tilde u_i(\tilde \sigma_i, \sigma_{-i}) + L_U(4L_t\delta_{\alpha}+\delta_{\tau}) $.
	\end{proof1}

	\begin{customthm}{\ref{prop:approx}}
		Let $s \in \Scal^d$ be an $\varepsilon$-BNE of the discrete game $G^d$ of a single-object auction that satisfies the assumptions 1-6. Let $\sigma \in \Scal$ be the strategy profile, where each $\sigma_i$ is the strategy induced by $s_i$. Then $\sigma$ is an $ \varepsilon + \Ocal(\delta_\alpha + \delta_\tau)$-BNE of the continuous game $G$.
	\end{customthm}
	
	\begin{proof1}
		Let $\sigma_i^*$ be a best response to $\sigma_{-i}$.
		Then $ \sigma_i^* $ induces a strategy $ \tilde s_i $ in the discrete game, which in turn induces a continuous strategy $ \tilde \sigma_i $.
		By Lemma \ref{lem:cont_cont} we get the following bound of the expected utility of the best response with respect to $ \tilde \sigma_i $.
		\begin{align*}
			\tilde u_i(\sigma_i^*,\sigma_{-i}) 
			&\leq \tilde u_i(\tilde\sigma_i,\sigma_{-i}) + L_U(4L_t\delta_{\alpha}+\delta_{\tau}) \\ 
			\intertext{Since $ (\tilde \sigma_i, \sigma_{-i}) $ is a strategy profile induced by the discrete profile $ (\tilde s_i, s_{-i}) $, we can use Lemma \ref{lem:disc_cont} and further obtain}
			&\leq \tilde u_i(\tilde s_i, s_{-i})+ L_U(4L_t\delta_{\alpha}+\delta_{\tau}) + L_U \delta_\tau. \\
			\intertext{Since $ (s_i,s_{-i}) $ is an $ \varepsilon $-equilibrium the previous term is bounded by }
			&\leq \tilde u_i(s_i, s_{-i}) + \varepsilon + L_U(4L_t\delta_{\alpha}+\delta_{\tau}) + L_U \delta_\tau
			\intertext{and applying again Lemma \ref{lem:disc_cont} to get a bound with respect to the induced profile $ \sigma $, we obtain}
			& \leq \tilde u_i(\sigma_i, \sigma_{-i}) + \varepsilon + L_U(4L_t\delta_{\alpha}+\delta_{\tau}) +2 L_U \delta_\tau.\\
			\intertext{Thus we get our claim that}
			\tilde u_i(\sigma_i^*,\sigma_{-i}) &\leq \tilde u_i(\sigma_i, \sigma_{-i}) + \varepsilon + \Ocal(\delta_\alpha + \delta_\tau).
		\end{align*}
	\end{proof1}
	\section{Expected utility functions based on different parametric assumptions.}\label{app:plot}
	
	In this appendix we provide plots of the expected utility function for different parametric assumptions of the bid function and different distributional assumptions. Figures \ref{fig:fpsb_linear} to \ref{fig:fpsb_sigmoid} illustrate ex-interim utilities for specific value draws ($o_i=0.7$) of a bidder. Plots for lower or higher valuations have similar properties.
	
	\begin{figure}[h]
		\centering
		\includegraphics[width=1.00\textwidth]{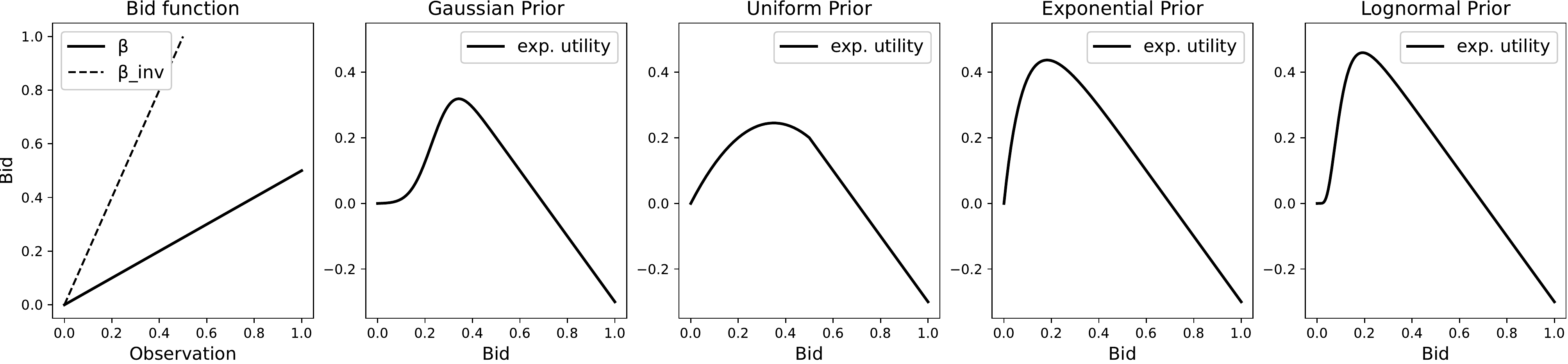}
		\caption{Expected utility with a linear bid function in a first-price sealed-bid auction.}
		\label{fig:fpsb_linear}
	\end{figure}
	
	\begin{figure}[h]
		\centering
		\includegraphics[width=1.00\textwidth]{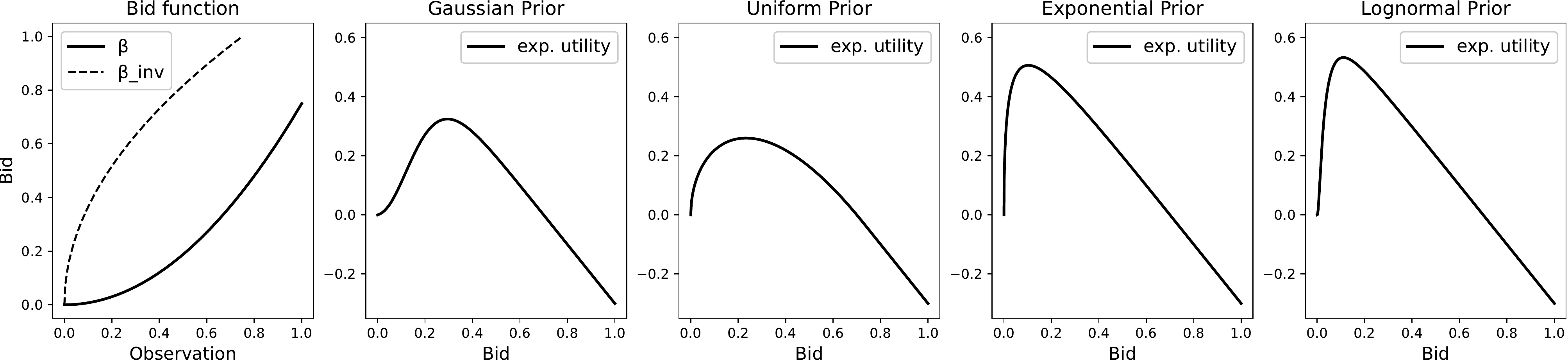}
		\caption{Expected utility with a convex bid function in a first-price sealed-bid auction.}
		\label{fig:fpsb_convex}
	\end{figure}

	\begin{figure}[h]
		\centering
		\includegraphics[width=1.00\textwidth]{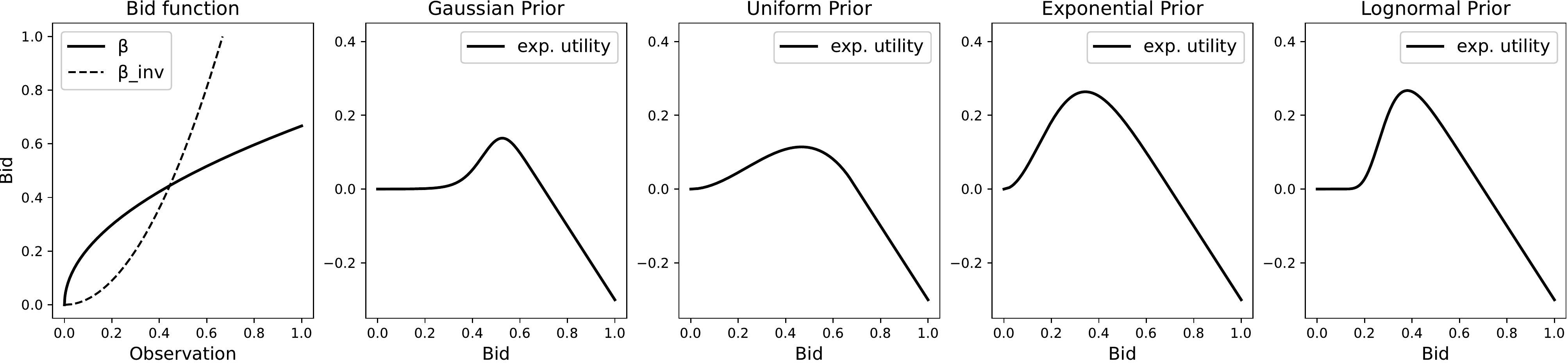}
		\caption{Expected utility with a concave bid function in a first-price sealed-bid auction.}
		\label{fig:fpsb_concave}
	\end{figure}
	
	\begin{figure}[h]
		\centering
		\includegraphics[width=1.00\textwidth]{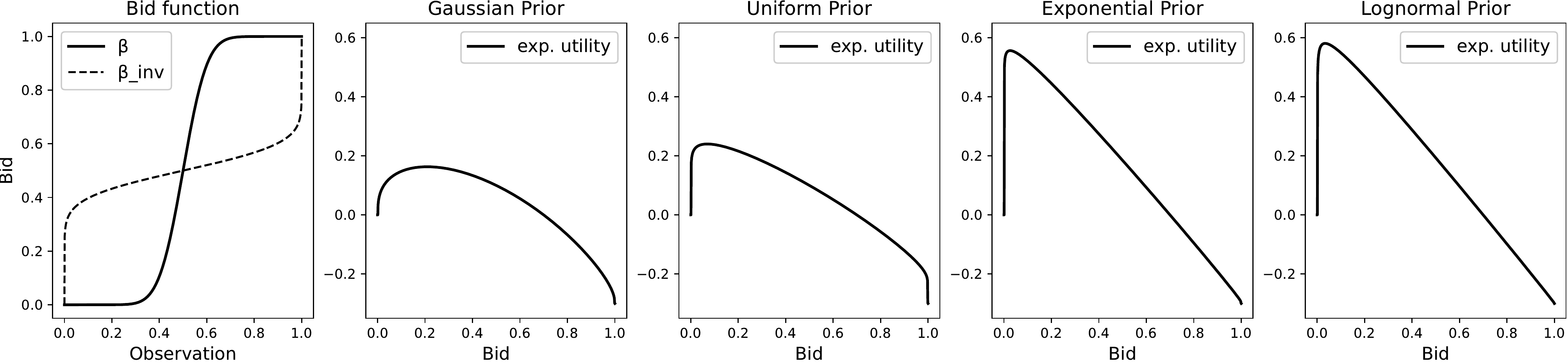}
		\caption{Expected utility with a sigmoid bid function in a first-price sealed-bid auction.}
		\label{fig:fpsb_sigmoid}
	\end{figure}

	\section{Monotonicity and Variational Stability in the Approximation Game}\label{app:variational}
	
	In this section we want to give evidence that variational stability and thereby monotonicity are not satisfied globally in our settings as discussed in Section \ref{sec:convergence}. For this we consider a simple example of a FPSB with two symmetric bidders. 
	
	Consider our approximation game $ \Gamma = (\Ical, \Scal^d, \tilde u) $ as defined in Definition \ref{def:approx_game}. Since the sets of discrete distributional strategies $ \Scal^d_i $ are compact, convex subsets of $ \R^{K \times L} $ and the utility functions $ \tilde u_i $ are linear and therefore concave in $ s_i $, we have a continuous, concave game as defined in \citep{mertikopoulos2019learning}. 
	\citet{rosen1965existence} refers to such games as n-person concave games.
	In this setting, Nash equilibria $s^* \in \Scal^d $ are precisely the solutions of the corresponding variational inequality $VI(F,\Scal^d)$
	\begin{equation}\label{eq:vi} \tag{VI}
		\langle F(s^*), s - s^* \rangle \leq 0, \quad  \forall s \in \Scal^d,
	\end{equation}
	with $ F = (F_i)_{i \in \Ical} $ and $ F_i(s) := \nabla_i \tilde u_i (s_i, s_{-i}) $. In the following, we want to give a short overview of the relevant definitions.
	
	% Monotonicity
	The game is said to satisfy the \textit{payoff monotonicity condition (MC)}, if
	\begin{equation}\label{eq:mc} \tag{MC}
		\langle F(s) - F(s'), s - s' \rangle =  \sum_i \langle \nabla_i \tilde u_i(s_i,s_{-i}) - \nabla_i \tilde u_i(s_i',s_{-i}'), s_i - s_i' \rangle \leq 0, \quad \forall s,s' \in \Scal^d
	\end{equation}
	with equality if and only if $ s = s' $. \citet{rosen1965existence} uses this property, which he calls diagonally strict concavity, and shows that if the game satisfies (MC), it admits a unique Nash equilibrium. In the literature on variational inequalities, this is also known as strict monotonicity \citep{facchinei2003}. 
	% Variational Inequalitey
	Furthermore, we say that a strategy profile $ s^* \in \Scal^d $ is \textit{variationally stable (VS)}, if there exists a neighborhood $ S \subseteq \Scal^d $ such that
	\begin{equation}\label{eq:vs} \tag{VS}
		\langle F(s), s-s^* \rangle = \sum_i \langle \nabla_i \tilde u_i(s_i,s_{-i}), s_i - s_i^* \rangle \leq 0, \quad \forall s \in S
	\end{equation}
	with equality if and only if $ s = s^* $. In terms of variational inequalities, global variational stable points are in the set of weak solutions of the corresponding variational inequality $VI(F,\Scal^d)$.
	\citet{mertikopoulos2019learning} extend results of monotone games and show that the weaker concept of (VS) suffices to get convergence for the no-regret algorithm dual averaging.
	
	We will now show that, even in the simplest setting, variational stability is not satisfied, and therefore convergence does not follow from these results. Let us consider a first-price sealed bid with two symmetric bidders and i.i.d. observations (valuations) $o \sim U([0,1]$. The observation and action space are discretized equally with $ K=L $ points, i.e., $ \Ocal_i^d = \Acal_i^d = \{0, \tfrac{1}{K-1}, \dots, \tfrac{K-2}{K-1}, 1\} $. Similar to our numerical experiments, we assume a tie-breaking rule where bidders win only if their bid is strictly greater than the opponents' bids. In that case, the discretized game has two symmetric, pure equilibria which basically correspond to $\beta_1(o) = \ulcorner \tfrac{o}{2}\urcorner $ and $ \beta_2(o) = \llcorner \tfrac{o}{2} \lrcorner $ \citep{rasooly2021}.
	
	\begin{figure}[h]
		\FIGURE
		{
			\includegraphics[width = .3\textwidth]{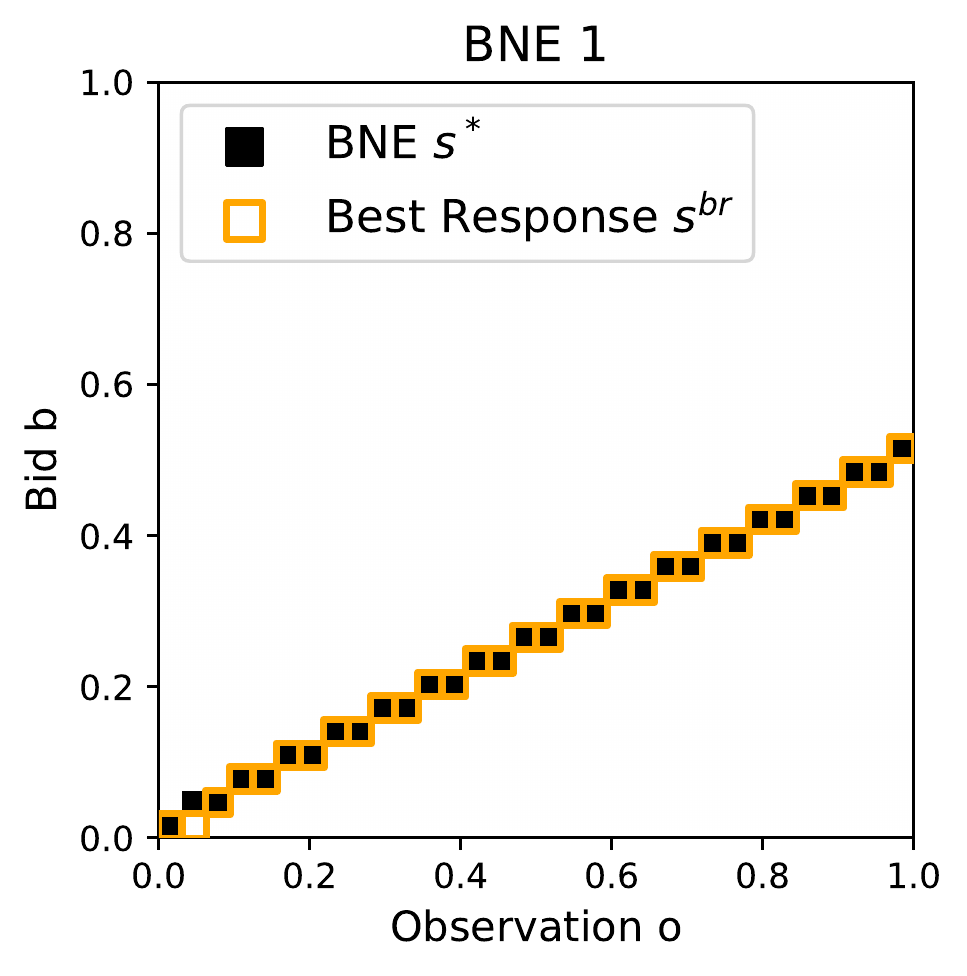}
			\includegraphics[width = .3\textwidth]{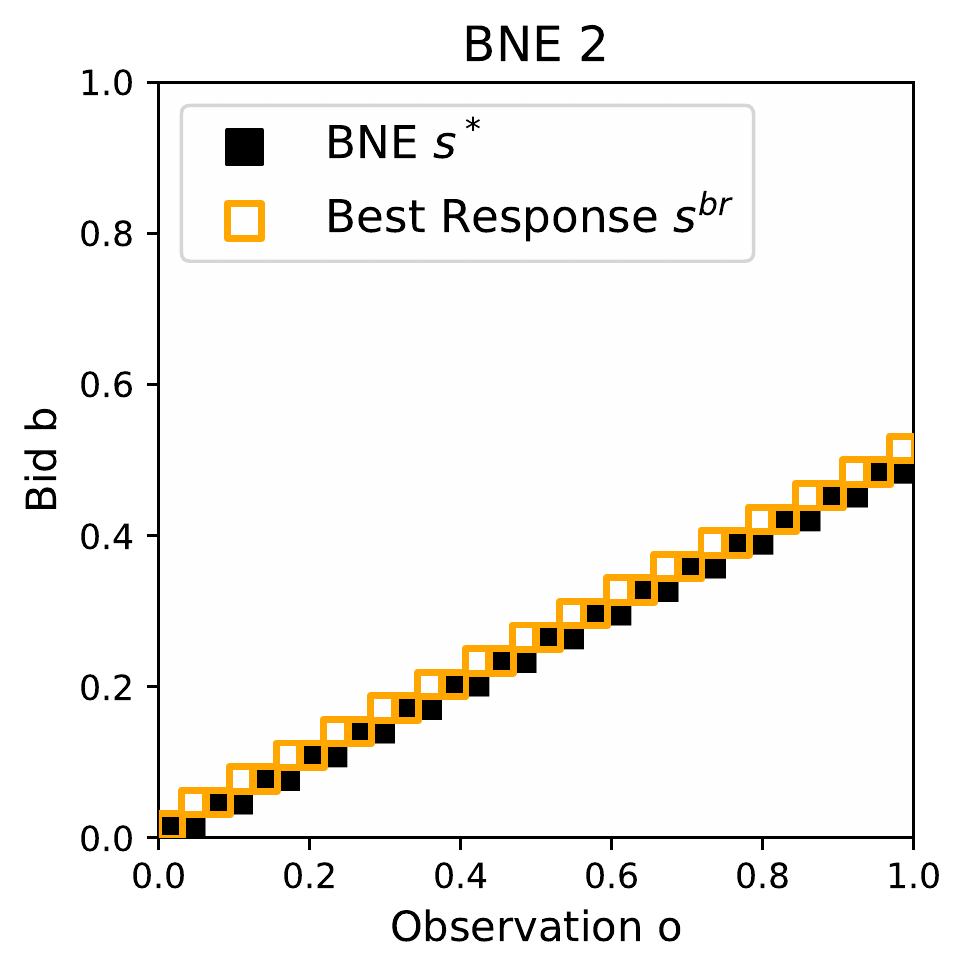}
			\includegraphics[width = .3\textwidth]{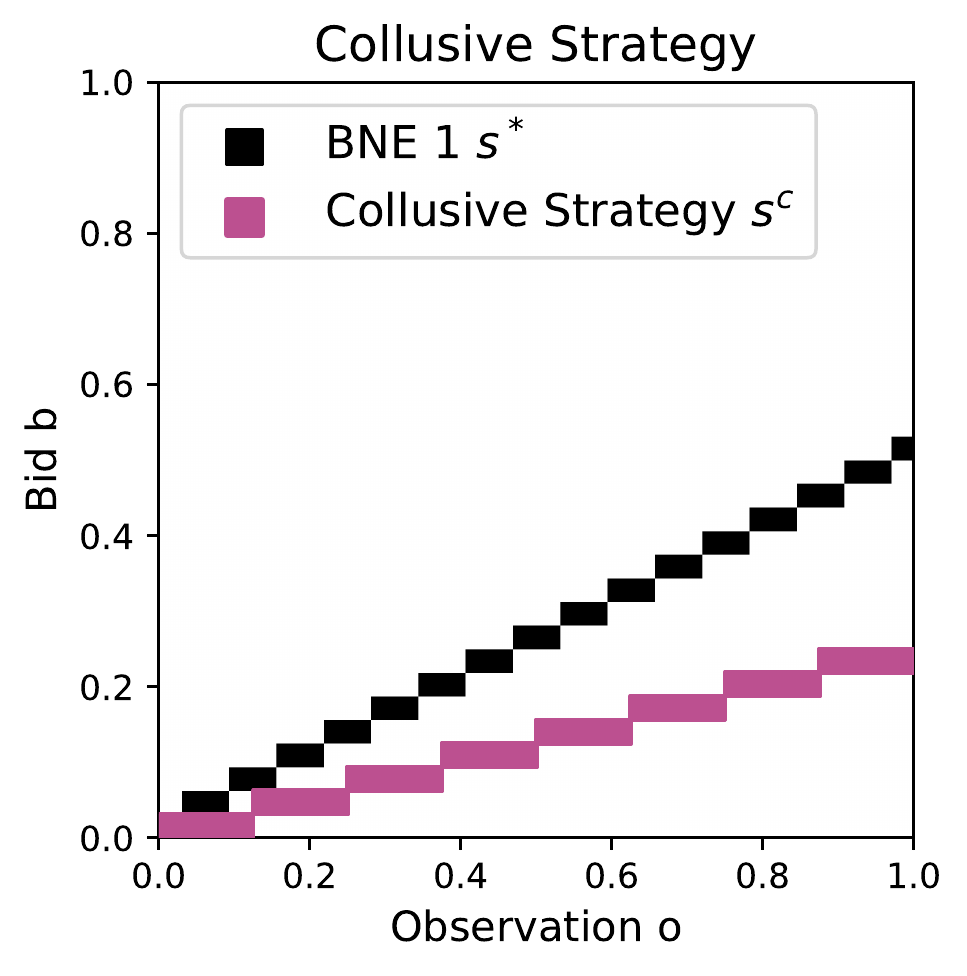}
		}
		{
			Counterexample for variational stability. \label{fig:example_vs}
		}
		{
			In the first two plots, we see the two discrete distributional BNE for the 2-player FPSB, which correspond to $ \beta_1 $ and $ \beta_2 $ indicated by the black squares. While the equilibrium strategies are obviously best responses to themselves, the yellow squares indicate alternative best responses, which makes the BNE non-strict.
			On the third plot we illustrate a BNE (black) and a collusive strategy (purple). For these strategies we observe that $ \tilde u_i(s_i^c,s_{-i}^c) > u_i(s_i^*,s_{-i}^c) $, i.e., unilaterally deviating from the collusive strategy profile to the BNE descreases the utility the bidder. This makes the collusive strategy profile a point, where VS w.r.t. the BNE is not satisfied.
		}
	\end{figure}
	
	First, we observe that there are two equilibria, which immediately proves that the game cannot satisfy (\ref{eq:mc}). Second, none of the BNE is globally variationally stable. If both bidders stick to the collusive strategy $ s_i^c $ (i.e., bid approximately $ \tfrac{1}{4} o_i $) each agent ends up with a higher utility, than in a situation where the agent deviates to a BNE $ s_i^* $. This means
	\begin{align*}
		\tilde u_i(s_i^*, s_{-i}^c) &< \tilde u_i (s_i^c, s_{-i}^c)
		\intertext{By symmetry, this stays true if we sum over $ i $. Since the utility $ \tilde u_i $ is linear in $ s_i $ the gradient does not depend on $ s_i $ and we can write}
		\sum_i \langle \nabla_i \tilde u_i(s_i^*, s_{-i}^c), s_i^* \rangle = \sum_i \langle \nabla_i \tilde u_i(s_i^c, s_{-i}^c), s_i^* \rangle &< \sum_i \langle \nabla_i \tilde u_i(s_i^c, s_{-i}^c), s_i^c \rangle.
		\intertext{Rearranging terms, we get an inequality which contradicts (\ref{eq:vs}) }
		\sum_i \langle \nabla_i \tilde u_i(s_i^*,  s_{-i}^c),  s_i^c - s_i^* \rangle &> 0.
	\end{align*}
	Therefore, the equilibrium $ s^* $ cannot be globally variationally stable. Note that this line of argument does not rely on the non-uniqueness of the equilibrium, or the specific tie-breaking rule. 
	We only use individual linearity of the bidder's utility function, which is a consequence of the discretization, and the fact that we can construct collusive strategies, where deviating to the BNE reduces the expected utility.
	
	Specifically, in this example, we can further show that both equilibria are not even locally variationally stable.
	In Figure \ref{fig:example_vs} we illustrate that the equilibrium strategies are not strict since the best responses are not unique. By linearity of the utility functions, the inequality (VS) is equal to zero for a BNE $ s^* $ and any convex combination $ s = \lambda s^* + (1-\lambda ) s^{br} $ of $ s^* $ and its best response $ s^{br} \neq s^* $. This means that either the BNE is not locally variationally stable (for every neighborhood we can choose $ \lambda  $ small enough), or the BNE and its best response are elements of some larger variationally stable set. But in the latter case, we know that this set has to be a convex set of equilibria \citep[Prop.~2.7]{mertikopoulos2019learning}. And since we can verify numerically that for instance, a convex combination of BNE2 and its best response BNE1 is not an equilibrium, this cannot be the case. 
	
	In conclusion, we have a setting which is not monotonic and not even locally variationally stable, but in which our methods still approximates the BNE.
	
\end{APPENDICES}
\end{document}

%% file: tables/interdependent_models.tex
\begin{table}[h]
	\TABLE
	{Types of interdependencies.\label{tab:learning-algorithms}}
	{\setlength{\tabcolsep}{0.3em}
		
	\newcolumntype{R}{>{\raggedright\arraybackslash}m{4.5cm}}%
	\newcolumntype{C}{>{\centering\arraybackslash}m{4cm}}%
	\begin{tabular}{
			R C C
		}
		\hline
		\up\down & Private $v$'s ($v = o$) & Common $v$'s (CV) \\
		\hline
		\up\down Independent $o$'s (PDF of $o$ is product of marginal PDFs) & Independent private values model (IPV) & Independent non-private values or common values\\
		\hline\
		\up\down Correlated $o$'s & Correlated or affiliated (APV) private values & Correlated non-private values \\
		\hline
	\end{tabular}}
	{}
\end{table}

%% file: tables/learning_algorithms.tex
\begin{table}[h]
	\TABLE
	{Overview of simultaneous online (SO) gradient updates.\label{tab:update-steps}}
	{\setlength{\tabcolsep}{0.3em}
	\begin{tabular}{
			|>{\centering\arraybackslash}m{2cm}
			|>{\centering\arraybackslash}m{3cm}
			|>{\centering\arraybackslash}m{10cm}
			|
		}
		\hline
		\up & Method & Update Rule \\
		\hline
		\up\down $ \text{SODA}_1 $ $(\text{SOMA}_1) $ & Dual Averaging + entropic regularizer & 
		$
			 (s_{i,t+1})_{k l} = (f^d_{o_i})_k \dfrac{(s_{i,t})_{k l} \exp \left( \eta_t (c_{i,t})_{k l}\right)}{\sum \limits_{l'} (s_{i,t})_{k l'} \exp ( \eta_t (c_{i,t})_{k l'})} \quad \forall k, l
		$\\
		\hline
		\up\down$ \text{SODA}_2 $ & Dual Averaging +  Euclid. regularizer & 
		$
			\begin{array}{rl}
				y_{i,t+1} &= y_{i,t} + \eta_t c_{i,t} \\
				s_{i,t+1} &= \argmax \{ \Vert s - y_{i,t+1} \Vert_2^2 \text{ s.t. } s \in \Scal_i\}
			\end{array}
		$ \\
		\hline
		\up\down$ \text{SOMA}_2 $ & Mirror Ascent +  Euclid. mirror map & 
		$ 
			s_{i,t+1} = \argmax \{ \Vert s - (s_{i,t} + \eta_t c_{i,t}) \Vert_2^2 \text{ s.t. } s \in \Scal_i\}
		$ \\
		\hline
		\up\down$ \text{SOFW} $ & Frank-Wolfe & 
		$
			\begin{array}{rl}
				\text{br}_{i,t+1} &= \argmax \{ \langle c_{i,t}, s \rangle \text{ s.t. } s \in \Scal_i\} \\
				s_{i,t+1} &= (1-\eta_t) s_{i,t} + \eta_t \text{br}_{i,t+1}
			\end{array}
		$ \\
		\hline
	\end{tabular}}
	{
		Note that agents want to maximize their utilities in our examples, which is why we change all update rules to ascent instead of descent methods.
		The gradient for each agent is denoted by $ c_i := \nabla_{s_i}u_i(s_i, s_{-i}) $.
		We use a non-increasing sequence of step sizes $ \{ \eta_t \} $ of the form $ \eta_t = \eta_0 t^{-\beta} $ for some $ \beta \in (0,1] $ for DA and MD and the commonly used step size $ \eta_t = \tfrac{2}{1+t} $ for Frank-Wolfe. 
	}
\end{table}

%% file: tables/interdep_results_cv.tex
\begin{table}[h]
	\TABLE
	{Results for the Common Value Model. \label{tab:interdep_results_cv}}
	{
		\begin{tabular}{ l l r c c }
			\hline
			\up \down Algorithm & step size & runtime & $ \mathcal L $ & $ L_2 $  \\
			\hline			
			\up 
			$ \text{SODA}_1 $ 				& $ \beta=0.50, \, \eta_0=100 $ & 10-13 min	& 0.007 (0.001) & 0.034 (0.000) \\
			$ \text{SODA}_2 $  				& $ \beta=0.05, \, \eta_0=1 $	& 14-16 min & 0.003 (0.000) & 0.019 (0.000) \\
			$ \text{SOMA}_2 $  				& $ \beta=0.50, \, \eta_0=50 $ 	& 7-9 s		& 0.003 (0.000) & 0.018 (0.000) \\
			$ \text{SOFW} $  				& - 							& 2-14 min	& 0.000 (0.001) & 0.196 (0.002) \\
			\down FP 						& - 							& 9-14 min	& 0.000 (0.001) & 0.439 (0.014) \\
			\down NPGA						& -								& 15 min	& 0.000 (0.000) & 0.009 (0.002)	\\
			\hline				
		\end{tabular}
	}
	{The mean (and standard deviation) of the approximated utility loss $ \mathcal L $ and $L_2$ distance, as well as the step size and runtime is reported.}
\end{table}

%% file: tables/interdep_results_av.tex
\begin{table}[h]
	\TABLE
	{Results for the Affiliated Values Model. \label{tab:interdep_results_av}}
	{
		\begin{tabular}{ l l r c c }
			\hline
			\up \down Algorithm & step size & runtime & $ \mathcal L $ & $ L_2 $  \\
			\hline
			\up 
			$ \text{SODA}_1 $ 				& $ \beta=0.5, \, \eta_0=100 $ 	& 15-16 s	& 0.002 (0.000) & 0.014 (0.000) \\
			$ \text{SODA}_2 $  				& $ \beta=0.5, \, \eta_0=1 $	& 11-12 s 	& 0.002 (0.000) & 0.012 (0.000) \\
			$ \text{SOMA}_2 $  				& $ \beta=0.5, \, \eta_0=1 $ 	& 11 s		& 0.002 (0.000) & 0.014 (0.000) \\ 
			$ \text{SOFW} $  				& - 							& 11 s		& 0.004 (0.001) & 0.020 (0.002) \\
			\down $ \text{FP} $  			& - 							& 12-13	s	& 0.005 (0.000) & 0.025 (0.001) \\
			\down NPGA						& -								& 15 min	& 0.002 (0.001) & 0.018 (0.009)	\\
			\hline
					
		\end{tabular}
	}
	{The mean (and standard deviation) of the approximated utility loss $ \mathcal L $ and $L_2$ distance, as well as the step size and runtime is reported.}
\end{table}

%% file: tables/llg_results_nz.tex
\begin{table}[h]
	\TABLE
	{Results for the local bidders in the LLG Model with  Nearest-Zero Rule. \label{tab:llg_nz}}
	{
		\begin{tabular}{ l c c c c c c  }
			\hline
			\up\down \multirow{2}{*}{Algorithm}  	&  \multicolumn{2}{c}{$ \gamma =  0.1 $ } &  \multicolumn{2}{c}{$ \gamma =  0.5 $ }  &  \multicolumn{2}{c}{$ \gamma =  0.9 $ }  \\ 
			\cline{2-7}
			\up\down  & $ \mathcal L $ & $ L_2 $ & $ \mathcal L $ & $ L_2 $  & $ \mathcal L $ & $ L_2 $ \\ 
			\hline			 
			\up	$ \text{SODA}_1 $ 	& 0.002 (0.000) & 0.022 (0.001) & 0.001 (0.000) & 0.022 (0.001) & 0.000 (0.000) & 0.025 (0.000) \\
			$ \text{SODA}_2 $ 		& 0.002 (0.000) & 0.021 (0.001) & 0.001 (0.000) & 0.024 (0.002) & 0.000 (0.000) & 0.025 (0.001) \\
			$ \text{SOMA}_2 $ 		& 0.002 (0.000) & 0.018 (0.002) & 0.001 (0.000) & 0.019 (0.001) & 0.000 (0.000) & 0.021 (0.000) \\
			SOFW					& 0.002 (0.000) & 0.018 (0.000) & 0.001 (0.000) & 0.023 (0.000) & 0.000 (0.000) & 0.034 (0.000) \\
			\down FP 				& 0.002 (0.000) & 0.021 (0.000) & 0.001 (0.000) & 0.023 (0.000) & 0.000 (0.000) & 0.028 (0.000) \\
			\down NPGA				& -				& - 			& 0.000 (0.000) & 0.011 (0.005) & -				& -				\\
			\hline				
		\end{tabular}
	}
	{We report the mean (and standard deviation) over ten runs for the utility loss $ \mathcal L $ and $ L_2 $ distance. $ \text{SODA}_1 $ takes 10-34 seconds, $ \text{SODA}_2 $ 1-6 seconds, and FP 31-39 seconds per run. All other methods run for less than 1 second.}
\end{table}

%% file: tables/llg_results_nvcg.tex
\begin{table}[h]
	\TABLE
	{Results for the local bidders in the LLG Model with Nearest-VCG Rule. \label{tab:llg_nvcg}}
	{
		\begin{tabular}{ l c c c c c c  }
			\hline
			\up\down \multirow{2}{*}{Algorithm}  	&  \multicolumn{2}{c}{$ \gamma =  0.1 $ } &  \multicolumn{2}{c}{$ \gamma =  0.5 $ }  &  \multicolumn{2}{c}{$ \gamma =  0.9 $ }  \\ 
			\cline{2-7}
			\up\down  & $ \mathcal L $ & $ L_2 $ & $ \mathcal L $ & $ L_2 $  & $ \mathcal L $ & $ L_2 $ \\ 
			\hline			 
		\up	$ \text{SODA}_1 $ 	& 0.001 (0.000) & 0.017 (0.001) & 0.001 (0.000) & 0.017 (0.001) & 0.001 (0.000) & 0.021 (0.001) \\
			$ \text{SODA}_2 $ 	& 0.001 (0.000) & 0.017 (0.000) & 0.001 (0.000) & 0.016 (0.000) & 0.000 (0.000) & 0.016 (0.000) \\
			$ \text{SOMA}_2 $ 	& 0.001 (0.000) & 0.015 (0.001) & 0.000 (0.000) & 0.014 (0.001) & 0.000 (0.000) & 0.016 (0.001) \\
	    	SOFW				& 0.001 (0.000) & 0.015 (0.000) & 0.000 (0.000) & 0.015 (0.000) & 0.000 (0.000) & 0.016 (0.000) \\
	    \down FP 				& 0.001 (0.000) & 0.019 (0.000)	& 0.001 (0.000) & 0.018 (0.000)	& 0.001 (0.000) & 0.019 (0.000)	\\
	    \down NPGA 				& -				& -				& 0.000 (0.000) & 0.016 (0.016)	& - 			& -	\\
			\hline				
		\end{tabular}
	}
	{We report the mean (and standard deviation) over ten runs for the utility loss $ \mathcal L $ and $ L_2 $ distance. $ \text{SODA}_1 $ takes for 8-16 seconds and FP up to 47 seconds to compute one strategy, while all other methods run for less than 2 seconds.}
\end{table}

%% file: tables/llg_results_nb.tex
\begin{table}[h]
	\TABLE
	{Results for the local bidders in the LLG Model with  Nearest-Bid Rule. \label{tab:llg_nb}}
	{
		\begin{tabular}{ l c c c c c c  }
			\hline
			\up\down \multirow{2}{*}{Algorithm}  	&  \multicolumn{2}{c}{$ \gamma =  0.1 $ } &  \multicolumn{2}{c}{$ \gamma =  0.5 $ }  &  \multicolumn{2}{c}{$ \gamma =  0.9 $ }  \\ 
			\cline{2-7}
			\up\down  & $ \mathcal L $ & $ L_2 $ & $ \mathcal L $ & $ L_2 $  & $ \mathcal L $ & $ L_2 $ \\ 
			\hline			 
			\up	$ \text{SODA}_1 $ 	& 0.001 (0.000) & 0.014 (0.001) & 0.001 (0.000) & 0.015 (0.002) & 0.001 (0.001) & 0.017 (0.001) \\
			$ \text{SODA}_2 $ 		& 0.001 (0.000) & 0.013 (0.000) & 0.000 (0.000) & 0.008 (0.000) & 0.000 (0.000) & 0.009 (0.001) \\
			$ \text{SOMA}_2 $ 		& 0.000 (0.000) & 0.012 (0.000) & 0.000 (0.000) & 0.008 (0.001) & 0.000 (0.000) & 0.009 (0.000) \\
			SOFW  					& 0.000 (0.000) & 0.013 (0.001) & 0.000 (0.000) & 0.009 (0.001) & 0.000 (0.000) & 0.012 (0.000) \\
			\down FP				& 0.001 (0.000)	& 0.017 (0.000) & 0.001 (0.000)	& 0.015 (0.000) & 0.001 (0.000) & 0.016 (0.000) \\
			\down NPGA				& -				& - 			& 0.001 (0.000)	& 0.021 (0.021) & - 			& - \\
			\hline				
		\end{tabular}
	}
	{We report the mean (and standard deviation) over ten runs for the utility loss $ \mathcal L $ and $ L_2 $ distance. $ \text{SODA}_1 $ takes 8-23 seconds and FP 31-39 seconds per run, while all other methods run for less than 2 seconds.}
\end{table}

%% file: tables/sa_results_gaussian.tex
\begin{table}[h]
	\TABLE
	{Results for the FPSB split-award auction with a truncated Gaussian prior. \label{tab:sa_gaussian}}
	{
		\begin{tabular}{ l l r c c }
			\hline
			\up\down Algorithm & step size & time & $ \mathcal L $ & $ L_2 $ \\ 
			\hline
			\up \down 
			$ \text{SODA}_1 $ 				& $ \beta=0.05, \, \eta_0=20 $ 		& 3-5 min	& -0.064 (0.001)	& 0.050 (0.010)	\\
			$ \text{SODA}_2 $  				& $ \beta=0.05, \, \eta_0=0.05 $	& 3-4 min 	& -0.077 (0.001)	& 0.067 (0.005) \\
			$ \text{SOMA}_2 $  				& $ \beta=0.50, \, \eta_0=0.05 $ 	& 7-6 min	& -0.086 (0.001)	& 0.100 (0.009) \\
			$ \text{SOFW} $  				& - 								& 7-11 min	&  0.031 (0.075)	& 0.029 (0.008) \\
			\down FP 						& - 								& 7-12 min	&  0.194 (0.024)	& 0.078 (0.006) \\
			\hline					
		\end{tabular}
	}
	{
	The mean (and standard deviation) of the approximated utility loss $ \mathcal L $ and $L_2$ distance (only for the 50\% share), as well as the step size and runtime are reported.
	}
\end{table}

%% file: tables/sa_results_uniform.tex
\begin{table}[h]
	\TABLE
	{Results for the FPSB split-award auction with a uniform prior. \label{tab:sa_uniform}}
	{
		\begin{tabular}{ l l r c c }
			\hline
			\up\down Algorithm & step size & runtime & $ \mathcal L $ & $ L_2 $ \\ 
			\hline
			\up \down 
			$ \text{SODA}_1 $ 				& $ \beta=0.05, \, \eta_0=20 $ 		& 2 min		& 0.009 (0.000)	& 0.024 (0.028)	\\
			$ \text{SODA}_2 $  				& $ \beta=0.05, \, \eta_0=0.05 $	& 2-3 min 	& 0.009 (0.000)	& 0.015 (0.000) \\
			$ \text{SOMA}_2 $  				& $ \beta=0.50, \, \eta_0=0.01 $ 	& 7-9 min	& 0.029 (0.002)	& 0.097 (0.016) \\
			$ \text{SOFW} $  				& - 								& 7-8 min	& 0.191 (0.032) & 0.075 (0.010) \\
			\down FP 						& - 								& 7-8 min	& 0.177 (0.031)	& 0.039 (0.008) \\
			\hline					
		\end{tabular}
	}
	{The mean (and standard deviation) of the approximated utility loss $ \mathcal L $ and $L_2$ distance (only for the 50\% share), as well as the step size and runtime are reported.}
\end{table}

%% file: tables/fpsb_risk.tex
\begin{table}[h]
	\TABLE
	{Results for risk-avers bidders in the FPSB auction with different risk parameter $ \rho $. \label{tab:risk}}
	{
		\begin{tabular}{ l c c c c c c  }
			\hline
			\up\down \multirow{2}{*}{Algorithm}  	&  \multicolumn{2}{c}{$ \rho =  0.5 $ } &  \multicolumn{2}{c}{$ \rho =  0.7 $ }  &  \multicolumn{2}{c}{$ \rho =  0.9 $ }  \\ 
			\cline{2-7}
			\up\down  & $ \mathcal L $ & $ L_2 $ & $ \mathcal L $ & $ L_2 $  & $ \mathcal L $ & $ L_2 $ \\ 
			\hline			 
			\up	$ \text{SODA}_1 $ 	& 0.001 (0.000) & 0.007 (0.000) & 0.001 (0.000) & 0.007 (0.000) & 0.001 (0.000) & 0.008 (0.000) \\
			$ \text{SODA}_2 $ 		& 0.001 (0.000) & 0.007 (0.000) & 0.001 (0.000) & 0.007 (0.000) & 0.001 (0.000) & 0.008 (0.001) \\
			$ \text{SOMA}_2 $ 		& 0.001 (0.000) & 0.007 (0.000) & 0.001 (0.000) & 0.007 (0.000) & 0.001 (0.000) & 0.008 (0.000) \\
			SOFW					& 0.001 (0.000) & 0.008 (0.000) & 0.001 (0.000) & 0.008 (0.000) & 0.001 (0.000) & 0.009 (0.000) \\
			\down FP 				& 0.002 (0.000) & 0.013 (0.000) & 0.002 (0.000) & 0.013 (0.000) & 0.003 (0.000) & 0.013 (0.001) \\
			\hline				
		\end{tabular}
	}
	{We report the mean (and standard deviation) over ten runs for the utility loss $ \mathcal L $ and $ L_2 $ distance. The runtime for all methods is less than 1 second per run.}
\end{table}

%% file: tables/discr_approx.tex
\begin{table}[h]
	\TABLE
	{Results for the FPSB with two bidders and different discretizations. \label{tab:discr_approx}}
	{
		\begin{tabular}{ l c c c c c }
			\hline
			\up\down $ K,L \quad $ & 16 	&32 & 64 & 128 & 256 \\ 
			\hline
			\up  	
			Utility Loss $ \mathcal L $	& 0.030 (0.001)	& 0.008 (0.000) & 0.002 (0.000) & 0.001 (0.000) & 0.001 (0.000) \\
			$ L_2 $ Distance			& 0.036 (0.001) & 0.018 (0.000) & 0.010 (0.001) & 0.008 (0.000  & 0.006 (0.000) \\
			\hline					
		\end{tabular}
	}
	{The mean (and standard deviation) of the approximated utility loss $ \mathcal L $ and $L_2$ distance over ten runs is reported.}
\end{table}